\documentclass[a4paper,fleqn,usenatbib]{mnras}

\usepackage{newtxtext,newtxmath}

\usepackage[T1]{fontenc}
\usepackage{ae,aecompl}

\usepackage{graphicx}	

\usepackage{amsmath}	
\usepackage{amssymb}	
\usepackage{epstopdf}	

\usepackage{soul}
\usepackage{pifont}

\usepackage{hyperref}

\newcommand{\gotheref}[1]{\hyperref[#1]{\scriptsize{\textcolor{blue}{[go there]}}}}
\newcommand{\gobackref}[1]{\hyperref[#1]{\scriptsize{\textcolor{blue}{[go back]}}}}

\newcommand{\cmark}{\ding{51}}%
\newcommand{\xmark}{\ding{55}}%
\newcommand{\LCDM}{$\Lambda$CDM}%

\newcommand{\rhalo}{r_\text{halo}}%

\newcommand{\voutflowWIM}{v_\text{outflow, WIM}}
\newcommand{\voutflowWNM}{v_\text{outflow, WNM}}
\newcommand{\vrad}{v_\text{r}}
\newcommand{\vvir}{v_\text{vir}}
\newcommand{\vcirc}{v_\text{circ}}
\newcommand{\vesc}{v_\text{esc}}

\newcommand{\rhogas}{\rho_\text{gas}}%

\newcommand{\etaWIM}{\eta_\text{WIM}}%
\newcommand{\etaHalpha}{\eta_{\text{H}\alpha}}%
\newcommand{\LHalpha}{L_{\text{H}\alpha\text{, broad}}}%
\newcommand{\vHalpha}{v_{\text{H}\alpha\text{, broad}}}%
\newcommand{\vbroad}{v_{\text{broad}}}%
\newcommand{\Moutflow}{M_{\text{outflow}}}%
\newcommand{\kappaCR}{\kappa_\parallel}%
\newcommand{\poutflow}{p_\text{outflow}}%
\newcommand{\vbroadWNM}{v_\text{WNM, broad}}%
\newcommand{\nuoutflow}{\tilde{v}_\text{outflow, WIM}}%
\newcommand{\aveZstar}{\left<Z_*\right>}%

\newcommand{\tablesymbol}[1]{\hspace{-0.3cm}\includegraphics[height=0.25cm]{Images/Symbols/#1.pdf}\hspace{-0.5cm}}

\newcommand{\HDthSfNoFb}{NoFb+thSf}
\newcommand{\HDthSfFb}{HD+thSf}
\newcommand{\HDthSfFbBoost}{HD+thSfBoost}

\newcommand{\HDSfNoFb}{NoFb}
\newcommand{\HDSfFb}{HD}
\newcommand{\HDSfFbBoost}{HD+Boost}
\newcommand{\MHDSfFb}{MHD}

\newcommand{\iMHDSfFb}{iMHD}
\newcommand{\RTSfFb}{RT}

\newcommand{\RTiMHDSfFb}{RTiMHD}

\newcommand{\RTnsCRiMHDSfFb}{RTnsCRiMHD}
\newcommand{\RTCRiMHDSfFb}{RTCRiMHD}
\newcommand{\fullphys}{`full-physics'}

\newcommand{\pandora}{Pandora}
\newcommand{\paperI}{Pandora I}
\newcommand{\ramses}{\textsc{ramses}}

\newcommand{\Msun}{\,\mathrm{M_\odot}}
\newcommand{\erg}{\,\mathrm{erg}}
\newcommand{\cm}{\,\mathrm{cm}}
\newcommand{\km}{\,\mathrm{km}}
\newcommand{\pc}{\,\mathrm{pc}}
\newcommand{\kpc}{\,\mathrm{kpc}}
\newcommand{\Kelvin}{\,\mathrm{K}}
\newcommand{\K}{\Kelvin}
\newcommand{\s}{\,\mathrm{s}}
\newcommand{\yr}{\,\mathrm{yr}}
\newcommand{\Myr}{\,\mathrm{Myr}}
\newcommand{\Gyr}{\,\mathrm{Gyr}}
\newcommand{\gram}{\,\mathrm{g}}

\newcommand{\kmps}{\,\km\,\s^{-1}}
\newcommand{\gcc}{\,\gram\,\cm^{-3}}

\newcommand{\rowlegend}{{\vspace{-0.1cm}\includegraphics[width=0.9\columnwidth]{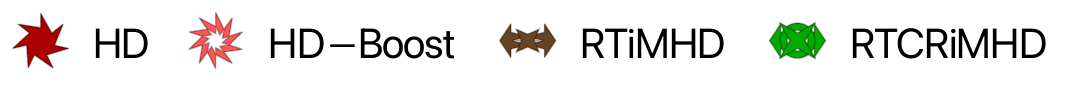}}\vspace{-0.3cm}}

\newcommand{\allrowlegendtwoshort}{{\vspace{-0.1cm}\includegraphics[width=\columnwidth]{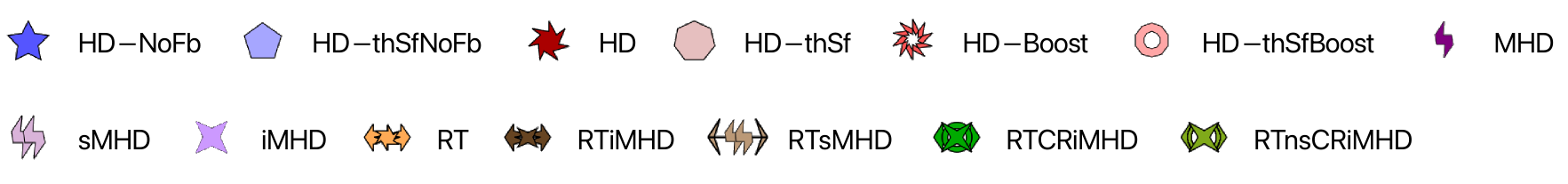}}\vspace{-0.3cm}}

\usepackage{scalerel,tikz}
\usetikzlibrary{svg.path}
\definecolor{orcidlogocol}{HTML}{A6CE39}
\tikzset{orcidlogo/.pic={
 \fill[orcidlogocol] svg{M256,128c0,70.7-57.3,128-128,128C57.3,256,0,198.7,0,128C0,57.3,57.3,0,128,0C198.7,0,256,57.3,256,128z};
 \fill[white] svg{M86.3,186.2H70.9V79.1h15.4v48.4V186.2z}
 svg{M108.9,79.1h41.6c39.6,0,57,28.3,57,53.6c0,27.5-21.5,53.6-56.8,53.6h-41.8V79.1z M124.3,172.4h24.5c34.9,0,42.9-26.5,42.9-39.7c0-21.5-13.7-39.7-43.7-39.7h-23.7V172.4z}
 svg{M88.7,56.8c0,5.5-4.5,10.1-10.1,10.1c-5.6,0-10.1-4.6-10.1-10.1c0-5.6,4.5-10.1,10.1-10.1C84.2,46.7,88.7,51.3,88.7,56.8z};
}}
\newcommand\orcid[1]{\href{https://orcid.org/#1}{\mbox{\scalerel*{
\begin{tikzpicture}[yscale=-1,transform shape]
\pic{orcidlogo};
\end{tikzpicture}
}{|}}}}

\title[Dwarf galaxy outflows and non-thermal physics]{The Pandora project. II: how non-thermal physics drives bursty star formation and temperate mass-loaded outflows in dwarf galaxies}

\author[Martin-Alvarez et al.]{Sergio Martin-Alvarez$^{1}$\thanks{E-mail: martin-alvarez@stanford.edu (SMA)}\orcid{0000-0002-4059-9850},
Debora Sijacki$^{2, 3}$,
Martin G. Haehnelt$^{2, 3}$,
Alice Concas$^{4}$,
Yuxuan Yuan$^{2, 3}$\orcid{0000-0001-6816-0682},
\newauthor
Roberto Maiolino$^{2, 5}$,
Risa H. Wechsler$^{1, 6, 7}$,
Francisco Rodr\'iguez Montero$^{8}$\orcid{0000-0001-6535-1766},
Marion Farcy$^{9}$,
Mahsa Sanati$^{10}$,
\newauthor
Yohan Dubois$^{11}$,
Joki Rosdahl$^{12}$,
Enrique Lopez-Rodriguez$^{13, 1}$\orcid{0000-0001-5357-6538},
and Susan E. Clark$^{1, 6}$\orcid{0000-0002-7633-3376}
\\
$^{1}$Kavli Institute for Particle Astrophysics \& Cosmology (KIPAC), Stanford University, Stanford, CA 94305, USA\\
$^{2}$Kavli Institute for Cosmology (KICC), University of Cambridge, Madingley Road, Cambridge CB3 0HA, UK\\
$^{3}$Institute of Astronomy, University of Cambridge, Madingley Road, Cambridge CB3 0HA, UK\\
$^{4}$European Southern Observatory, Karl-Schwarzschild-Strasse 2, D-85748 Garching bei Muenchen, Germany\\
$^{5}$Cavendish Laboratory, University of Cambridge, 19 J. J. Thomson Ave., Cambridge CB3 0HE, UK\\
 $^{6}$Department of Physics, Stanford University, Stanford, CA 94305, USA\\
$^{7}$SLAC National Accelerator Laboratory, Menlo Park, CA 94025, USA\\
$^{8}$Kavli Institute for Cosmological Physics (KICP), The University of Chicago, IL 60637, USA\\
$^{9}$Institute of Physics, Laboratory for Galaxy Evolution, EPFL, Observatoire de Sauverny, Chemin Pegasi 51, 1290 Versoix, Switzerland\\
$^{10}$Department of Physics, University of Oxford, Keble Road, Oxford OX1 3RH, UK\\
$^{11}$Institut d’Astrophysique de Paris, UMR 7095, CNRS and Sorbonne Universit\'{e}, 98 bis boulevard Arago, 75014 Paris, France\\
$^{12}$Univ Lyon, Univ Lyon1, Ens de Lyon, CNRS, Centre de Recherche Astrophysique de Lyon UMR5574, F-69230 Saint-Genis-Laval, France\\
$^{13}$Department of Physics \& Astronomy, University of South Carolina, Columbia, SC 29208, USA\\
}

\date{MNRAS, submitted}

\pubyear{2025}

\begin{document}

\newcommand{\papertitle}[2]{\textbf{#1 \citep{#2}}.\newline}

\label{firstpage}
\pagerange{\pageref{firstpage}--\pageref{lastpage}}
\maketitle

\begin{abstract}
Dwarf galaxies provide powerful laboratories for studying galaxy formation physics. Their early assembly, shallow gravitational potentials, and bursty, clustered star formation histories make them especially sensitive to the processes that regulate baryons through multi-phase outflows. Using high-resolution, cosmological zoom-in simulations of a dwarf galaxy from \textit{the Pandora suite}, we explore the impact of stellar radiation, magnetic fields, and cosmic ray feedback on star formation, outflows, and metal retention. We find that our purely hydrodynamical model without non-thermal physics {--} in which supernova feedback is boosted to reproduce realistic stellar mass assembly {--} drives violent, overly enriched outflows that suppress the metal content of the host galaxy. Including radiation reduces the clustering of star formation and weakens feedback. However, the additional incorporation of cosmic rays produces fast, mass-loaded, multi-phase outflows consisting of both ionized and neutral gas components, in better agreement with observations. These outflows, which entrain a denser, more temperate ISM, exhibit broad metallicity distributions while preserving metals within the galaxy. Furthermore, the star formation history becomes more bursty, in agreement with recent JWST findings. These results highlight the essential role of non-thermal physics in galaxy evolution and the need to incorporate it in future galaxy formation models.
\end{abstract}

\definecolor{brown}{rgb}{0.5, 0.3, 0.0}
\definecolor{orange}{rgb}{0.8, 0.5, 0.0}


\section{Introduction}
\label{s:Introduction}
Dwarf galaxies are the primary dwellers of small dark matter haloes, characterised by their pervasive abundance and shallow gravitational potential wells. An in-depth understanding of dwarf galaxies remains a pressing concern, as they provide a unique probe of our $\Lambda$ cold dark matter (\LCDM) model via their abundance, distribution topology on large scales as well as internal structural properties \citep[e.g.,][]{Garrison-Kimmel2014, Onorbe2015, Sawala2016, Bullock2017, Geha2017, Jethwa2018}. In a paradigm where galaxy formation proceeds hierarchically, dwarf galaxies constitute the fundamental building blocks of their more massive counterparts. Consequently, dwarf systems not only serve as simple laboratories to investigate the processes of star formation and baryonic feedback, but also allow exploring the physics at play during the formation of the first galaxies in our Universe \citep{Geha2012, Rey2020, Gelli2020, Santistevan2020, Sanati2020, Gutcke2022, Koudmani2022, Martin-Alvarez2023, Sanati2024, Kim2024preprint}. Their often initially rapid stellar mass growth, shallow gravitational potential wells, simplified environments, and pristine gas reservoirs particularly devoid of metals make them unique candidates to investigate the cycle of baryons in galaxies and the intricate interrelation between star formation, galactic winds, and galaxy metallicities.

Star formation is one of the fundamental processes regulating galaxy formation and evolution, not only by converting gas into stars, but also by enriching the ISM with metals.
Runaway star formation is believed to be largely thwarted by internal galaxy feedback processes, allowing its self-regulation. In dwarf galaxies, the primary drivers of this feedback include stellar radiation \citep{Rosdahl2015b, Geen2015b, Geen2017, Grudic2021, Martin-Alvarez2023}, stellar winds \citep{Agertz2013, Fichtner2024}, and supernova (SN) explosions \citep[e.g.,][]{White1978, Springel2003b, Dubois2008, Hopkins2012a, Teyssier2013, Rosdahl2017, Smith2019, Martin-Alvarez2023}. 

In addition to the local suppression of star formation, these feedback processes are the engines of the baryon cycle, expelling gas into the circumgalactic medium (CGM) and even the intergalactic medium (IGM) \citep{Brooks2007, Chisholm2015, Heckman2015, Hayward2017, Zheng2024}. Crucially, these galactic outflows will shape the properties of the CGM gas as their mass-loading, energetics, and metal enrichment influence the hydrodynamical evolution of CGM gas. This, in turn, will determine whether heating or cooling processes dominate, and the timescales for the eventual gas re-accretion that further fuels the baryonic and star formation cycle \citep{Keres2005, Ocvirk2008, Tumlinson2011, Ford2014, Tumlinson2017, Kocjan2024}.

While observational measurements of galactic outflows are a challenging endeavour (e.g., see a review by \citealt{Veilleux2005} for a discussion on UV absorption lines, low S/N, kinematic complexity and line saturation; and more recently \citealt{Veilleux2020} for cool outflows), they reveal the ejection of gas from the ISM through winds featuring warm and cold thermodynamical components, for which it is possible to characterise key properties such as the column density and velocity \citep{Steidel2010, Heckman2015, Concas2019, ForsterSchreiber2019, Schroetter2019, Schroetter2021, Cherrey2025}. These observations further estimate the dimensionless mass-loading ratio, defined as the rate of outflowing mass per rate of star formation.
Higher temperature outflow gas phases inherently entrain higher specific thermal energies, dominating the outflows' energy density \citep{Kim2018}. These hot outflows are often probed through X-ray observations, and their detections are often limited to starburst galaxies \citep{Strickland2000, Strickland2004, Yamasaki2009, Bogdan2013}. Importantly, in these systems, significant cold ($T < 10^3\,\Kelvin$) outflows are also present \citep{Leroy2015, Chisholm2016, Fluetsch2019, Fluetsch2021, Lopez-Rodriguez2021, Lopez-Rodriguez2023b}.
In smaller mass galaxies, hot outflows are mostly absent. When present, they are primarily attributed to active galactic nuclei \citep{Mezcua2016}. Instead, these galaxies frequently feature more moderate temperature winds, typically seen through an ionized component with low to moderate velocity, (possibly) non-escaping outflows \citep{Chisholm2017, McQuinn2019, Concas2022, Marasco2023}.
These multi-phase, metal-enriched galactic outflows also act as sources for the complex thermodynamical and chemical state of the CGM  \citep[e.g.,][]{Werk2016, Stern2016, Concas2022, Choi2024, Zheng2024}.

Together, these processes result in a complex and evolving picture where galaxy formation simulations need to reproduce not just the observed galaxy properties but the thermal and chemical state of the baryon cycle that sets the properties of the CGM.
The complexity and computational cost of large-scale cosmological simulations lead to shortcomings such as limited resolution, ineffective feedback, and inadequate modelling of some of the relevant physical processes \citep[e.g.,][]{Katz1992, Dubois2014, Vogelsberger2014, Schaye2015, Pillepich2018a, Dave2019}. These are often alleviated through parametric calibration and/or sub-grid prescriptions for galactic winds, such as hydrodynamically decoupled wind particles \citep[see][ for a review]{Vogelsberger2020}. Various galaxy outflow properties -- the proportion of retained gas, entrained metals, distance of ejection -- are either set or significantly affected by these modelling choices, with the mass-metallicity relation being a notable example \citep{Tremonti2004, Dave2017, Bahe2017, DeRossi2017, Torrey2019, Dubois2021}.
Even in models featuring considerably higher resolution, some degree of artificial feedback enhancement is frequently required to reproduce the scaling relation between dark matter halo mass and stellar mass \citep{Rosdahl2018, Smith2019, Dubois2021, Shen2025preprint}.
This often leads to single-phase, exclusively hot outflows from galaxy formation simulations that may be at odds with the observed multi-phase winds \citep{Concas2019, Bennett2025}.
Here we investigate accounting for well-known non-thermal physics as a promising solution, focusing on three main components and their associated processes: magnetic fields, cosmic rays, and stellar ionizing radiation followed with full, on-the-fly radiative transfer.

Magnetic fields play a role in the baryon cycle even prior to the process of star formation, modifying the gas distribution across phases \citep{Evirgen2019, Kortgen2019} and supporting molecular clouds against collapse which drives higher star formation clustering \citep{Martin-Alvarez2020, Robinson2024}. Magnetic fields also influence the star formation efficiency \citep{Federrath2012, Zamora-Aviles2018}, modify outflow gas dynamics, by, for example, affecting the mixing layer \citep{Shukurov2018, Gronnow2018, vandeVoort2021, Lee2025preprint}, and influence the CGM thermal instability \citep{Ji2018}. Magnetised outflows have been detected down to dwarf galaxy masses \citep{Taziaux2025}. This makes polarimetric observations of galaxy outflows a unique dynamical tracer, where the combination of radio and far-infrared observations enable studying large-scale and diffuse gas (radio) as well as small-scale and dense gas (far-infrared) \citep{Borlaff2023, Lopez-Rodriguez2023, Martin-Alvarez2024a, Maglione2025}. These observations allow probing the galaxy-halo interface in detail \citep{Krause2020, Lopez-Rodriguez2023b}, and reveal important kinematic information of outflows along the plane of the sky \citep{Stein2020, Heald2022, Stein2025}.

Through its complex interactions with the surrounding medium, radiation is a pervasive contributor to the overall feedback landscape in galaxies: young, massive stars ionize their surroundings and pre-process gas through photo-evaporation. This can lead to more efficient SN explosions \citep{Geen2015b, Sartorio2021}. These effects also make radiation an important source of early stellar feedback, halting and even suppressing ongoing star formation in molecular clouds depending on their mass \citep{Dale2012, Krumholz2014chapter, Agertz2020, Menon2023}, and may even lead to the suppression of inflows onto the galaxy, leading to starvation of the smallest galaxies \citep{Katz2020}. Some studies show radiation is capable of directly driving winds from molecular clouds \citep{Menon2023}. Growing evidence however suggests that these small-scale effects may not lead to a stellar mass reduction in cosmological galaxy formation simulations. Instead, stellar radiation may lead to a minor star formation rate increase \citep{Martin-Alvarez2023}, as reduced star formation clustering decreases the mass-loading of galaxy outflows \citep{Rosdahl2015b, Smith2021}\footnote{See \citet{Emerick2018} for the opposite effect, in low mass galaxies.}.

The potential of cosmic rays (CRs) as an effective pressure source further supporting galactic outflows has been long recognised \citep[e.g.,][]{Pfrommer2007, Jubelgas2008, Booth2013, Salem2014a, Girichidis2018, Dashyan2020, Buck2020, Hopkins2021a, Farcy2022, Curro2024, Farcy2025}. This is due to their softer adiabatic index, lower energy losses away from dense ISM regions, and their ability to enhance SN momentum deposition even at small scales deep within the ISM \citep{Diesing2018, Curro2022}. Numerical simulations have also found CRs capable of suppressing star formation in galaxies \citep{Jubelgas2008, Pakmor2016, Commercon2019, Hopkins2020, Nunez-Castineyra2022, Simpson2023, Curro2024}. While CR propagation in galaxies is still not fully understood, and often simplifying assumptions are adopted to model their diffusion and streaming, and further improvements of their numerical modelling are required \citep{Armillotta2021, Hopkins2022a, Hopkins2022b, Girichidis2022, Thomas2023, Ponnada2024, Butsky2023, Armillotta2024}, their 
distinct effects have established CRs as another fundamental player in the understanding of galaxy outflows and warrant further study. 

The \pandora~suite of high-resolution cosmological zoom-in simulations was introduced in \paperI~\citep{Martin-Alvarez2023}. That work employed the 17 zoom-in simulations in the suite to show that, in addition to the independent effects described above, the combination of stellar radiation, CRs, and magnetic fields leads to a complex interplay with pervasive effects across multiple galaxy properties (such as the stellar mass--halo mass relation, mass--size relation, kinematics, cusp--core halo profiles), and observables (such as optical appearance, resolved kinematics, galaxies colour--magnitude diagram, radio synchrotron emission). \pandora~has also been employed to investigate how different combinations of non-thermal physics influence Ly$\alpha$ and LyC escape and observables \citep{Yuan2024}. Pandora serves as a pathfinder for our upcoming Azahar simulations (\citealt{Martin-Alvarez2025aas}; Martin-Alvarez et al. in prep.) featuring a larger cosmological volume of approx. $(10\, \text{cMpc})^3$, and which have been already employed to study various high-redshift galaxy formation puzzles \citep[e.g.,][]{Witten2024, Yuan2025, Dome2025}. 
In this work, we employ a subset of models from the \pandora~suite, focusing on radiation and CRs to understand how the interplay of different non-thermal physics shapes star formation, galaxy outflow properties, and the resulting metal enrichment of the galaxy.
The manuscript is structured as follows: Section~\ref{s:Methods} provides a brief introduction to the \pandora~suite, with Section~\ref{ss:SimsSuite} summarizing the subset of models studied here. We report our main results in Section~\ref{s:Results}, investigating star formation and SN feedback (Section~\ref{ss:SFH_and_SN}), outflow properties (Section~\ref{ss:outflow_profiles_and_PDF}), briefly comparing outflows with observations (Section~\ref{ss:outflow_observables}), and concluding with an analysis of metal enrichment of the galaxy and enriched gas flows (Section~\ref{ss:metal_enrichment}). Our main conclusions are outlined in Section~\ref{s:Conclusions}.

\section{Methods}
\label{s:Methods}
\subsection{The \pandora~simulation suite}

The galaxy formation simulations analysed in this work are a subset of the cosmological zoom-in \pandora~suite \citep{Martin-Alvarez2023}. In this section we summarise the basic characteristics of the \ramses~\citep{Teyssier2002} code used to generate the simulations, and the \pandora~models studied. 

The \pandora~dwarf galaxy (halo virial mass $M_\text{vir} (z = 0) \sim 10^{10} \Msun$; with initial conditions introduced by \citet{Smith2019} under the label `\textit{Dwarf1}' and re-generated for \ramses) forms in a zoom-in sub-volume of $2.5$~cMpc in width. In this region, the dark matter and stars  collisionless particles have mass resolutions of $m_\text{DM} = 1.5 \cdot 10^{3} \Msun$ and $m_\text{*} = 400 \Msun$, respectively. 

The adaptive grid is allowed to refine the zoom sub-volume down to a full-cell size of $\Delta x \sim 7$~physical pc. Cell refinement is triggered when the total mass of a resolution element, measured as $(\Omega_{\rm m} / \Omega_{\rm b})\,m_\text{baryons} + m_\text{DM}$, exceeds $8 \times m_\text{DM}$. Here, $\Omega_{\rm m} = 0.3065$ and $\Omega_{\rm b} = 0.0483$ are the cosmological matter and cosmological baryonic matter densities, respectively. Refinement is also triggered when a resolution element fulfills the Jeans criterion $\Delta x > \lambda_J / 4$. The Jeans criterion is disabled for cells with a comoving size $\Delta x > 1.8$~ckpc, to avoid excessive refinement in the low-density regions of the simulation. In this expression, $\lambda_J$ is the local Jeans length.

Our magneto-hydrodynamics (MHD) models employ the \ramses~ constrained transport (CT) solver by \citet{Teyssier2006} and \citet{Fromang2006}, and super-comoving coordinate implemented in \citet{Martin-Alvarez2018}. This CT method numerically guarantees down to double precision the solenoidal constraint for the magnetic field ($\vec{\nabla} \cdot \vec{B} = 0$). Our radiative transfer (RT) models employ the {\sc ramses-rt} implementation by \citet{Rosdahl2013} and \citet{Rosdahl2015a}. The radiative flux budgets for each stellar particle are determined according to the age and metallicity-dependent spectral energy distributions of the {\sc bpass}v2.0 models \citep{Eldridge2008,Stanway2016}. The overall RT configuration is similar to that of the {\sc sphinx} simulations \citep{Rosdahl2018, Rosdahl2022} that match well the reionization history of the Universe. The models with CRs employ the CR solver described by \citet{Dubois2016} and \citet{Dubois2019}, modelling CRs as an energy density representing $\sim$GeV protons. It accounts both for CR streaming (with streaming velocity equal to the Alfv\'en speed), anisotropic diffusion with a diffusion coefficient $\kappaCR = 3 \cdot 10^{28} \cm^2 \text{s}^{-1}$ \citep{Ackermann2012,Salem2016,Pfrommer2017a,Cummings2016}, and Coulomb and hadronic losses~\citep{Guo2008}. In the models studied here, magnetic fields and CRs are seeded by SN explosions, as described below. Further details of the RT, CR and MHD implementations, and their employed configurations are provided in \citet{Martin-Alvarez2023}. 

All \pandora~models studied here employ our MTT (magneto-thermo-turbulent) star formation prescription (presented by \citealt{Kimm2017}; extended to MHD in \citealt{Martin-Alvarez2020}). In short, cells at the highest level of refinement \citep{Rasera2006} are allowed to spawn new stellar particles when their gravitational pull is larger than the local MTT pressure support. They form stars according to the local MTT properties \citep{Padoan2011, Federrath2012} of the gas employing a Schmidt law \citep{Schmidt1959}.

The models we study all include mechanical SN feedback \citep{Kimm2014, Kimm2015}, which is injected into its hosting cell, and 48 (active) cell neighbours, according to an equal-solid-angle weighting. Each event injects (mass, momentum and energy) by selecting for each neighbour cell one out of two possible feedback modes: a momentum-conserving mode, injecting the SN terminal momentum, or an energy-conserving mode (Sedov), injecting the SN energy. Which solution is employed is determined by whether the ratio of swept-up vs. ejected mass is above a given threshold \citep[eq.~7,][]{Kimm2015}. The SN-specific energy is $\varepsilon_\text{SN} = E_\text{SN} / M_\text{SN}$, with $E_\text{SN} = 10^{51} \erg$ and $M_\text{SN} = 10\,\Msun$ {--} except for \HDSfFbBoost, which features $\varepsilon_\text{SN, Boost} = 4\,\varepsilon_\text{SN}$. In the studied models including MHD, these explosions inject 1\% of their energy into the magnetic component ($\gtrsim 10^{-5}$~G when injected at scales of $\sim\!\!10$~pc; reproducing the magnetisation observed at the scale of SN remnants; \citealt{Parizot2006}) through two small-scale toroids with arbitrary orientations. This implementation reproduces magnetic fields in observed galaxies \citep{Martin-Alvarez2021, Martin-Alvarez2024a, Dacunha2025}. Additional details of this implementation are provided in Appendix~A of \citet{Martin-Alvarez2021}. Similarly, in the studied model including CRs, 10\% of the SN energy (extracted from its mechanical energy budget) is injected into the CR energy, motivated by observations \citep{Morlino2012, Helder2013}, and is a standard value employed by similar studies \citep[e.g.][]{Pfrommer2017b, Butsky2018, Curro2024}.

\subsection{A representative subset of models} 
\label{ss:SimsSuite}


\begin{table*}
\centering
\caption{Subset of simulations studied in this work. From left to right, we indicate model symbol and label, whether the model includes magnetic fields (MHD), radiative transfer (RT) and cosmic rays (CR), the stellar feedback configuration, and further details regarding the configuration of the simulation. From top to bottom, simulations account for additional baryonic physics. The full sample of \pandora~models is provided in \citet{Martin-Alvarez2023}.}
\label{table:setups}

\begin{tabular}{l l l l l l l l l}
\hline
& Simulation & MHD & RT & CR & Stellar feedback & Further details \\
\hline
\tablesymbol{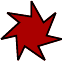} & \HDSfFb & \xmark
& \xmark & \xmark & Mech & Standard hydrodynamical simulation \\

\tablesymbol{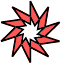} & \HDSfFbBoost & \xmark
& \xmark & \xmark & Boosted Mech & Calibrated feedback through boosted SN: $2 E_\text{SN}$, $0.5 M_\text{SN}$. \\

\tablesymbol{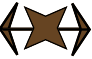} & \RTiMHDSfFb & \cmark
& \cmark & \xmark & Radiation + MagMech & Including radiation and magnetic fields. SN inject $E_\text{mag,SN}$.\\

\tablesymbol{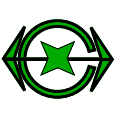} & \RTCRiMHDSfFb & \cmark
& \cmark & \cmark & Radiation + CRMagMech & \pandora~`full physics' model; SN inject $E_\text{mag,SN}$ + $E_\text{CR,SN}$.\\
\hline
\end{tabular}
\end{table*}

In this work, we restrict our analysis to a subset of the \pandora~simulations. The subset has been selected to provide a representative view of the main physical effects at play in the interplay between star formation and galaxy outflows. The four selected models are summarised in Table~\ref{table:setups}, and correspond to: i) the standard hydrodynamics model characteristic of traditional simulations (\HDSfFb), ii) an `enhanced' SN feedback representative of calibrated hydrodynamical models (\HDSfFbBoost; with SN specific energy increased by $4\times$), iii) a model including the important ISM non-thermal physics of radiation and magnetic fields (\RTiMHDSfFb), and iv) the \pandora~full-physics simulation incorporating CRs as well (\RTCRiMHDSfFb). These are the models highlighted as models of interest by \citet{Martin-Alvarez2023}.

To identify the galaxy of interest in the simulation, we select its host halo at $z = 8$, and determine its exact position through a shrinking spheres algorithm \citep{Power2003}. We determine its position in subsequent snapshots (both forward and backward in time) by selecting the innermost 500 particles and finding their updated centre-of-mass, followed by the recursive application of our shrinking spheres method. For each snapshot, each galaxy is associated with its hosting halo, extracted from a halo catalogue generated using {\sc halomaker} \citep{Tweed2009}. When multiple progenitors are identified, we only employ the most massive one for our analysis here. This method, and its on-the-fly application to \ramses~cosmological simulations will be further detailed in Martin-Alvarez et al. (in prep). The main galaxy measurements are performed in the central region of the halo ($r < 0.2\;r_\text{halo}$, with $r_\text{halo}$ the virial radius of the halo), as described in the corresponding measurement descriptions provided in Section~\ref{s:Results}.

\section{Results}
\label{s:Results}

\subsection{A qualitative picture of SN-driven outflows accounting for different physical processes}
\label{ss:Qualitative_view}

\newcommand{\frontwidth}{0.48\columnwidth}%
\begin{figure*}
    \centering
    \includegraphics[width=\frontwidth]{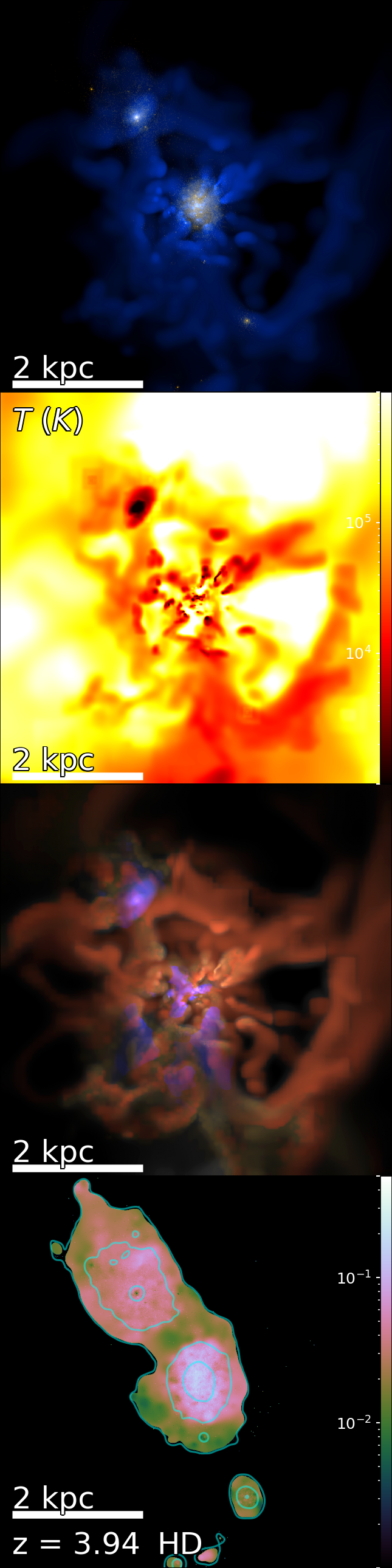}%
    \includegraphics[width=\frontwidth]{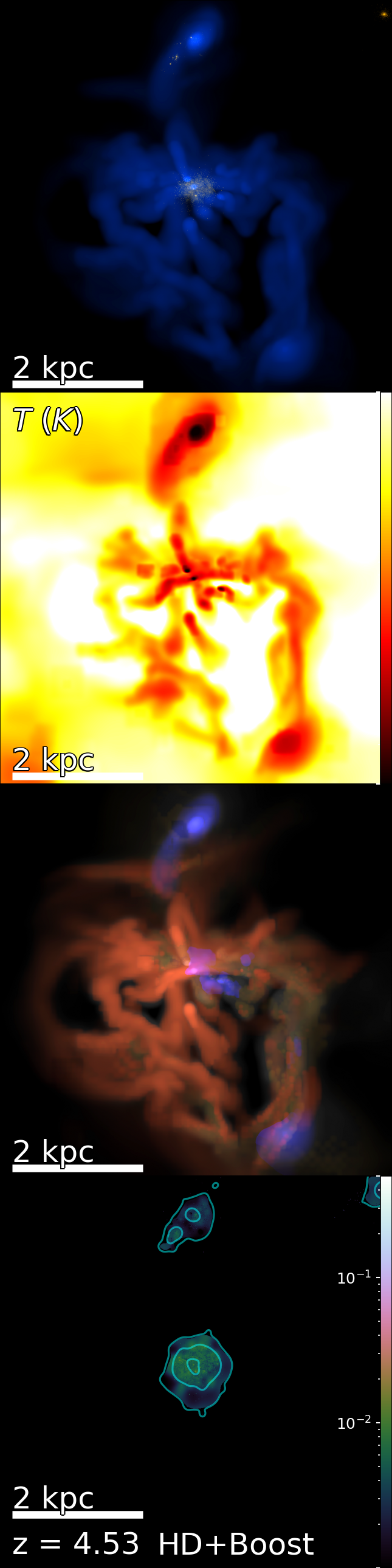}%
    \includegraphics[width=\frontwidth]{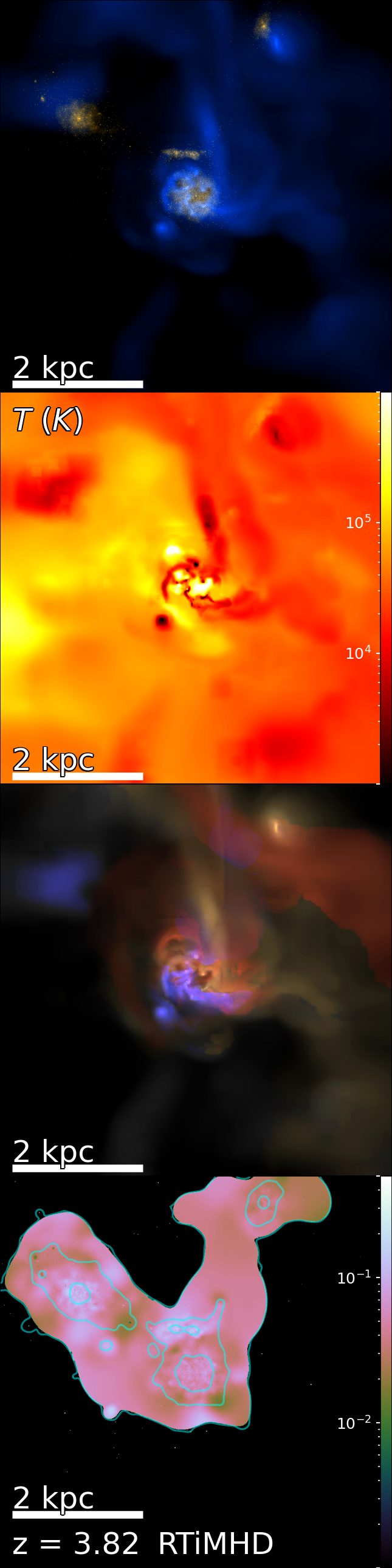}%
    \includegraphics[width=\frontwidth]{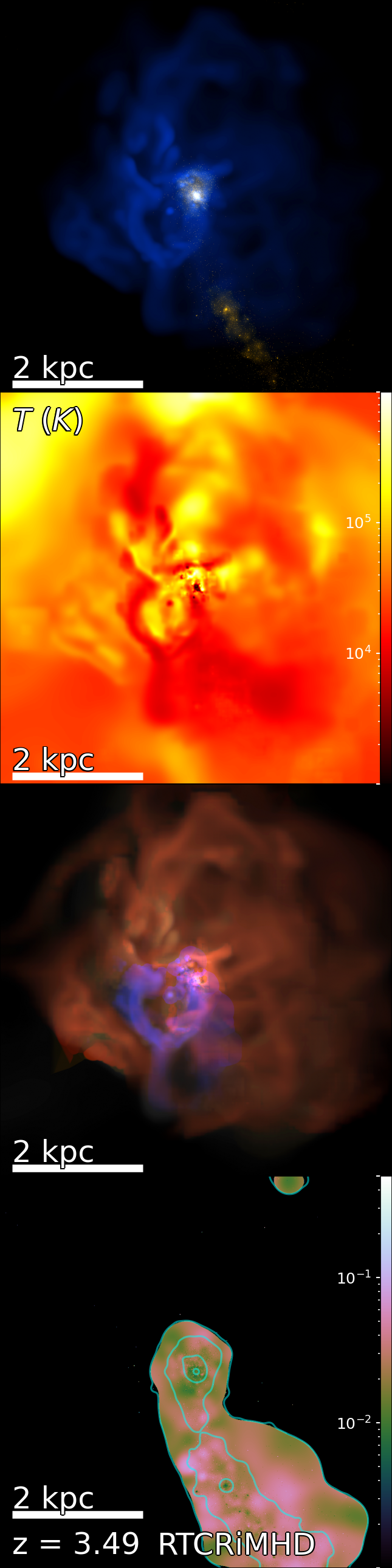}%
    \caption{Projected views centred on the \pandora~dwarf illustrating outflow events across our four studied models. The area displayed in each panel approximately encompasses the central galactic region ($0.2\,\rhalo$), 5\,$\kpc$ across. Each of the columns, from left to right, corresponds to the following models: standard hydrodynamics (\HDSfFb), calibrated SN feedback hydrodynamics (\HDSfFbBoost), radiative transfer and MHD (\RTiMHDSfFb), and our `full-physics' simulation with radiation, CRs and MHD (\RTCRiMHDSfFb).
    {\bf (Top row panels)} Mass-weighted densities for the gas (blue) and stellar (gold) components.
    {\bf (Second row panels)} Gas temperature maps.
    {\bf (Third row panels)} Gas density flows separated according to radial velocity ($\vrad$) into inflows (blue; $\vrad < -\vvir = 12.5\,\kmps$) and outflows (golden; $\vrad > \vvir$, and red for faster outflows $\vrad > 40\,\kmps$). For further visual guidance, in this row we display the full gas density field using a gray scale. 
    {\bf (Bottom row panels)} Mass-weighted stellar metallicity. To guide the eye, the overlaid contours display the $[10^{5}, 10^{6}, 10^{7}, 10^{8}]\,\Msun\,\kpc^{-2}$ stellar surface density isocontours.
    Galactic outflows in \HDSfFb~and \HDSfFbBoost~appear hotter and with rugged and defined shock structures. Including radiation reduces the outflows, whereas CRs lead to temperate and more homogeneous outflows, as well as complex metallicity topology.}
    \label{fig:front_image}
\end{figure*}

Fig.~\ref{fig:front_image} provides a qualitative comparison of the four different \pandora~dwarf galaxy simulation models, illustrating the impact of varying physical processes on SN-driven outflows. Each panel is 5~$\kpc$ across, approximately encompassing the central galactic region ($0.2\,\rhalo$, with $\rhalo$ the virial radius of the dark matter halo). The columns, from left to right, correspond to the main models we study here: standard hydrodynamics (\HDSfFb), calibrated SN feedback hydrodynamics (\HDSfFbBoost), RT and MHD (\RTiMHDSfFb), and our `full-physics' simulation incorporating radiation, CRs, and MHD (\RTCRiMHDSfFb).
In the top row we show the gas (blue) and stellar (gold) mass-weighted mean densities.
The second row of the figure displays the gas temperature.
The third row shows gas density flows, separated according to the radial velocity of each cell ($\vrad$) into inflows (blue) and outflows (gold and red), with the full gas density field displayed in greyscale for context. Gas is selected as inflowing ($\vrad < -\vvir$) or outflowing ($\vrad > \vvir$ as gold tones; $\vrad > 40\, \kmps$ further colored in red), where $\vvir \sim 12.5\,\kmps$ is the virial velocity of the halo. 
Panels in the bottom row display the mass-weighted metallicity in solar units ($Z_{
\odot} = 0.012$; \citealt{Schneider2012}). We include isocontour lines to guide the eye to the location of the main galaxy and satellites. To avoid spurious variations of metallicity measurements in regions of undersampled (or extremely low) stellar mass, we set the map to null values outside the outermost contour (i.e., for stellar surface densities lower than $10^{5}\,\Msun\,\kpc^{-2}$). 

The distribution of gas around the galaxy varies notably across our models. In the \HDSfFb~model, we observe relatively large and hot outflow bubbles ($\sim$$1 - 2\,\kpc$) confined by dense gas shells with irregular and turbulent surfaces. These bubbles contain the bulk of the outflowing gas mass. The properties of outflow bubbles are even more pronounced in the \HDSfFbBoost~case, with bubbles that reach higher temperatures and larger sizes, despite a lower star formation rate (SFR). The bubbles exhibit a complex substructure, resembling multiple bursts of SN-driven winds that are carving multiple holes in the CGM. Conversely, the high temperature bubbles in the \RTiMHDSfFb~model do not extend significantly beyond the galaxy (e.g., third panel of the \RTiMHDSfFb~column, seen as outflowing, high-density regions), and due to the presence of the radiation field the gas surrounding the galaxy has a smooth appearance. The significantly more diffuse stellar component in this model (top panel; see also Fig.~4 in \citealt{Martin-Alvarez2023}) leads to a lower clustering of stellar feedback, discussed further in Section~\ref{ss:SFH_and_SN}. The \RTCRiMHDSfFb~model shows outflowing bubbles comparable in size to those in the \HDSfFb~model, but with smoother surfaces due to the presence of the non-thermal components. This \RTCRiMHDSfFb~model also has a temperate and outflowing diffuse envelope surrounding the galaxy, suggesting a scenario where outflows are initially driven by SNe and subsequently expelled further by cosmic ray pressure in a more isotropic fashion. Interestingly, stellar metallicity distribution is very sensitive to the different physics models, with the dwarf galaxy in the \HDSfFbBoost~simulation being extremely metal poor (see further Section~\ref{ss:outflow_observables}), while in the \RTCRiMHDSfFb~simulation dwarf and its surroundings are enriched to $Z_* \sim 0.1\,Z_{\odot}$, and exhibit complex metal distribution and gradients. 

\subsection{Bursty and concentrated star formation drives mass-loaded outflows}
\label{ss:SFH_and_SN}

\begin{figure*}
    \centering
    \includegraphics[width=1.9\columnwidth]{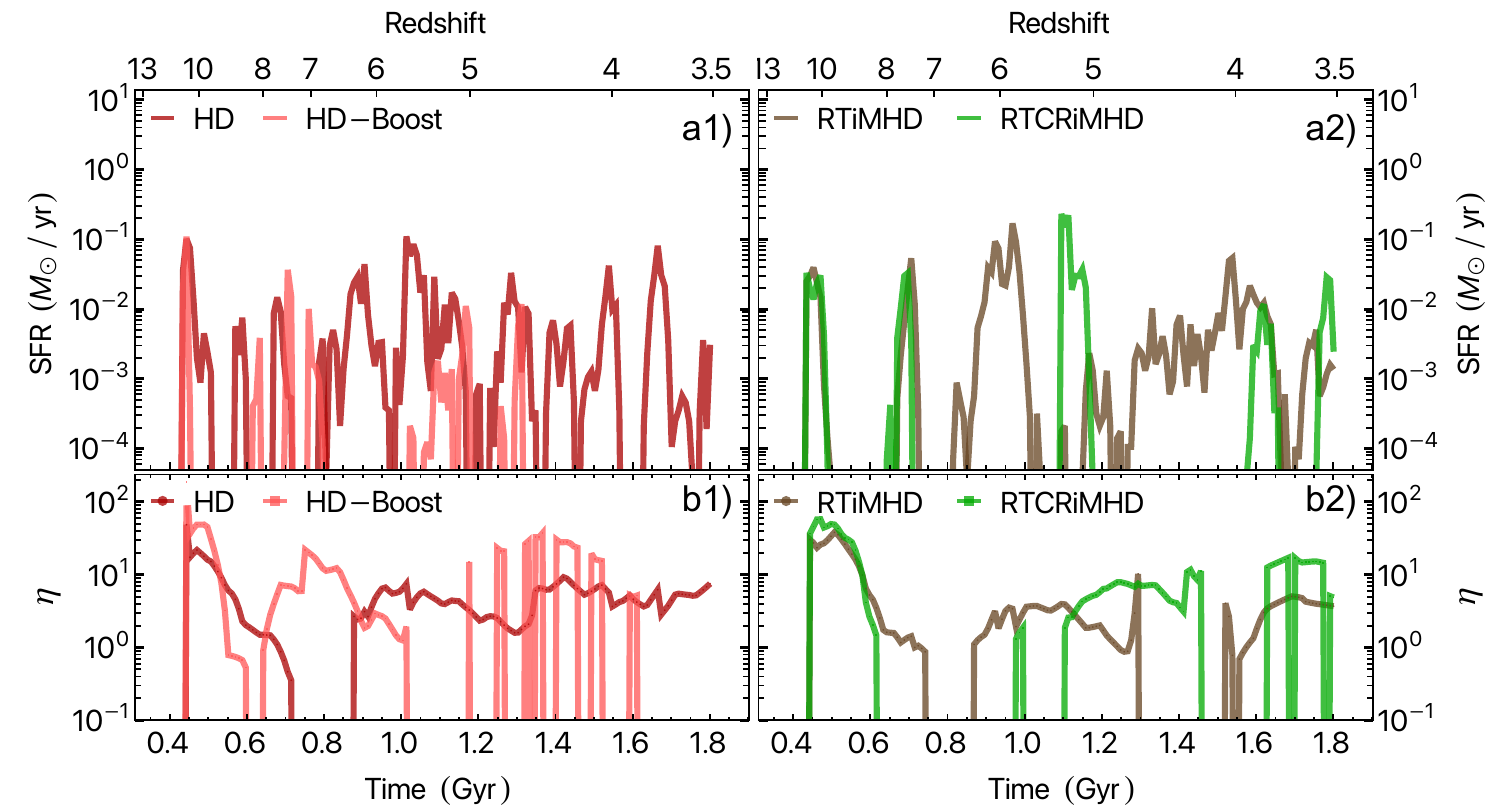}\\
    \vspace{0.5cm}
    \hspace{1.0\columnwidth}\includegraphics[width=1.0\columnwidth]{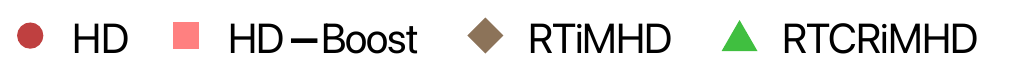}\\
    \vspace{-1.cm}
    \includegraphics[width=1.9\columnwidth]{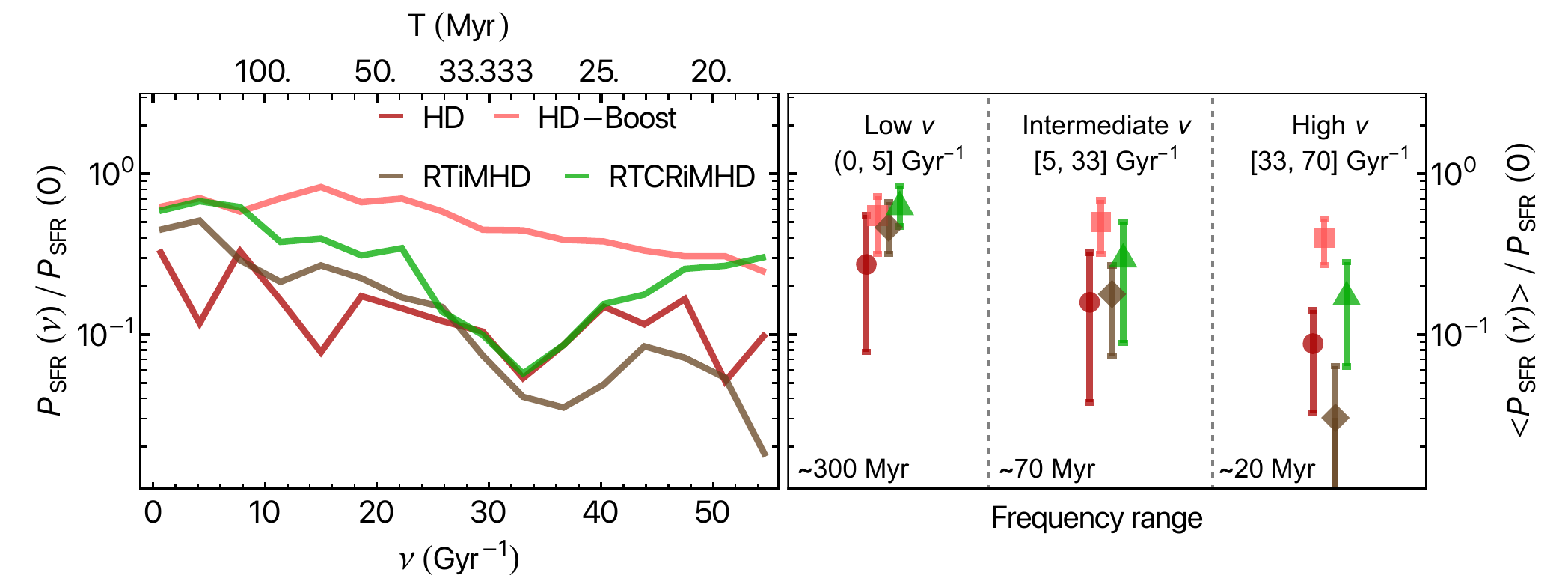}\\
    \caption{{\bf (Top row)} Star formation rate (SFR) 
    for our hydrodynamical models \HDSfFb~and \HDSfFbBoost~(panel a1), and for our more complex models including stellar radiation and magnetic fields \RTiMHDSfFb, as well as CRs, \RTCRiMHDSfFb~(panel a2).  
    {\bf (Central row)} Outflowing gas mass-loading factor ($\eta = \dot{M}_\text{gas} / $ SFR; see text for further details) radially ejected through a spherical shell 
    positioned at $0.5\,\rhalo$.
    Including radiation leads to a more continuous SFR, whereas its combination with CRs leads to a more bursty star formation history.
    The calibrated feedback \HDSfFbBoost~and full physics \RTCRiMHDSfFb~simulations attain the highest outflow mass-loading factors. The \RTiMHDSfFb~has the lowest $\eta$ values, albeit those remain comparable to the \HDSfFb~scenario.
    {\bf (Bottom row)} Fourier transform of the star formation history for the same models, normalised to the zero-frequency mode (left panel). The average power is separated into the three frequency regimes corresponding to long, intermediate and short timescales (right panel).}
    \label{fig:SFH}
\end{figure*}

Before we proceed to qualitatively characterise the properties of galaxy outflows across our different models, we investigate the interrelation between the star formation histories (SFHs) and the galactic outflows in the \pandora~simulation suite. To this effect, we show in Fig.~\ref{fig:SFH} the SFR and outflowing gas mass-loading factors ($\eta = \dot{M}_\text{gas} / \text{SFR}$) across different simulation models. The top row shows the SFR for our hydrodynamical models (panel a1), and for the more complex models including stellar radiation and magnetic fields (panel a2), both with and without CRs. The bottom row presents the outflowing gas mass-loading factor $\eta$ through a spherical shell with a thickness of 200~pc positioned at $0.5\,\rhalo$%
\footnote{Due to the rapid variability of star formation in dwarf galaxies, to compute the $\eta$ values we smooth the star formation rate. Here we assume a $\sigma \sim 0.042\,\text{Gyr}$ (corresponding to FWHM of $\sim$100~Myr), further including a $10^{-4} \Msun \, \yr^{-1}$ floor to avoid artificially large $\eta$ during periods of negligible SFR. To consider gas outflowing, we require a cell to have a positive radial velocity $\vrad > \vvir$, which reduces spurious contributions to galaxy outflows. 
We also include a uniform $\Delta t=0.2\,\rm Gyr$ delay for the $\eta\,(0.5\,\rhalo)$ calculations, which corresponds to a local maximum in the cross-correlation of SFR and outflows, and is equal to the approximate time for a gas outflow to travel from the galaxy centre to $0.5\,\rhalo$ at $
\sim$$2\,\vvir$ {--} approximately our measured mass-weighted average velocity of outflows.
}.

The onset of significant star formation is comparable across all models, and occurs at $\sim 0.4$~Gyr. The models with stellar radiation have a somewhat lower initial SFR (up to $0.03\,\Msun \, \yr^{-1}$ instead of $0.1\,\Msun \, \yr^{-1}$), as the stellar radiation acts as a source of early stellar feedback. This initial star formation event continues for $\sim$$100\,\Myr$ in the \HDSfFb~model, whereas it is rapidly halted by the `boosted' SN feedback in the \HDSfFbBoost~model, and by stellar radiation and SNe in the \RTiMHDSfFb~and \RTCRiMHDSfFb~models.
After a second star formation event at $t \sim 0.7\,\Gyr$, the evolution of the models starts to differ more significantly. We focus on our hydrodynamical models first (panel a1). Despite its rapid variation on timescales of $\lesssim50\,\Myr$ (frequencies of $f_\text{SFR} \sim (50 \Myr)^{-1}$), the \HDSfFb~model maintains an approximately continuous SFR. The \HDSfFbBoost~model exhibits a notably different SFH, characterised by an `on' and `off' SFR. Its star formation proceeds through three main extended bursts of star formation, with a typical duration of $\lesssim 200-300$~Myr. The \RTiMHDSfFb model (panel a2) shows a `flatter' and uninterrupted star formation during the $z = 7$ to $z = 6$ and $z = 5$ to $z = 4$ periods. Its star formation resembles that of the \HDSfFb~model, but with much weaker short-time-scale fluctuations.
The \RTCRiMHDSfFb~model has a particularly episodic SFH, with sharp bursts of star formation separated by extended quiescent periods. Its episodes of star formation differ from the \HDSfFbBoost~case by being more cyclic and concentrated in time. 
To quantify these variations in further detail, the bottom left panel of Fig.~\ref{fig:SFH} shows the Fourier transform of each SFH, normalised by their corresponding 0-th mode (proportional to the total stellar mass) to account for the differing stellar masses at the end of the studied period. We identify three main regimes of interest in the frequencies of SFR fluctuations: long timescales of $\sim$$300\,\Myr$, intermediate timescales of $\sim$$70\,\Myr$, and short timescales of approximately $\lesssim 20\,\Myr$. The average power in each regime is shown in the bottom right panel of the same figure. The
Fourier transform of the star formation history confirms a more episodic SFH for the \RTCRiMHDSfFb~model that rapidly transitions from quiescent to star forming, and the role of radiative transfer (especially notable in \RTiMHDSfFb) in smoothing out the star formation variability on short timescales. 
The observed behaviour of the \RTCRiMHDSfFb~model is attributed primarily to the combined effect of the non-thermal pressures from CRs and magnetic fields. Their pressure support allows gas clouds to grow in mass until they become supercritical, leading to their rapid collapse and concentrated star formation. We note that this effect is likely sensitive to the value of $\kappaCR$ employed - e.g., \citet{Commercon2019} and \citet{Dashyan2020} attain higher ISM gas densities for higher $\kappaCR$, potentially leading to increased burstiness.

As expected, these variations in the overall shape and properties of the SFHs across our models have important effects on the mass-loading factors measured for our dwarf galaxy. The \HDSfFb~simulation has approximately sustained outflows, with low mass-loading factors ranging around $\sim 1 - 5$. The \HDSfFbBoost~model reaches the highest mass-loading factors we measure despite its low SFR (and its lack of concentrated SN events), due to the enhanced energy deposition per SN in this model. The \RTiMHDSfFb~model has the lowest mass-loading factors, and during its multiple periods of low or no star formation, the mass-loading tends to $\eta \rightarrow 0$. The \RTCRiMHDSfFb~simulation has also high mass-loading factors, on the order of $\sim$$10$, and comparable to the \HDSfFbBoost~model. Notably, these values are reached with significantly lower energy injection, underscoring the efficiency of its episodic star formation and CR-driven outflows. The different properties associated with the outflows emerging from each model are discussed in Section~\ref{ss:outflow_profiles_and_PDF}.

\begin{figure}
    \centering
    \includegraphics[width=\columnwidth]{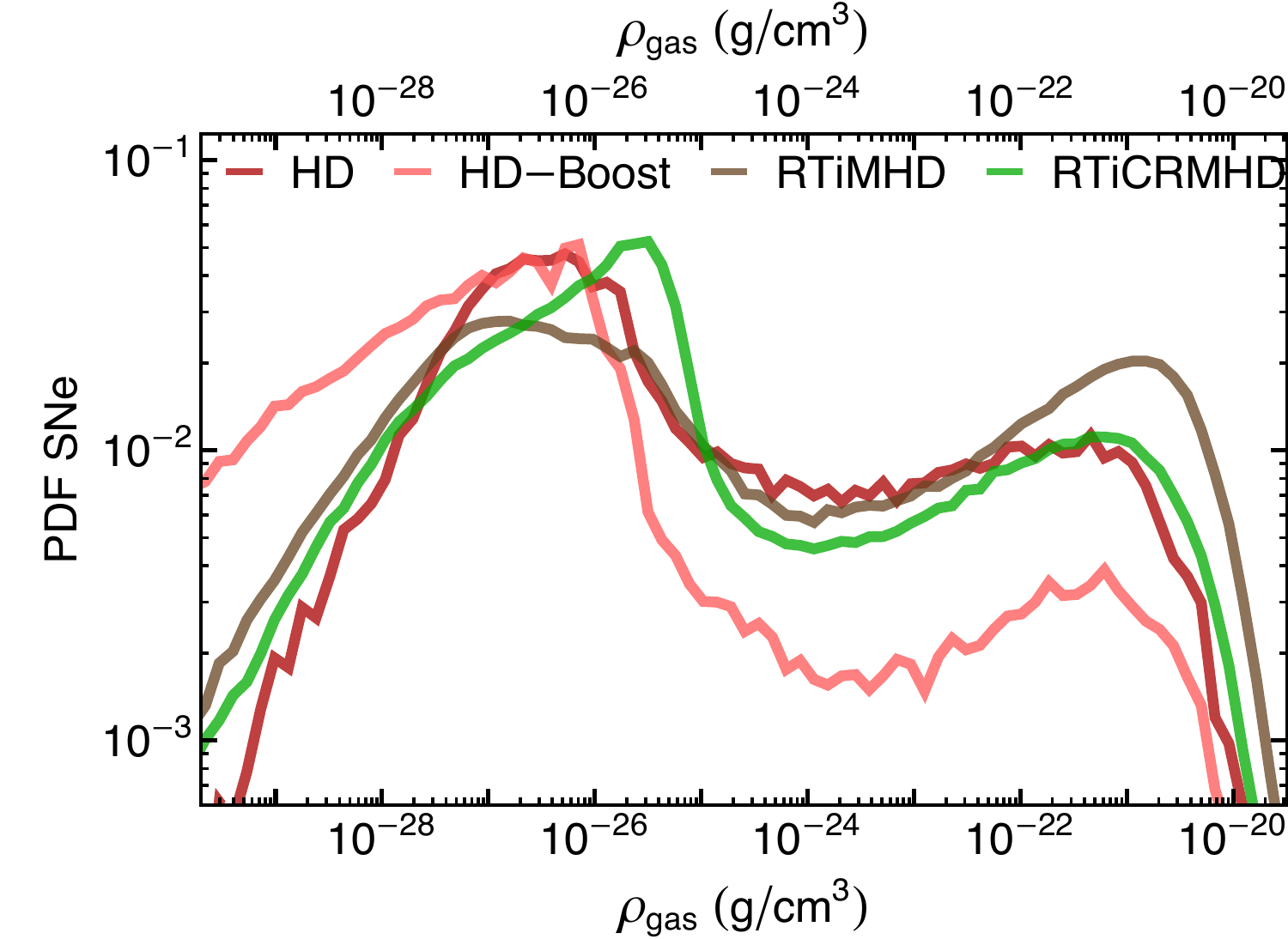}\\
    \includegraphics[width=\columnwidth]{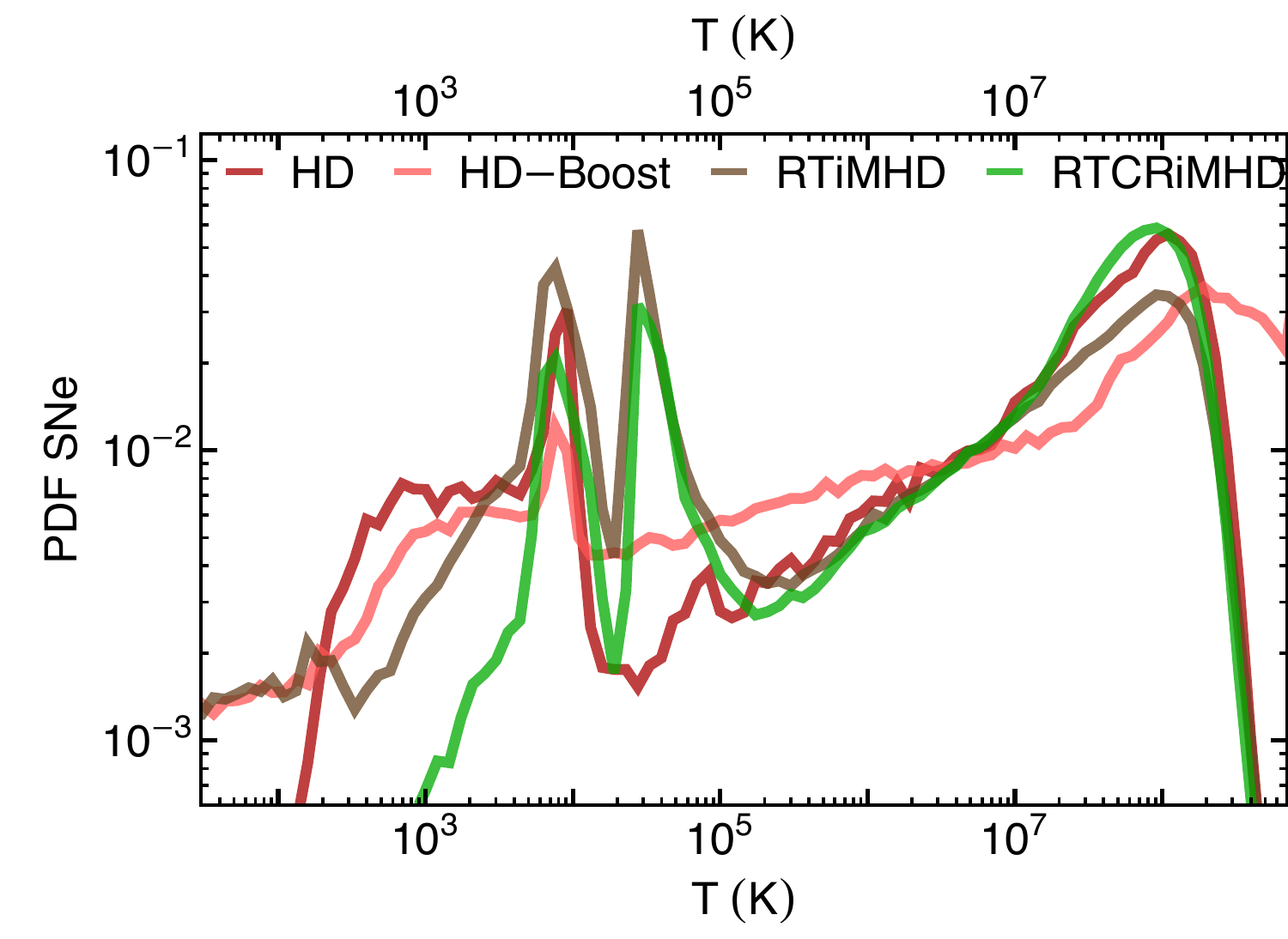}\\
    \caption{PDF of the small-scale environmental thermodynamical properties where SN events take place in each of the models: gas density (top panel) and gas temperature (bottom panel). Across all simulations, most SNe take place at densities $\rho \sim 10^{-26} \gcc$. Different galaxy formation physics leads to notably different  fractions of SN events taking place at high densities ($\rho > 10^{-24} \gcc$). Early feedback in the form of stellar radiation significantly increases the number of SNe at high densities by reducing star formation (and consequently SN) clustering, whereas `boosted' SN feedback leads to their drastic reduction, due to effectiveness of few events in dispersing dense gas. The \RTCRiMHDSfFb~simulation shows the strongest temporal SN clustering, while maintaining the proportion of events taking place in dense, photo-heated regions.}
    \label{fig:SN_PDF}
\end{figure}

The correlation between star formation and subsequent outflows is driven by SN events, with their effectiveness influenced by their small-scale ISM environments, which will also affect the gas phases entrained in galaxy outflows.
We show the thermodynamical properties of the gas cells where SN events take place in Fig.~\ref{fig:SN_PDF}. Panels display the probability density functions (PDFs) for all the SNe events in each simulation (from $z = 8$ to $z = 3.3$), both for the hosting cells gas density (top panel) and temperature (bottom panel). 

Overall, the gas density distribution of cells hosting SN events is bimodal, with a dominant peak at low densities of $\sim 10^{-27} \gcc$ {---} $10^{-26} \gcc$, and a secondary peak at high densities of  $\sim10^{-21} \gcc$. This bimodal distribution is physically driven: stars are formed in dense regions where, in the absence of any feedback process, they explode as SNe with some delay after their formation. This occurs inside or in the vicinity of dense clouds with $\rho_{\text{gas}} \gtrsim 10^{-22} \gcc$. Subsequent SN events originating from the same localized star formation event will typically occur within and around the sites of prior SNe events, which are very hot ($T \approx 10^7$ {---} $10^8$ K), leading to a tail towards lower densities as supernova remnants expand, and ultimately leading to a peak at densities of $\sim 10^{-27} \gcc$. Notably, it is precisely within these lower density, higher temperature gas where SNe feedback becomes most effective \citep{Naab2017, Rodriguez-Montero2022}. While \citet{Kimm2014} have demonstrated that our mechanical prescription alleviates overcooling and recovers the expected momentum injection \citep{Thornton1998}, there are additional numerical considerations to note in a realistically simulated ISM.
For example, \citet{Martizzi2015} and \citet{Kim2015} have shown that the presence of ISM inhomogeneities does not significantly affect the terminal momentum injected by SNe, hence supporting the use of momentum-injecting solutions in galaxy formation modelling \citep{Hopkins2018a, Smith2018, Shimizu2019}. \citet{Kim2015} have also studied shocks propagation, showing that they reach further distances along low density channels. The vast majority of SN events in our simulations take place at $\Delta x < 10\,\pc$ ($\sim$90~\%; and 80\% for \HDSfFbBoost), with virtually all the remaining events taking place at $\Delta x < 15\,\pc$ and low gas densities $\rhogas < 10^{-24}\,\gcc$. We estimate $\gtrsim 80\%$ of events use our adiabatic mode. Note that these ratios, derived from central host cells, may slightly underestimate the proportion of snowplow events; although this is alleviated by the simulation code requirement for contiguous cells not being more than one level lower resolution, and our refinement strategy of resolving the local Jeans length with at least 4~cells.
For SNe in dense media, a higher preservation of terminal momentum at the end of the snowplow phase is expected for higher resolutions, with \citet{Gentry2020} estimating an increase of $1.4\times$ if our models featured resolutions of $1\,\pc$. 

In the \HDSfFb~model, most SN events occur at the low gas density peak, with a relatively flat distribution from intermediate densities of $\rhogas \sim 10^{-24} \gcc$ to $10^{-21} \gcc$, and virtually no events taking place at higher densities. The \HDSfFbBoost~model significantly suppresses the number of SN events at densities $\gtrsim 10^{-25} \gcc$ (approximately $5\times$ less frequent than in the \HDSfFb~simulation). Such suppression of low-effectiveness and rapidly cooling environments, combined with a higher energy deposition per SN ($4 \times$ higher than the canonical value), leads to much more effective feedback in the \HDSfFbBoost~simulation. Additionally, the tail of SNe taking place at the lowest densities is much flatter, with \HDSfFbBoost~being the only model featuring a significant number of SN events in the $\rhogas \lesssim 10^{-28} \gcc$ regime, resulting from the higher energy deposition effectively inflating larger bubbles and clearing the gas to lower densities. The above described density PDFs translate almost directly to the temperature PDFs of the \HDSfFb~and \HDSfFbBoost~models, where the low gas density peak directly corresponds to high temperature environments pre-processed by previous SNe ($T \sim 10^7 {-} 10^8\,\Kelvin$) and the high density events take place in the thermally stable warm ionized medium with $T \sim 10^4\,\Kelvin$, or inside colder star forming regions with $T \lesssim 10^3\,\Kelvin$.
Incorporating stellar radiation in the \RTiMHDSfFb~model results in the lower concentration of star formation, and a subsequent reduction of both temporal and spatial SN clustering. Although radiation locally reduces the gas density prior to the occurrence of the first SN events \citep{Walch2015}, the reduced clustering we measure is consistent with the observed increase of SNe events at high gas densities. 
Radiation provides an effective early feedback that halts ongoing star formation in large gas clouds, photo-heating them to $3 \times 10^4\, \Kelvin$. In the \RTiMHDSfFb~simulation, the increased proportion of SN events at high densities predominantly take place at this photo-heating temperature of $T \sim 3 \times 10^4\, \Kelvin$ \citep{Grudic2022}, which will have implications for the properties of the gas entrained in outflows.
Finally, the combination of stellar radiation with CRs in the \RTCRiMHDSfFb~model leads to a gas density PDF of SNe host cell which is comparable to that in the \HDSfFb~simulation. Once again, a higher temporal clustering of star formation, and the subsequent SNe drive an increase of the proportion of explosions taking place at lower densities and high temperatures, while maintaining a large proportion of events taking place in dense, photo-heated regions. Furthermore, the inclusion of CRs leads to a slight increase of (diffuse) ISM densities in the presence of additional non-thermal pressure terms \citep{Dashyan2020}, and the low density peak of SN events is displaced to slightly higher densities of  $10^{-26} {-} 10^{-25} \gcc$ instead of slightly below.

The reduction of SN feedback efficiency through the inclusion of radiation in our simulations is well-explained by the distribution of SN events across the different ISM phases and environments combined with a steady star formation. We also find large variations of the SNe host cell PDFs for our `boosted' SN feedback model, but its increased efficiency is most likely the result of a considerably higher energy deposition per SN event, with the lower number of events at higher densities being secondary. While our `full-physics' model has some observable variations in the SNe host cell PDFs, the episodic star formation and the properties of the outflowing gas are the main culprits for its higher outflow efficiency. We will investigate the physical properties of such outflowing gas in the next section.

\subsection{Dense and temperate CR-driven galactic outflows}  
\label{ss:outflow_profiles_and_PDF}

\begin{figure*}
    \centering
    \includegraphics[width=2.1\columnwidth]{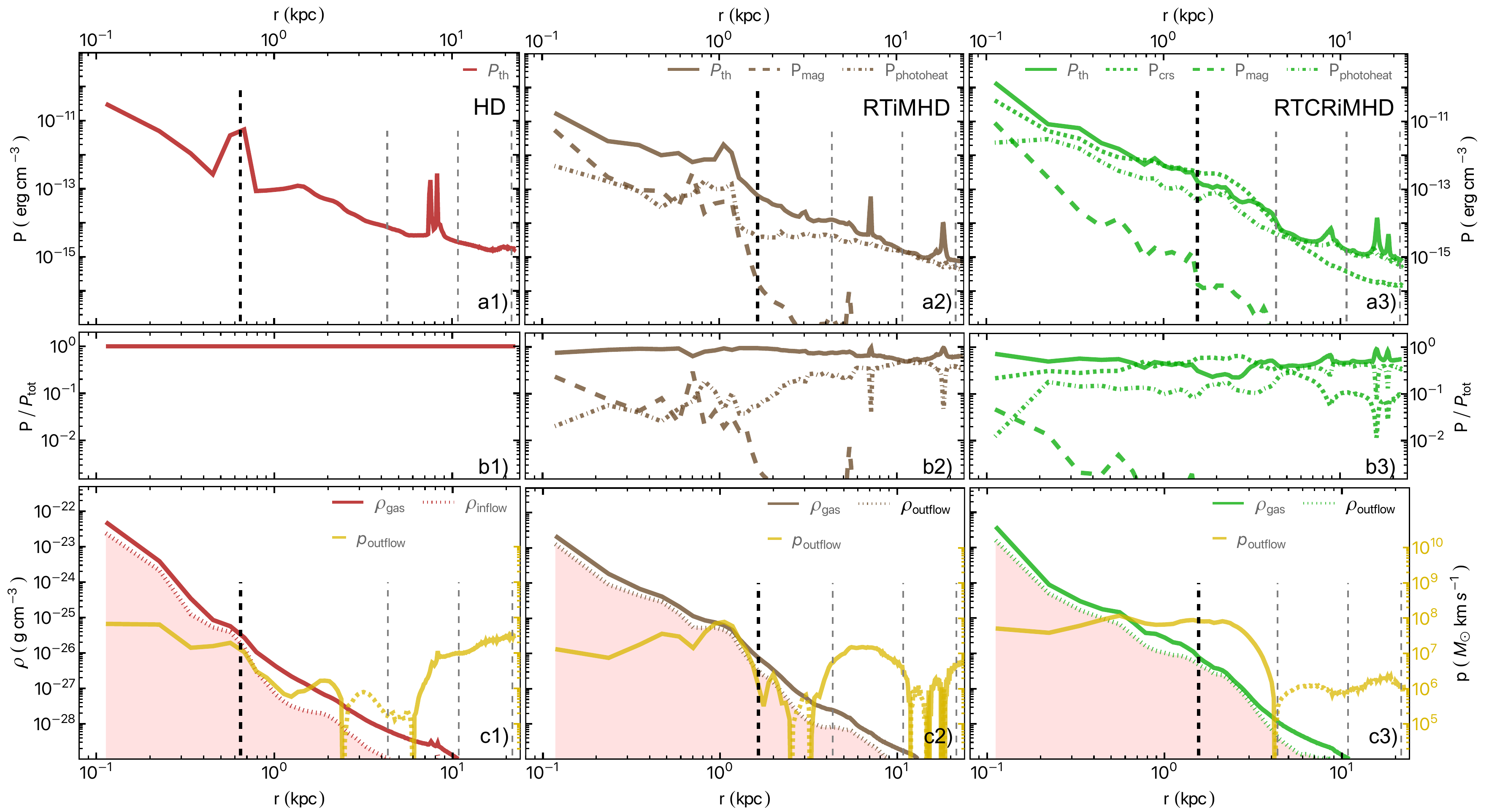}\\
    \caption{
    Radial profiles for various outflow-related quantities, centred on the \pandora~dwarf. Profiles correspond to star formation episodes at $z = 3.5$, and are  representative of times with non-negligible star formation (Appendix~\ref{ap:profile_variability}). The \RTCRiMHDSfFb~model in particular exhibits larger variations due to its burstier star formation.
    {\bf (Top row)} Pressure profiles for the thermal (solid), magnetic (dotted), radiative (dot-dashed), and cosmic ray (dashed) components.
    {\bf (Central row)} Fractional pressures for the same components as shown in the top row.
    {\bf (Bottom row)} Density profiles, separated into total (solid line) and outflowing components (dotted line). We overlay the radial profile of the net outflowing momentum using golden solid lines, displayed as dashed for radii when the net momentum is inflowing.
    We include additional vertical lines in each panel corresponding to twice the half-mass radii of the stellar component (thick black dashed line, $2\, r_{*}$), and the 0.2, 0.5 and 1.0 $\rhalo$ (gray dashed lines). 
    In the models without CRs, outflowing gas (e.g., $\sim 1\,\kpc$ in panel b2) is thermally supported (panels a1 and a2). Conversely, outflows in the \RTCRiMHDSfFb~model ($\sim 3\,\kpc$ in panel b3) are dominated by CR pressure (panel a3). Overall, in the \RTCRiMHDSfFb~simulation, thermal pressure dominates across most radii {-} although with important contributions from the other pressure components. CRs dominate in the outflow bubbles and provide significant support at $r \lesssim 3\,r_{*}$. Magnetic pressure is only globally important inside the galaxy, whereas the radiative photo-heating pressure becomes more important at large radii, driving the thermal pressure through photo-heating.
    }
    \label{fig:OutflowProfs}
\end{figure*}

To review how outflow-related quantities and their pressure contributions are distributed around the \pandora~dwarf galaxy, we show radial profiles representative of star formation episodes at $z = 3.5$ in Fig.~\ref{fig:OutflowProfs}. The top row of panels show the pressure profiles for different physical components: thermal (solid), magnetic (dotted), radiative (dot-dashed), and cosmic ray (dashed), whenever included. We provide further details on their computation in Appendix~\ref{ap:pressures}. To illustrate how representative these pressure profiles are, and study their variability in high-vs-low SFR regimes, we discuss time-aggregated radial pressure profiles in Appendix~\ref{ap:profile_variability}. The central row shows the fractional contribution to the total pressures for each of the pressure components shown in the top panels. The bottom row of panels display the density profiles, with solid lines representing the total density and dotted lines indicating the outflowing gas. The radial profile of the net outflowing momentum is overlaid with golden solid lines, and shown as dashed curve at radii where the net momentum is inflowing. Vertical lines in all panels mark radii of interest: a proxy for the size of the galaxy at twice the stellar half-mass radius (thick black dashed line) and the 0.2, 0.5, and 1.0 fractions of $\rhalo$ (gray dashed lines). 

The pressure profiles in the top row provide useful insights into the mechanisms supporting gas at different galactocentric distances, and serve as proxies for the pressure gradients that drive outflows \citep[see e.g.,][]{Curro2024}. The thermal pressure profiles approximately follow power laws (roughly following $\propto r^{-2}$) up to $\sim 0.5\,\rhalo$, with distinct, sharp peaks attributable to substructures. Overall, thermal pressures are comparable between the \HDSfFb~and \RTiMHDSfFb~models, while the \RTCRiMHDSfFb~simulation has a higher thermal pressure especially in the central region. Across all models, the thermal component dominates the pressure budget at all radii $r > 0.1 \kpc$, except some narrow spatial regions where $P_\text{CR}$ dominates. When reviewing other pressure components through their fractional contribution (middle row), magnetic pressure only dominates locally in high density regions of the ISM. This suggests that magnetic fields do not directly affect outflowing gas support, and intervene primarily through their contribution to star formation regulation. The photo-heating pressure dominates the contribution from stellar radiation. We find this pressure only dominates the local pressure in the regions where massive stars from recent star formation events are located. However, at larger radii outside the galaxy it is comparable to the thermal pressure, and is responsible for increasing the thermal pressure through its heating effects, featuring a higher energy budget%
\footnote{Note that at pressure equality, relativistic fluid energies such as the radiative one have a higher energy budget ($u_{\mathrm{rad}} = 3\,P_{\mathrm{rad}} = 3\,P_{\mathrm{CR}}$), whereas the thermal ($u_{\mathrm{th}} = \tfrac32 P_{\mathrm{th}}$) and magnetic ($u_{\mathrm{mag}} = P_{\mathrm{mag}}$) components have a $P / u$ ratio closer to unity.}.
 Interestingly, in our \RTCRiMHDSfFb~simulation, the photo-heating and CR pressures are comparable up to $\sim0.2\,\rhalo$, and moderately dominated by the CRs. The CR pressure is in fact the second most significant pressure source within $r < 0.2\,\rhalo$. The CR pressure also dominates in gas outflow fronts, particularly evident at $r \sim 3 {-} 4\,\kpc$ which correlates with the edge of the $\poutflow$ and outflowing gas density (panel c3). Beyond this radius, the CR pressure drops back to the underlying pressure profile. When the pressure profile is computed without the contribution from the outflowing gas the higher pressure of CRs at $r \sim 3\,\kpc$ is not present, and the non-outflowing CR pressure profile scales as an approximate power-law out to the edge of the dark matter halo ($r \in [1, 15]\,\kpc$). The bottom panels of Fig.~\ref{fig:OutflowProfs} provide further information about the relationship between the pressure profiles and gas outflows. The galactic outflows in the \HDSfFb~simulation, reflected by peaks in both density and momentum, show relatively unchanged pressures. The only exception is the SN-triggered pressure peak at $r \sim 0.6 \kpc$ accelerating a new outflow. In contrast, the \RTCRiMHDSfFb~outflows are characterized by concentrations of CR pressure, which decreases sharply beyond the outflowing momentum shells (e.g., $r \sim 2\,\kpc$), in agreement with outflow-injected CRs in the halo and CGM. By providing additional support against the pull from the gravitational potential, both the photo-heating and CRs pressures will promote denser outflows that will reach larger distances from the galaxy. This is reflected in higher net outflowing momenta at $r \lesssim 1\,\kpc$ in these two models ($\poutflow \sim 10^7 \Msun\,\kmps$ for \RTiMHDSfFb; $\poutflow \sim 10^8 \Msun\,\kmps$ for \RTCRiMHDSfFb) when compared with \HDSfFb~($\poutflow \sim 10^6 - 10^7 \Msun\,\kmps$). These pressure distributions and their properties are broadly preserved during episodes of non-negligible star formation, but vary at times of quiescence (sSFR $< 1\,\Gyr^{-1}$), especially for models with burstier star formation (Appendix~\ref{ap:profile_variability}).

\newcommand{\PDFwidth}{1.5}%
\begin{figure*}
    \centering
    \includegraphics[width=\PDFwidth\columnwidth]{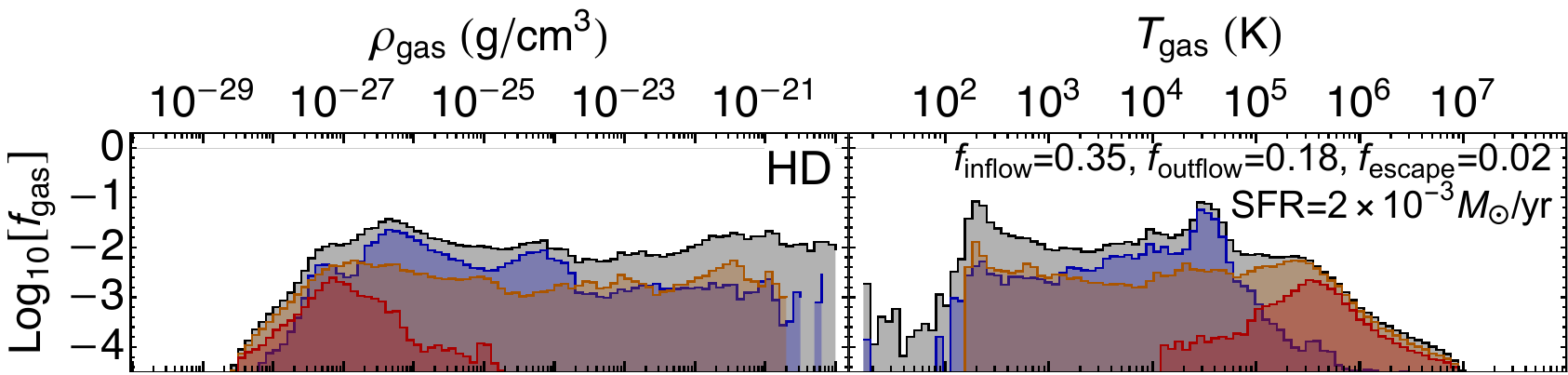}\\
    \includegraphics[width=\PDFwidth\columnwidth]{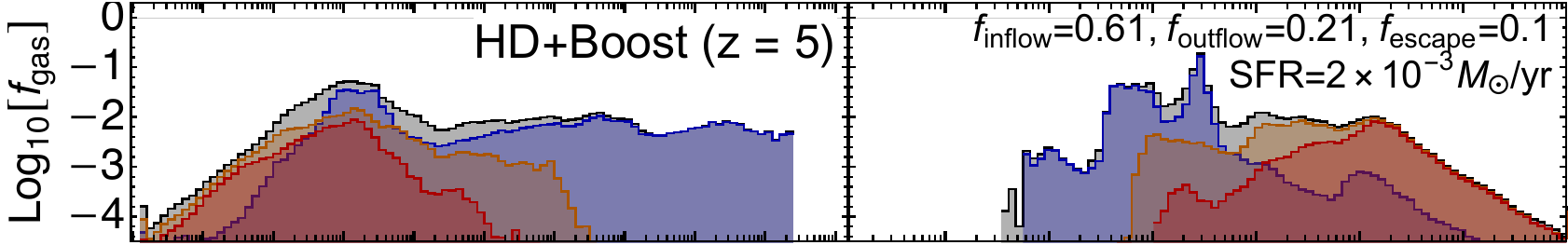}\\
    \includegraphics[width=\PDFwidth\columnwidth]{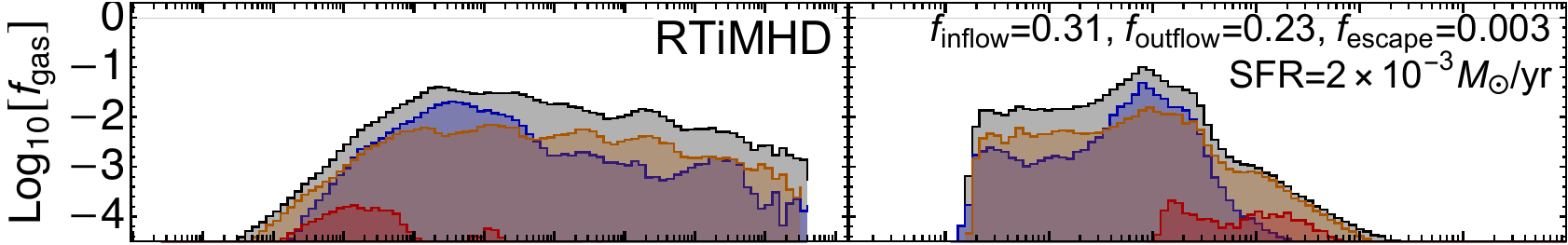}\\
    \includegraphics[width=\PDFwidth\columnwidth]{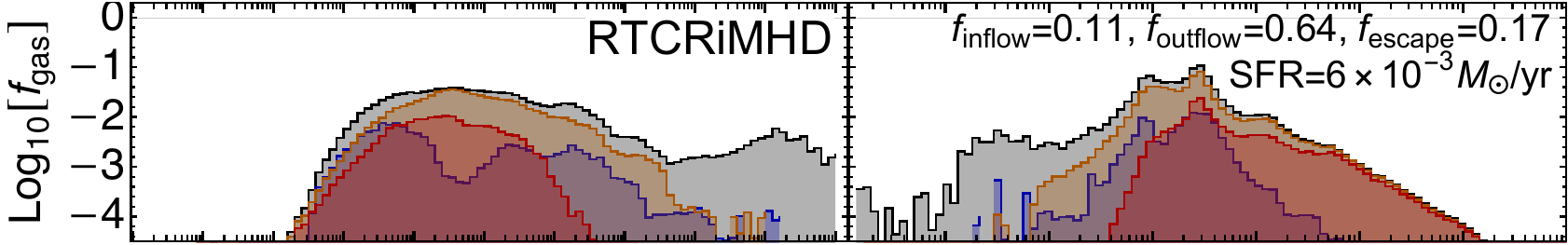}\\
    \vspace{-0.07cm}
    \includegraphics[width=\PDFwidth\columnwidth]{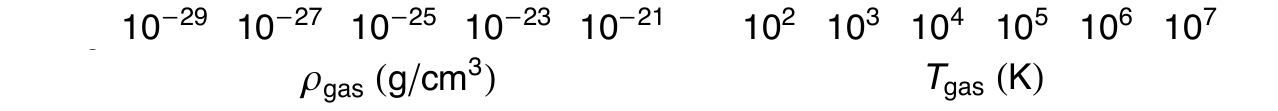}\\
    \includegraphics[width=0.8\columnwidth]{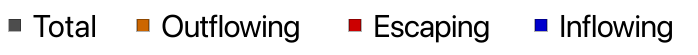}\\
    \vspace{-0.2cm}
    \caption{Gas mass fraction distribution functions as a function of gas density (left column) and gas temperature (right column) for the central galactic region ($r < 0.2\,\rhalo$). Distributions are shown for snapshots of non-negligible star formation activity in the galaxy, at $z = 3.5$ (except for \HDSfFbBoost, see main text). Total gas fraction (gray histograms), inflowing gas fraction (blue histograms, $v_\text{gas,radial} < \vvir \sim 12.5\kmps$), outflowing gas fraction (orange histograms, $v_\text{gas,radial} > \vvir$) and escaping gas fraction (red histogram, $v_\text{gas,radial} > \vesc (r = 0.2\,\rhalo) $) are shown for our four simulation models (top to bottom rows). Each temperature panel lists the SFR of each galaxy during the past 100~Myr, and the total mass fraction of the inflowing, outflowing and escaping gas component. The \RTCRiMHDSfFb~simulation has lower inflow rates as well as higher outflow and escaping rates per given SFR. The outflowing gas is also colder and denser, particularly notable in the escaping gas regime.}
    \label{fig:OutflowProps}
\end{figure*}

To review the physical properties of galaxy outflows in the context of the multi-phase ISM, we show in Fig.~\ref{fig:OutflowProps} the distribution of mass within the central galactic region ($r < 0.2\,\rhalo$) at times of non-negligible star formation. Gas mass is separated according to the flow properties. The panels show the gas mass fraction distribution as a function of gas density (left column) and gas temperature (right column). The gray histograms depict the overall gas mass fraction distribution for each simulation, and additional coloured histograms show the mass fractions of inflowing (blue histograms, $v_\text{gas,radial} < -\vvir \sim - 12.5\,\kmps$ to exclude gas with relatively low velocities), outflowing (orange histograms, $v_\text{gas,radial} > \vvir$), and escaping (red histograms). Escaping gas is selected according to $v_\text{gas,radial} > \vesc (r = 0.2\,\rhalo)$, with escape velocities $\vesc \sim 33 {-}37 \,\kmps$. Note that the mass of outflowing gas is a fraction of the total mass, and the mass of escaping gas is a fraction of the mass of outflowing gas.
From top to bottom, panels correspond to the \HDSfFb, \HDSfFbBoost, \RTiMHDSfFb, and \RTCRiMHDSfFb~simulations at $z \sim 3.5$ (except for \HDSfFbBoost\footnote{Due to lack of significant star formation past $z \sim 5$ in \HDSfFbBoost, at $z = 3.5$ the model does not have any prominent outflows. This is primarily due to the absence of any significant SN feedback for a prolonged time \citep{Agertz2009b}. Therefore, we opt to present its PDF during the star formation event at $z = 5$, when it features an SFR comparable to the other models at $z = 3.5$.
}, shown during its star formation event at $z = 5$), but selected to be representative of the dwarf galaxy across times when its SFR $\neq 0$. Additional integrated information about the current state of the galaxy is provided in the top right corner of the temperature panels: the SFR of each galaxy over the past 100~Myr, and the integrated fraction of gas mass in different outflowing phases.

We focus first on the left-hand panels, where gas densities typically range from $\rhogas \sim 10^{-28} $ to $\sim 10^{-20}\,\gcc$. In \HDSfFb~and \HDSfFbBoost~the mass fraction PDF is approximately uniform over all gas densities, with a moderate peak around $\rhogas \sim 10^{-27} - 10^{-26}\gcc$. Including radiative transfer in our \RTiMHDSfFb~model leads to further gas accumulation in the diffuse gas phase with $\rhogas \sim 10^{-26} - 10^{-24}\,\gcc$, and a reduced proportion of gas at higher densities. The PDF in the `full-physics' \RTCRiMHDSfFb~model exhibits the same behaviour as in the model including radiation  but has  a cut-off at low densities instead of the long tail seen for \HDSfFb, \HDSfFbBoost~and \RTiMHDSfFb. As shown by the red histograms, this tail towards lower densities is dominated by escaping gas, which typically exhibits an approximately log-normal shape peaking at $\rhogas \sim 10^{-26}\,\gcc$. However, in the \RTCRiMHDSfFb~model, the escaping gas features higher densities that reach the intermediate range ($\rhogas \sim 10^{-26} - 10^{-24}\,\gcc$), with comparable fall-offs towards both higher and lower densities. This is due to the denser and smoother outflows produced by CRs (\citealt{Girichidis2018}; see also the projection in Fig.~\ref{fig:front_image}). The \RTCRiMHDSfFb~simulations has gas densities $\rhogas < 10^{-28}\,\gcc$ only at very specific times, such as shortly after its SF burst at $z \sim 5$. Additionally, our \fullphys~simulation is the only model that features escaping gas at $\rhogas > 10^{-24}\,\gcc$. At even higher densities, gas with positive radial velocities becomes only outflowing (rather than escaping; orange histograms).

The right column in Fig.~\ref{fig:OutflowProps} shows the distribution of gas mass fraction as a function of temperature, where most of the mass sits between $10^2\, \K \lesssim T \lesssim 10^7\, \K$. The PDFs indicate a significant fraction of cold, dense gas in star forming regions ($T \sim 10^2\,\K$; e.g., \HDSfFb, \RTCRiMHDSfFb), a concentration of gas around the thermally stable $T \sim 10^4\, \K$, and a long tail towards high temperatures above $T \gtrsim 10^5\, \K$. Including stellar radiation leads to a smoother distribution between ($T \sim 10^4 {-}\,3 \cdot 10^4\,\K$), accentuated by hydrogen photo-ionization and photo-heating. Stellar radiation also reduces the proportion of gas at $T < 100\, \K$ which is present at almost all times with non-zero SFR, due to rapid evaporation of giant gas clouds from early stellar feedback. Including CRs increases the proportion of gas at $T \lesssim 500\,\K$, with a much lower fraction of low temperature gas being inflowing at $T \lesssim 10^{4}\,\K$.

Across all our models, we find most gas at temperatures above $10^6\, \Kelvin$ to be escaping (i.e., $\vrad > \vesc$) as this gas is heated by SN explosions. Including radiation reduces the proportion of escaping gas, and entrains a higher relative proportion of gas at lower temperatures ($T \sim 10^5 \,\K$). Combining stellar radiation with CRs yields a large proportion of outflowing and escaping gas. Notably, the escaping gas also has significantly lower temperatures, with some of the gas heated by stellar radiation now entrained in escaping outflows ($T \sim 3\cdot10^4\,\K$), leading to a more multi-phase nature of the outflows. We find for the \RTCRiMHDSfFb~model an average $\sim15\%$ (and a median of $\sim5\%$) of the outflow mass to be neutral gas, although there are large variation across cosmic time (with some snapshots having neutral-mass dominated outflows) . This is
in better agreement with multi-phase and neutral gas observations \citep[e.g.,][albeit mostly of more massive galaxies]{Chen2010, Fluetsch2019, Schroetter2019, Veilleux2020, Romano2023, DEugenio2023}. This result is particularly notable when comparing with the \HDSfFbBoost~case, for which the outflows feature exceptionally high temperatures, and are virtually dominated by the hot and ionized gas phase. The mass-loading in different phases is further discussed and compared with observations in the next section. 

\subsection{Comparing the properties of the \pandora~suite with observations}  
\label{ss:outflow_observables}

\subsubsection{Warm and ionized outflow mass-loadings and velocities}

\begin{figure}
    \centering
    \includegraphics[width=\columnwidth]{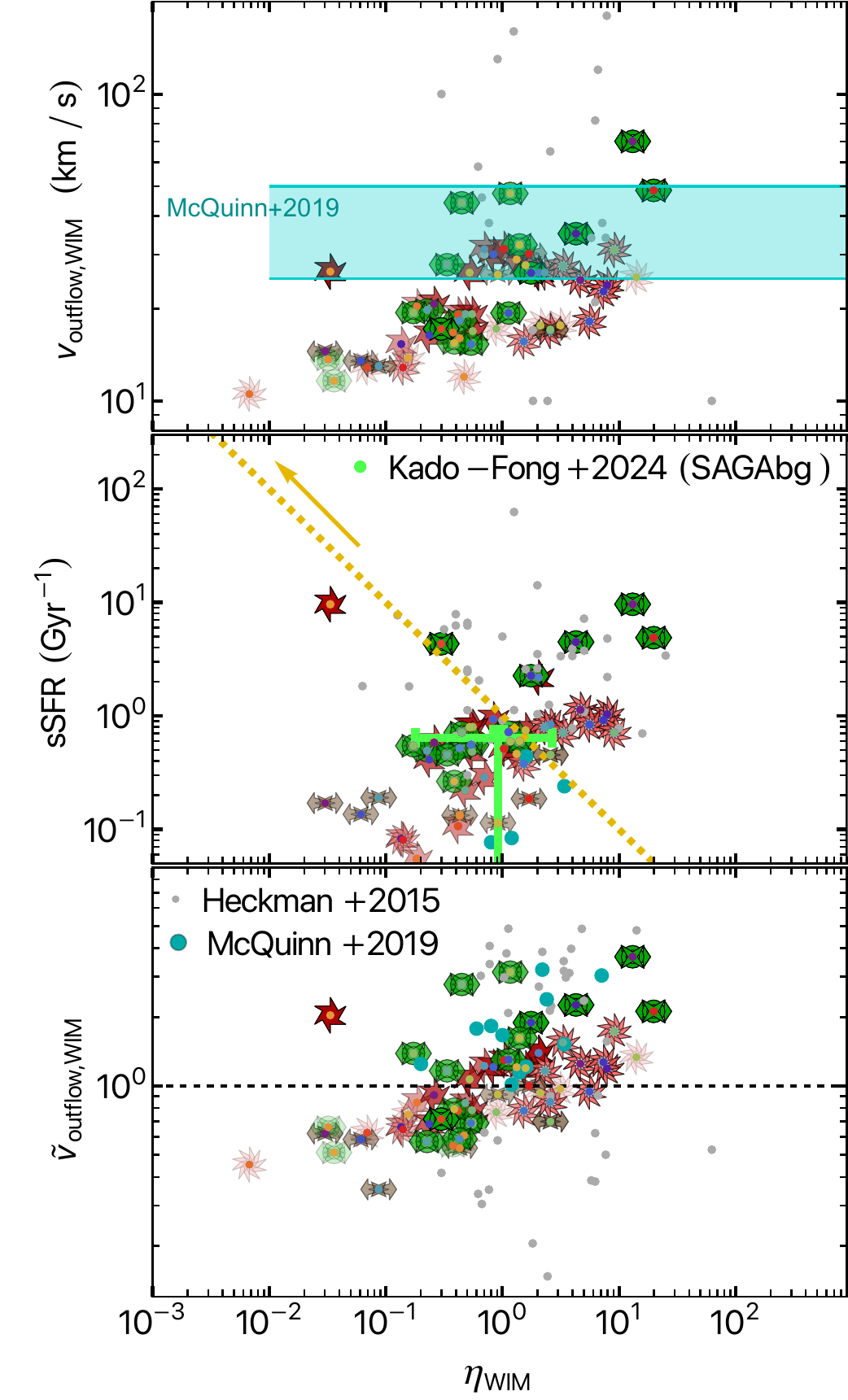}\\
    \rowlegend
    \caption{
    Intrinsic (simulated) properties of the warm ionized outflows for the Pandora dwarf galaxy. From top to bottom, panels display the following quantities as a function of the WIM mass-loading factor: the mass-weighted WIM gas outflow velocity, $\voutflowWIM$, the specific SFR (sSFR) over the last 100 Myr; and a dimensionless velocity ratio $\nuoutflow = \voutflowWIM / \vcirc$. Different data-points from our simulations correspond to different snapshots in the interval of $3.5 < z < 5$. Snapshots featuring considerably lower star formation rates are displayed with lower opacity symbols. The gold coloured line in the central panel indicates increasing SFR for a fixed outflow and stellar mass. We overlay as coloured circles the redshift of the models, ranging from high redshift (blue) to low redshift (red). We find no clear systematic trends across the studied period. The inclusion of radiation which acts as an early stellar feedback process results in slower and weaker outflows. However, when CR feedback is incorporated as well, WIM has much larger outflow velocities and mass-loading factors.
    Fig.~\ref{fig:ObsLikeOutflows} shows comparable quantities, measured instead following an observation-like procedure for H$\alpha$ emission.}
    \label{fig:OutflowVsObs}
\end{figure}

We now turn to investigate how the different outflow properties we measure in the simulations compare with the observed properties of ionized gas. The separation into ionized and neutral gas is computed through a simple separation according to the hydrogen ionization fraction $x_\text{HII}$: neutral gas density, $\rho_\text{HI} = \rho_\text{gas} (1 - x_\text{HII})$, and ionized gas density, $\rho_\text{HII} = \rho_\text{gas} x_\text{HII}$.
In Fig.~\ref{fig:OutflowVsObs}, where we focus on intrinsic quantities as directly calculated from the simulations, we show how the mass-loading factor of warm ionized medium (WIM) varies across our models during the $z \in \left[5, 3.5\right]$ interval\footnote{For observation-like analysis of the WIM and WNM outflows see Section~\ref{sss:ObsLikeOutflows} and \ref{sss:WNMOutflows}.}. We classify cells to be within the WIM and WNM temperature range when they fulfill $200\,\Kelvin < T \leq 10^6\,\Kelvin$, with their gas mass contributing proportionally to their ionization fraction. From top to bottom, panels show the mass-weighted WIM gas outflow velocity $\voutflowWIM$, the specific SFR (sSFR) measured in a time window of $100$~Myr, and a dimensionless velocity ratio $\nuoutflow$. 
This quantity measures the relevance of outflow velocities versus the characteristic circular velocity, $\vcirc$, of the system as $\nuoutflow = \voutflowWIM / \vcirc$. We measure $\voutflowWIM$ and the $\etaWIM$ factor at $0.2\,\rhalo$. Each data point corresponds to a different snapshot, displayed with higher opacities when the SFR is higher. We also indicate the redshift evolution by including coloured circles in each of our data points, ranging from high redshift (blue; $z \sim 5$) to low redshift (red; $z \sim 3.5$). These provide some guidance on whether any systematic trends emerge across the studied period. We find that there are no apparent trends beyond those driven by the temporal variations of the SFR.

The panels show outflows to have the highest mass-loading factors in the \HDSfFbBoost~and \RTCRiMHDSfFb~models, typically with $\etaWIM \sim 1-10$ for moderate star formation events. The \RTiMHDSfFb~model has the lowest mass-loading factors overall, with $\etaWIM \sim 0.1 -1$. These mass-loading factors reflect our previous results, where radiation dampens the mass-loading of the outflows while CRs increase it. The mass-weighted warm ionized gas outflow velocities, $\voutflowWIM$, shown in the first panel, vary significantly across our different models. The \HDSfFb~and \HDSfFbBoost~simulations have outflows with relatively narrow velocity ranges, $\voutflowWIM \sim 15 - 30\,\kmps$. The velocities in the \RTiMHDSfFb~model with radiation are reduced to $\sim$$15\,\kmps$. The \RTCRiMHDSfFb~model has the highest mass-weighted outflow velocities, at times reaching $\sim 70\,\kmps$. 
Perhaps with the exception of the \RTiMHDSfFb~model, all the models show reasonable agreement with the outflow velocities inferred by the local observations by \citet{Heckman2015} and those assumed by \citet{McQuinn2019}, included in the figure for reference\footnote{We note here that our simulations are of a dwarf galaxy at high redshift, while these observations are of local dwarfs. This qualitative comparison is intended for systems of similar mass and SFR, and does not inform us about the cosmological evolution of dwarf galaxies.}.

The second panel of Fig.~\ref{fig:OutflowVsObs} shows the relation between $\etaWIM$ and the sSFR, where all our models feature some degree of the expected positive correlation between increasing sSFR and $\etaWIM$ {--} a larger fraction of mass is ejected as the relative importance of the ongoing SFR with respect to the galaxy mass increases. To aid interpretation we include a dashed golden line of constant stellar and outflow mass, with the arrow indicating increasing SFR. Overall, all our models display this positive correlation of higher outflow rates for higher sSFR. The \HDSfFb~and \HDSfFbBoost~models have comparable sSFR values, typically of the order of $1\,\text{Gyr}^{-1}$ whenever WIM outflows are present, and lower otherwise. The \RTiMHDSfFb~has somewhat lower sSFR values $\lesssim 0.5\,\text{Gyr}^{-1}$ and outflow rates. \RTCRiMHDSfFb~displays higher WIM outflow mass-loading factors, as well as higher sSFR, spanning a larger dynamical range, attributed to its higher star formation burstiness. This supports the connection between high star formation burstiness, feedback and galactic outflows \citep{Carniani2024}, and poses non-thermal physics as an important regulator of their interplay. Notably, the bulk of our galaxy measurements is in very good agreement with the inferred mass-loading factors by \citet{Heckman2015} as well as \citet{Kado-Fong2024} which employ background galaxies (SAGAbg) observed by the SAGA Survey \citep{Geha2017}. 

As shown in \citet{Martin-Alvarez2023}, our different galaxy formation models lead to distinct predictions for the dynamical masses and sizes of the resulting galaxy. Mass and size of the galaxy are thereby related to the circular velocity as $\vcirc = \sqrt{G M / r}$. To estimate the relevance of outflow velocity per system, we propose the following diagnostic: a dimensionless $\nuoutflow (r = 0.2\,\rhalo) = \voutflowWIM / \vcirc (r)$ against the dimensionless mass-loading factor, shown in the bottom panel of Fig.~\ref{fig:OutflowVsObs}. We include a dashed horizontal black line at $\nuoutflow = 1$ to separate the low and high $\nuoutflow$ regimes. This plot highlights how the models studied here separate in the $\nuoutflow - \etaWIM$ parameter space: at low mass-loading factors ($\etaWIM < 1$), all models are comparable, with outflowing gas moving at relatively low characteristic speeds ($\nuoutflow \sim 0.8$). However, as $\etaWIM$ increases, we find outflows with comparatively different characteristic velocity ratio $\nuoutflow$. The \RTiMHDSfFb~simulation not only typically has the lowest mass-loading, but also the slowest $\nuoutflow$. In agreement with our previous findings, \HDSfFb~reaches intermediate values for both the high end of $\nuoutflow$ and $\etaWIM$. Notably, while both \HDSfFbBoost~and \RTCRiMHDSfFb~reach the highest $\etaWIM$ values, \RTCRiMHDSfFb~has the highest $\nuoutflow$ overall. Hence, the $\nuoutflow (\etaWIM)$ parameter space predicts different outflow properties for thermal versus non-thermal galaxy formation models allowing to observationally differentiate models. 

\subsubsection{Observation-like measurement of ionized outflows}
\label{sss:ObsLikeOutflows}

\begin{figure}
    \centering
    \includegraphics[width=\columnwidth]{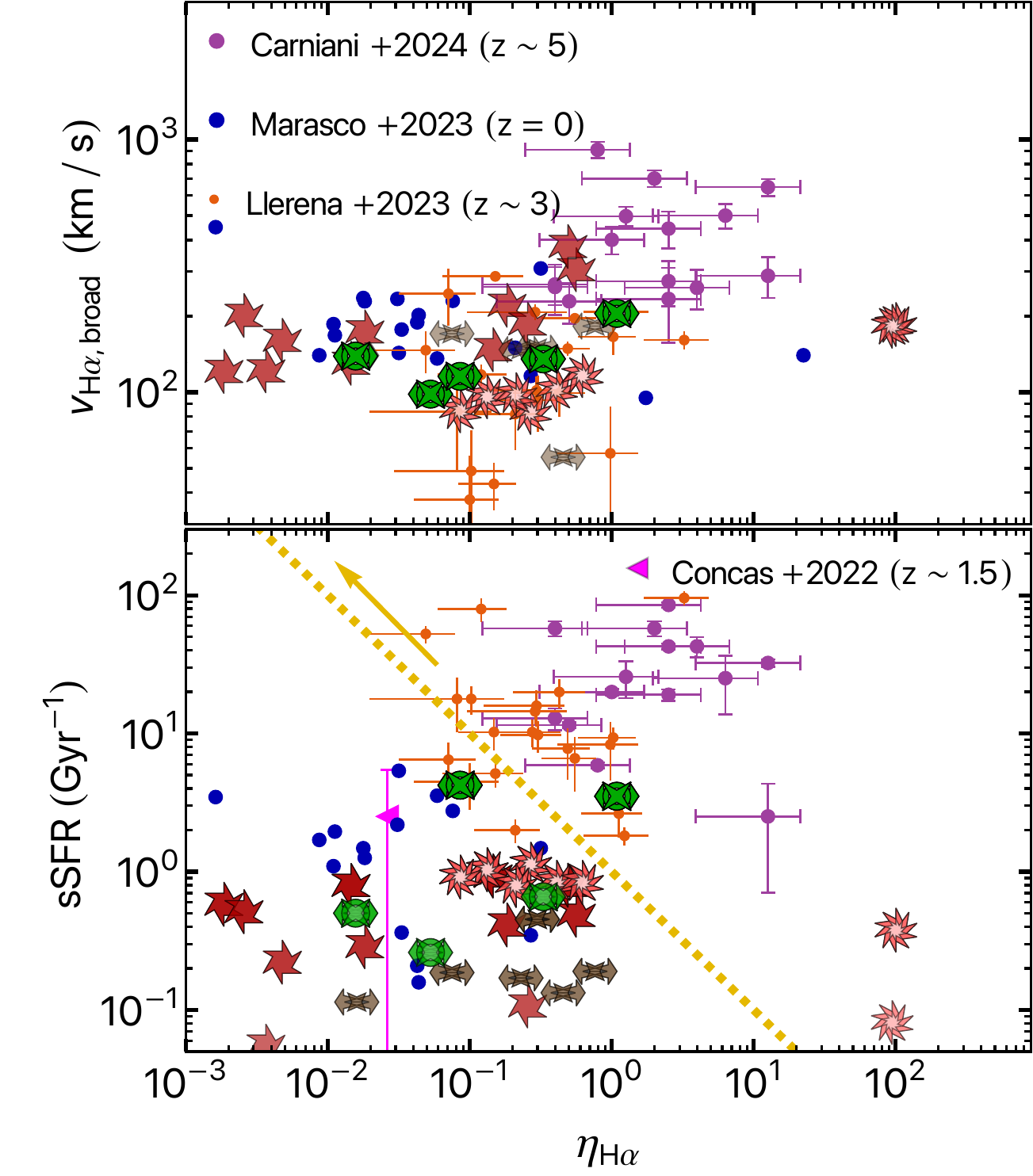}
    \caption{Observation-like properties of the ionized outflows for our simulated galaxy, as measured from synthetic H$\alpha$ line profiles. The top panel displays the outflow velocity $\vHalpha$ measured from the broad component of the line, a proxy for the $95^\text{th}$ percentile of the WIM velocity; whereas the bottom panel shows the sSFR over the last $100$~Myr. Both are shown as a function of the observationally inferred mass-loading factor, $\etaHalpha$ (see Appendix~\ref{ap:obs_like_measurements} for estimation of $\vHalpha$ and $\etaHalpha$). Our measurements are compared with observational results inferring outflow properties through the methodology we emulate, spanning from low ($z = 0$, \citealt{Marasco2023}; $z \sim 1.5$, upper limit/non-detection for the low-stellar mass bin by \citealt{Concas2022}), to high ($z \sim 3$, \citealt{Llerena2023}; $z \sim 5$, \citealt{Carniani2024}) redshift. Overall, we find most of our simulated models to be in broad agreement with observations for the overlapping range of stellar masses.}
    \label{fig:ObsLikeOutflows}
\end{figure}

Various observational studies extract the properties of galaxy outflows from the analysis of emission line profiles \citep[e.g.,][]{Fluetsch2021, Marasco2023, Carniani2024}. By modelling these as a combination of Gaussian components, the broad component can be identified as associated with the gas outflowing from galaxies. The integral of its emission (or in some models, only a fraction of its phase space, e.g., \citealt{Marasco2023}) is then related to the outflowing mass $M_\text{outflow}$, and its FWHM to the outflow velocity, which we label $v_\text{broad}$. These two quantities and the estimate for the size of the galaxy $r_\text{galaxy}$, allow measuring the outflow rate as
\begin{equation}
    \dot{M}_\text{outflow} = \frac{M_\text{outflow} v_\text{broad}}{r_\text{galaxy}}.
    \label{eq:obs_outflow_rate}
\end{equation}
In this section, we follow a simple procedure aimed to follow such studies to estimate observation-like outflow properties. We outline the key steps, and highlight the most important caveats. A more detailed discussion about these is provided in Appendix~\ref{ap:obs_like_measurements}. 

For each galaxy, we compute the H$\alpha$ emission (estimated following \citealt{Katz2022b}) line profile accordingly to the line-of-sight velocity for 12 uniformly-distributed directions. We fit a double Gaussian to the line profiles, and associate the broad component with the outflowing gas. Following \citet{Rupke2005}, the outflow velocity $v_\text{broad}$ is frequently defined as
\begin{equation}
    v_\text{broad} = \frac{\text{FWHM}_\text{broad}}{2}+\Delta v,
    \label{eq:Rupke2005}
\end{equation}
where $\text{FWHM}_\text{broad}$ is the full-width-at-half-maximum of the broad component, and $\Delta v$ is the difference between the two components mean velocity. This velocity is often associated to the $95^\text{th}$ percentile of the velocity distribution (which we confirm in Fig.~\ref{fig:BroadVs95}), but does not necessarily correspond to the outflowing gas mass-weighted velocity (see Appendix~\ref{ap:obs_like_measurements}). The outflow mass is extracted from the total luminosity of the broad component, $\LHalpha$, following the relation employed by \citet{Concas2022} and \citet{Marasco2023}
\begin{equation}
    M_\text{outflow} = 3.2 \cdot 10^{5} \frac{\LHalpha}{10^{40}\,\text{erg}\,\text{s}^{-1}} \left(\frac{100\,\text{cm}^{-3}}{n_e}\right)\, \text{M}_\odot,
    \label{eq:obs_outflow_mass}
\end{equation}
where $n_e$ is the electron number density. Here, we follow observations and assume $n_e \sim 300\,\text{cm}^{-3}$, although we note that the typical outflow densities in our simulations are lower and that $n_e$ has a clear radial gradient\footnote{This assumptions about $n_e$ may bias our $M_\text{outflow}$ measurements low, and is further discussed in Appendix~\ref{ap:obs_like_measurements}.}. Finally, we set $r_\text{galaxy}$ as twice the half-mass radius $r_{*}$.

Following this procedure, we obtain the mass-loading factors, $\etaHalpha$, and compare this with the observation-like outflow velocity, $\vHalpha$, and the sSFR\footnote{For consistency we maintain the sSFR measured in the same time window of $100$~Myr, but note that measuring the SFR over shorter time periods may reveal higher peaks of SFR, particularly for higher burstiness models. This would shift the simulated data points along the golden dashed line in the $\text{sSFR} - \eta$ panels as indicated by the arrow. Detailed comparisons with different observations may require adjusting the timescale accordingly to the specific observational SFR tracer.} in Fig.~\ref{fig:ObsLikeOutflows}. Overall, we find $\vHalpha$ to be similar across our models, with values in reasonable agreement with both the observations at low-redshift by \citet{Marasco2023} and at high-redshift ($z \sim 3$) by \citet{Llerena2023}. The velocities inferred by \citet{Carniani2024} at $z \sim 5$ are somewhat higher ($\gtrsim 200\,\km\,\s^{-1}$), although their stellar masses and sSFR are also higher ($M_* \gtrsim 5\cdot10^{7}\Msun$ compared with our $M_* \lesssim 10^{7}\Msun$). We however stress that our mock outflow velocities $\vHalpha$ and their trend with the mass-loading factor are not an accurate reflection of mass-weighted WIM outflows as directly calculated from our simulations\footnote{The \HDSfFb~model appears to have somewhat higher $\vHalpha$ than the other models. We also note that \HDSfFb~also has a higher amount of valid double Gaussian fits.}. All of our models compare reasonably well with observations in the sSFR - $\etaHalpha$ relation. By reaching higher sSFR values in some snapshots, the \RTCRiMHDSfFb~model is also comparable with some higher redshift observations, although further sampling than provided by our simulations will be required. 
Building up on this apparent agreement for most of our models with observations of ionized outflows, we highlight the importance of forward-modelling the simulation results to bridge the gap when comparing with observations. This enables to better constrain the models as well as to infer the `true' outflow properties from observations. Doing this causes non-negligible variations that reduce our mass-loading factors and increase the outflow velocities with respect to our results in Fig.~\ref{fig:OutflowVsObs}. In future work we plan to employ more sophisticated forward-modelling techniques, and expand on this analysis with the larger sample of galaxies from the Azahar simulations \citep{Martin-Alvarez2025aas}. 

\begin{figure}
    \centering
    \includegraphics[width=\columnwidth]{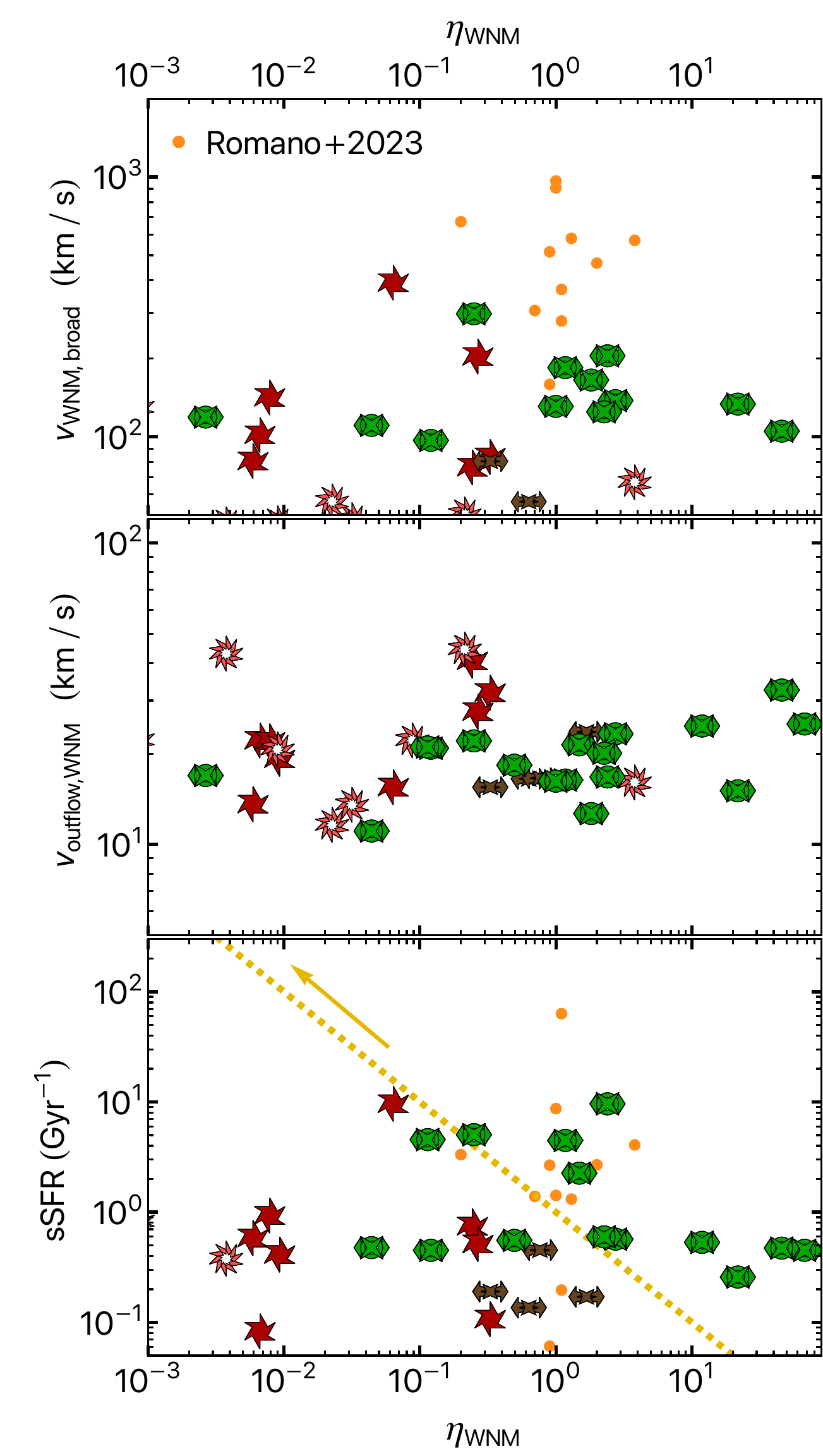}\\
    \rowlegend
    \caption{From top to bottom, panels display $\vbroadWNM$ (a proxy for the $95^\text{th}$ percentile of the WNM velocity distribution), the mass-weighted WNM outflow velocity $\voutflowWNM$, and the sSFR. All quantities are shown as a function of the WNM mass-loading factor $\eta_\text{WNM}$, directly measured from the simulations. The \RTCRiMHDSfFb~model is capable of driving temperate outflows with considerably higher neutral gas mass-loading factors than its purely hydrodynamical counterparts, and is in good agreement with the observations by \citet{Romano2023}.}
    \label{fig:OutflowWNMhalo}
\end{figure}

\subsubsection{Neutral outflow mass-loadings and velocities}
\label{sss:WNMOutflows}
As shown above, galaxy winds in the presence of CRs entrain a larger proportion of denser and more temperate gas \citep[see also e.g.][]{Girichidis2018, Dashyan2020, Hopkins2021a, Farcy2022, Curro2024}. We analyse the capability of ejecting warm neutral medium (WNM) outflows from the simulated dwarf galaxy in Fig.~\ref{fig:OutflowWNMhalo}, which shows from top to bottom, the $\vbroadWNM$, the mass-weighted WNM outflow velocity, and the sSFR as a function of the WNM mass-loading factor. The \RTCRiMHDSfFb~model is particularly effective at driving WNM outflows, with $\eta_\text{WNM} \sim 10^0 - 10^1$, significantly higher than for our purely hydrodynamical simulations (\HDSfFb~and \HDSfFbBoost). Beyond the pressure gradients associated with non-thermal processes {--} which provide a mechanism to drive denser gas outflows without heating them to higher temperatures {--} the RT-driven local photo-heating of gas (bottom panel of Fig.~\ref{fig:SN_PDF}) also increases the fraction of partially ionized gas at $T \lesssim 3 \cdot 10^4\, \Kelvin$ which is entrained in outflows. In the \RTCRiMHDSfFb~model our dwarf galaxy exhibits a higher fraction of events with larger $\vbroadWNM$ values, which are in reasonable agreement with the observations by \citet{Romano2023}. While $\vbroadWNM$ is comparable or somewhat higher for the \RTCRiMHDSfFb~model, the mass-weighted outflow velocity in this model is lower than in the thermally-dominated feedback models, reflecting a higher proportion of the ejected mass that is moving at lower velocities. Finally, by examining the sSFR - $\eta_\text{WNM}$ relation, we find that the combination of burstiness in the \RTCRiMHDSfFb~model with its higher $\eta_\text{WNM}$ leads to a good agreement with the observations by \citet{Romano2023}, highlighting how observations of neutral outflows can help discriminate between models featuring different non-thermal physical processes.

Finally, we note that we measure only a very small proportion of cold neutral outflows (used as a proxy for molecular outflows; $T \leq 200\,\Kelvin$), and mostly only present in the \RTCRiMHDSfFb~model. Whether such outflows in dwarf galaxies, which already feature low CO detections \citep{Schruba2012}, may only be driven by AGN \citep{Mezcua2016}, or are even not expected \citep{Barfety2025} remains to be understood.

\subsubsection{Galaxy stellar metal enrichment and gas outflow metallicities}
\label{ss:metal_enrichment}

\begin{figure*}
    \centering
    \includegraphics[width=1.205\columnwidth]{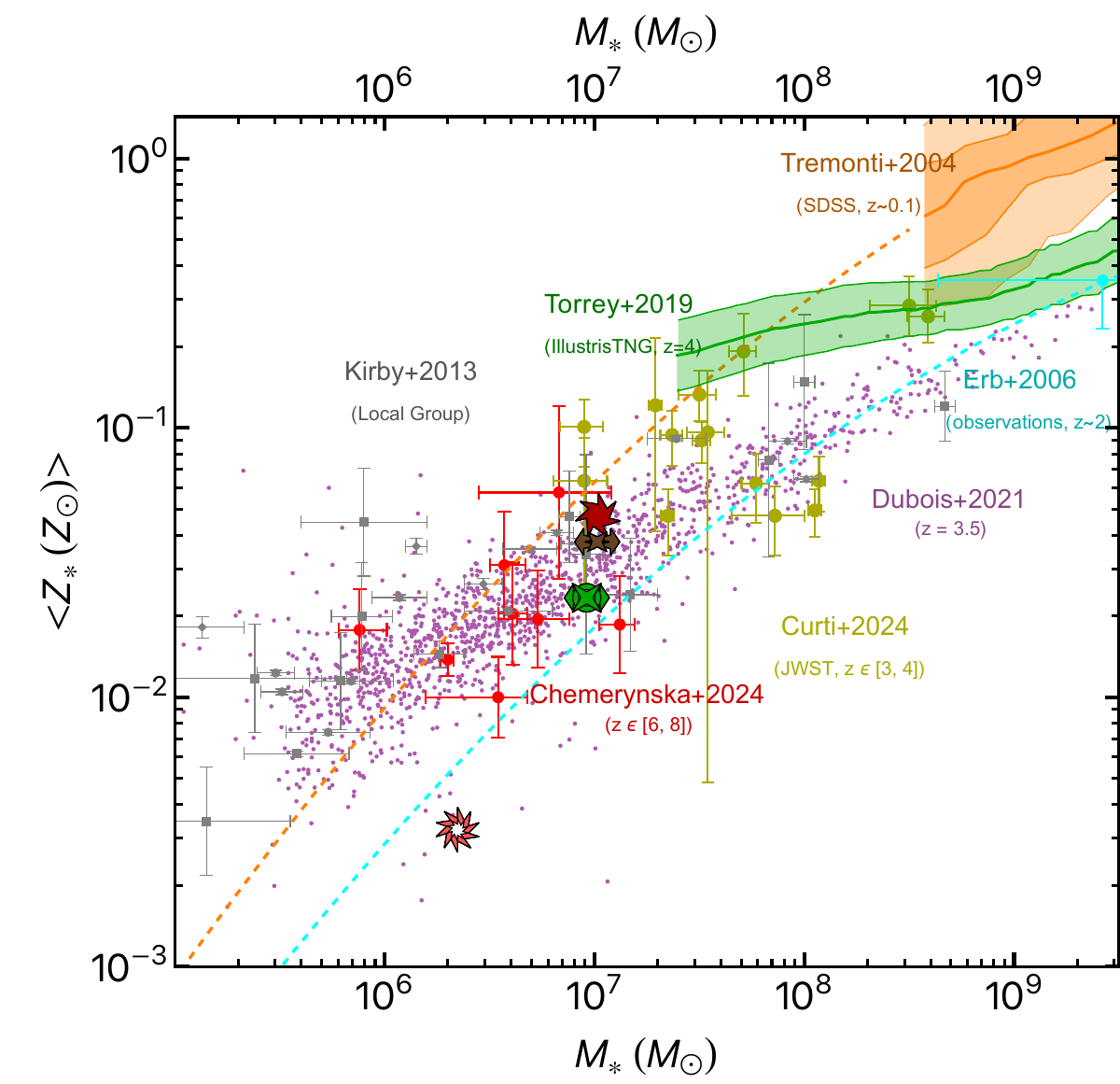}%
    \hspace{0.1cm}%
    \includegraphics[width=0.8\columnwidth]{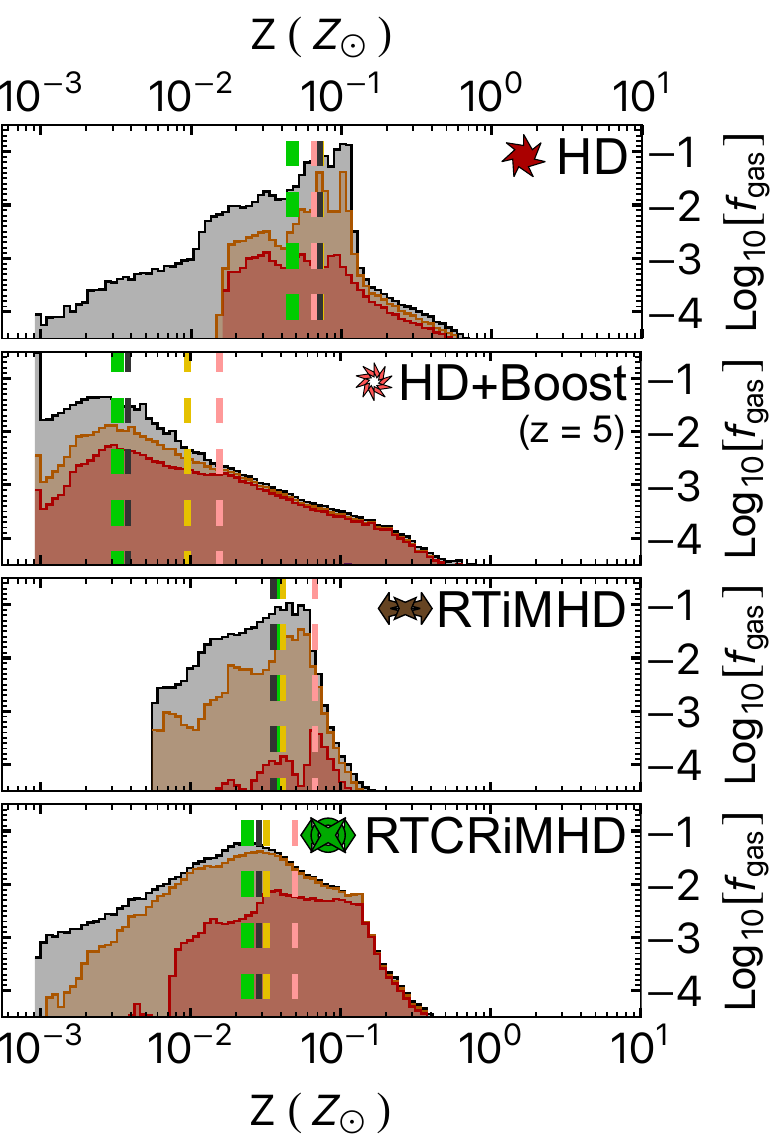}\\
    \caption{{\bf (Left panel)} Stellar mass--stellar metallicity relation of the \pandora~dwarf galaxy at $z = 3.5$ for our different models. We show for comparison local dwarf galaxies observations (gray points) by \citet{Kirby2013} as well as higher redshift observations employing \textit{JWST} by \citet{Curti2024} at $3 < z < 4$ (yellow points) and \citet{Chemerynska2024} at $ 6 < z < 8$ (red points). We include SDSS data of galaxies with larger stellar masses at low \citep{Tremonti2004} (orange bands) and high \citep[$z \sim 2$,][]{Erb2006} redshift (cyan points and cyan dashed line). We also show data for the NewHorizon \citep[$z = 3.5$, violet data points; ][]{Dubois2021} and IllustrisTNG \citep[$z = 4$, green band; ][]{Torrey2019} simulations for comparison.
    Most of our models provide a reasonable match to observations, except the `boosted' feedback model \HDSfFbBoost.
    {\bf (Right column panels)} Gas mass metallicity PDFs, with gas (black distribution) separated into outflowing (orange) and escaping (red) as done in Fig.~\ref{fig:OutflowProps}. Vertical dashed lines correspond to the average stellar (green), gas (black), outflowing (gold) and escaping (pink) metallicities. The \RTCRiMHDSfFb~simulation has a large fraction of escaping gas which is highly metal enriched, as well as a significant outflowing component that has a broad range of metallicities.}
    \label{fig:Metallicities2}
\end{figure*}

Another important galactic property that is notably affected by galactic outflows is metal enrichment, as the primary source of enrichment is the SN feedback that is also responsible for accelerating such outflows. The left panel of Fig.~\ref{fig:Metallicities2} shows how the stellar metal enrichment of the simulated galaxies vary across our models, whereas the right column of panels provides further detail on the outflowing gas metallicity. Focusing first on the left panel, we show the average stellar metallicity for the \pandora~suite compared with Local Group observations \citep{Kirby2013}, higher redshift \textit{JWST} observations of gas metallicities in large dwarf galaxies by \citet{Curti2024} (we restrict the comparison to their data at $3 < z < 4$) and smaller dwarfs (at $ 6 < z < 8$) by \citet{Chemerynska2024}. We show extrapolated metallicities for low-redshift \citep{Tremonti2004} and high-redshift \citep{Erb2006} SDSS observations. See also e.g., \citet{Zahid2013} and \citet{Curti2020} for additional mass–metallicity relations across redshifts. We also include predictions at $z \sim 3.5$ from the high-resolution cosmological simulation NewHorizon \citep{Dubois2021}, and from IllustrisTNG \citep{Torrey2019} to contextualise our simulations. Due to the expected lack of growth and enrichment evolution of the galaxy down to $z \sim 0$ (\citealt{Martin-Alvarez2023}; see also low-$z$ measurements in Appendix~\ref{ap:metallicities_allmodels}), comparisons to low redshift data are insightful. Such comparison is particularly relevant when considering recent findings by \citet{Curti2024}, indicating low-mass galaxies exhibit metallicities comparable to local analogues such as 'Blueberry' and 'Green Pea' galaxies \citep{Yang2017a, Yang2017b}.

All of our models have stellar metallicities that match observational data, with the exception of the model with enhanced feedback, \HDSfFbBoost, which leads to metallicities significantly lower than observed, as most of its metals are ejected from the galaxy. This emphasizes that stellar metallicity is a powerful diagnostic tool for distinguishing between different galaxy formation models. We briefly describe how the larger sample of \pandora~models occupies this enrichment space in Appendix~\ref{ap:metallicities_allmodels}. 
The right-hand panels of Fig.~\ref{fig:Metallicities2} showcase the metallicity distribution of the gas in the galaxy (black), separated into outflowing (orange) and escaping (red) gas following the same approach as for Fig.~\ref{fig:OutflowProps}. We include vertical dashed lines corresponding to the average metallicities of the stellar component (green line), total gas component (black line), the outflowing gas (golden line) and the escaping gas (pink line). 
Across most models, the average stellar metallicity is comparable to that of the outflowing gas, with the notable exception of the \HDSfFbBoost~simulation. This disparity emerges from the too efficient ejection of metals from the galaxy after 'boosted' SN feedback events, which in turn restricts the increase of the stellar metallicity, leading to the disagreement with observations. This over-efficient ejection of metals is showcased by the PDF of escaping gas, which contains virtually all gas with $Z > \aveZstar$ in \HDSfFbBoost. Instead, the models featuring $Z_*$ comparable to observations are capable of retaining some of their enriched gas with $Z \sim \aveZstar$. 
The \HDSfFb~and \HDSfFbBoost~ models have comparable metallicity distributions shapes for outflowing and escaping gas at $Z > \aveZstar$. This is due to their explosive, single acceleration mechanism through SNe. In contrast, the non-thermal physics models have more differentiated outflowing and escaping gas PDFs at $Z \gtrsim \aveZstar$. This is not only due to non-thermal pressures contributing to gas acceleration, but also to different local small-scale environmental properties around SN events (Fig.~\ref{fig:SN_PDF}). In addition to having denser and more temperate escaping outflows, \RTCRiMHDSfFb~escaping outflows also have a heterogeneous mixture of different metallicities, interestingly also containing a non-negligible fraction of low-metallicity material. Our findings confirm metal enrichment observations of dwarf galaxies as valuable diagnostics to distinguish not only outflow driving mechanisms, but also the physics of galaxy formation models more generally. In upcoming work (Martin-Alvarez et al. in prep.) we will address how different galaxy formation models affect the spatial distribution of such metallicities (Fig.~\ref{fig:front_image}, where observables such as metallicity gradients having the potential to further constrain dwarf galaxy formation models \citep[e.g.,][]{Fu2024a, Fu2024b}. Complementarily, galaxy outflows from dwarf galaxies drive the observational signatures of their CGM enrichment \citep{Werk2014, Prochaska2017, Li2020, Zheng2024}, serving as a further constraint on galaxy formation models with different outflow characteristics. 

\section{Conclusions}
\label{s:Conclusions}
In this work, we investigate the interplay between star formation, stellar feedback, galaxy outflows, and metal enrichment in dwarf galaxies. We analyse a representative subset of the \pandora~galaxy formation models \citep{Martin-Alvarez2023}, focusing on a high-resolution ($\Delta x \sim 7,\pc$) cosmological zoom-in simulation of a dwarf galaxy with halo mass $M_\text{vir}(z = 0) \approx 10^{10}\, \Msun$. The models studied include: a standard hydrodynamic simulation (\HDSfFb), a 'boosted' SN feedback model (\HDSfFbBoost), a model with magnetohydrodynamics and stellar radiation via radiative transfer (\RTiMHDSfFb), and a `full-physics' simulation incorporating stellar radiation, magnetic fields, and cosmic rays (\RTCRiMHDSfFb).

We examine how the star formation history --- and in particular, its burstiness --- influences the galaxy's ability to launch outflows, and how this is modulated by the local environments of SN explosions. We further study how different forms of non-thermal pressure shape the properties of outflowing and escaping gas. Finally, we compare our simulation results to observations of galaxy-scale outflows and metal enrichment. Our main findings can be summarized as follows:

\begin{enumerate}
    \item The star formation history of the dwarf galaxy is highly sensitive to the included physics. In the standard hydrodynamic model (\HDSfFb), star formation proceeds continuously with mild SN-driven variability. Adding radiation smooths these short-timescale fluctuations by suppressing small-scale clustering. When cosmic rays are included, star formation becomes episodic and strongly bursty—more so even than in the `boosted' SN feedback model. These results align with recent \textit{JWST} observations that reveal bursty, mini-quenched, and rejuvenated star formation in low-mass galaxies \citep[e.g.,][]{Looser2024, Endsley2024, Dome2025, Baker2025, Witten2025}.
    
    \item The distribution of SN explosion environments reveals a density and temperature bimodality across models, reflecting explosions in both dense star-forming clouds and low-density, pre-processed gas. In the `boosted' SN model, even a few clustered explosions efficiently disperse dense gas, reducing the number of SNe in high-density regions. In contrast, the inclusion of radiative transfer reduces SN clustering, leading to more explosions in dense and photo-heated environments. The `full-physics' model shows a similar distribution to the standard case (\HDSfFb), but with SNe occurring at slightly higher densities in the diffuse ISM while retaining the characteristic photo-heating temperature peak.

    \item Thermal pressure dominates the overall gas support across the halo, but cosmic rays provide significant additional pressure within the central region ($r < 0.2,r_{\rm halo}$), especially during strong outflow events. Beyond this radius, radiation pressure becomes increasingly important and contributes to thermal heating in the circumgalactic medium. Magnetic pressure is generally subdominant but can locally dominate in dense star-forming clouds within the galaxy’s innermost regions.
    
    \item Cosmic ray feedback drives dense, fast, and mass-loaded outflows that include a significant neutral gas component and span a broader range of temperatures and metallicities. Despite injecting less total energy than the `boosted' SN model, this `full-physics' run produces more efficient outflows—highlighting the critical role of cosmic rays in shaping multi-phase winds.
    Future observations that can constrain the multi-phase nature of outflows both from local and high redshift dwarfs will be instrumental in validating our predictions.

    \item Observable properties of ionized gas outflows—particularly the relation between mass-loading factor and normalized mass-weighted outflow velocity (e.g., $v_{\rm out}/v_{\rm circ}$)—provide a powerful diagnostic for distinguishing between feedback models.
    Under our simple observation-like analysis -- employing H$\alpha$ line profiles -- all our models are in reasonable agreement with observations \citep{Concas2022, Marasco2023, Llerena2023, Carniani2024}.
    Among our simulations, the `full-physics' model best matches observed trends in both local and high-redshift galaxies, and especially for observations of neutral outflows. This suggests that cosmic ray feedback plays a key role in shaping realistic outflows.
        
    \item Except for the `boosted' SN model, all \pandora~simulations yield stellar metallicities consistent with observations of both local and high-redshift dwarf galaxies \citep{Kirby2013, Chemerynska2024, Curti2024}. The `full-physics' model produces denser, more temperate outflows with a wide range of metallicities---from $\sim 10^{-3} Z_\odot$ to $\sim 0.4 Z_\odot$---revealing the chemically complex nature of dwarf galaxy outflows and their significant contribution to CGM enrichment.    
\end{enumerate}

Our findings underscore the fundamental importance of incorporating non-thermal physics—stellar radiation, magnetic fields, and cosmic rays—into models of galaxy formation. These processes are essential for regulating star formation, driving multi-phase outflows, and shaping the chemical enrichment of galaxies.

The emerging picture from our simulations is one in which non-thermal pressures and early stellar feedback critically shape the smallest galactic systems. This has exciting implications for upcoming multi-wavelength observatories such as JWST, the upcoming extremely large telescopes (ELT, GMT, TMT), SKAO, ngVLA and Roman, which will enable more detailed studies of the kinematics, thermodynamics, and metal content of galaxy outflows. In particular, measurements of outflow velocities and metallicities across gas phases offer a sensitive means to distinguish between competing models of galaxy formation and feedback.
Our results show that forward-modelling the observational process leads to better agreement on a face-value comparison with observed trends than using intrinsic simulation quantities. This highlights how sophisticated synthetic observation analysis will be fundamental to understand these upcoming panchromatic observations. 
To extend this work, our upcoming high-resolution Azahar cosmological simulation suite (\citealt{Martin-Alvarez2025aas}; Martin-Alvarez et al., in prep.) will systematically explore a broad range of galaxy masses, incorporating these new physical prescriptions. These simulations will provide new insight into how radiative transfer, cosmic rays, and magnetic fields shape galaxy evolution across cosmic time.

\section*{Acknowledgements}
The authors kindly thank the referee for their careful consideration of this manuscript, and their insightful comments and suggestions, which have highly improved the quality of this manuscript.
This research was supported by the Kavli Institute for Particle Astrophysics and Cosmology.
We would like to thank Dalya Baron, Jaeyeon Kim, and Sal Fu for useful comments and discussions. S.M.A. is supported by a Kavli Institute for Particle Astrophysics and Cosmology (KIPAC) Fellowship, and by the NASA/DLR Stratospheric Observatory for Infrared Astronomy (SOFIA) under the 08\_0012 Program. SOFIA is jointly operated by the Universities Space Research Association, Inc. (USRA), under NASA contract NNA17BF53C, and the Deutsches SOFIA Institut (DSI) under DLR contract 50OK0901 to the University of Stuttgart. S.M.A also acknowledges visitor support from the Kavli Institute for Cosmology, Cambridge, where part of this work was completed. 
D.S. acknowledges support by European Research Council Starting Grant 638707 `Black holes and their host galaxies: coevolution across cosmic time'. D.S. and M.H. additionally acknowledge support from the Science and Technology Facilities Council (STFC; grant number ST/W000997/1).
M.F. acknowledges funding from the Swiss National Science Foundation (SNF) via a PRIMA Grant PR00P2 193577 `From cosmic dawn to high noon: the role of black holes for young galaxies'.
M.S. acknowledges the support from the Swiss National Science Foundation under Grant No. P500PT\_214488. 
F.R.M. was supported by funds provided by the Kavli Institute for Cosmological Physics at the University of Chicago through an endowment from the Kavli Foundation.

This work used the DiRAC@Durham facility managed by the Institute for Computational Cosmology on behalf of the STFC DiRAC HPC Facility (www.dirac.ac.uk). The equipment was funded by BEIS capital funding via STFC capital grants ST/P002293/1, ST/R002371/1 and ST/S002502/1, Durham University and STFC operations grant ST/R000832/1. DiRAC is part of the National e-Infrastructure. 
Some of the computing for this project was performed on the Sherlock cluster. We would like to thank Stanford University and the Stanford Research Computing Center for providing computational resources and support that contributed to these research results.
This work was performed using resources provided by the Cambridge Service for Data Driven Discovery (CSD3) operated by the University of Cambridge Research Computing Service (www.csd3.cam.ac.uk), provided by Dell EMC and Intel using Tier-2 funding from the Engineering and Physical Sciences Research Council (capital grant EP/P020259/1), and DiRAC funding from the Science and Technology Facilities Council (www.dirac.ac.uk).

\section*{Data availability}
The data employed in this manuscript is to be shared upon reasonable request contacting the corresponding author.
\bibliographystyle{mnras}
\bibliography{references, references2}

@article{Rupke2005,
    title = {{ Outflows in Infrared‐Luminous Starbursts at z < 0.5. II. Analysis and Discussion }},
    year = {2005},
    journal = {The Astrophysical Journal Supplement Series},
    author = {Rupke, David S. and Veilleux, Sylvain and Sanders, D. B.},
    number = {1},
    month = {9},
    pages = {115--148},
    volume = {160},
    publisher = {American Astronomical Society},
    url = {https://iopscience.iop.org/article/10.1086/432889 https://iopscience.iop.org/article/10.1086/432889/meta},
    doi = {10.1086/432889/FULLTEXT/},
    issn = {0067-0049},
    arxivId = {astro-ph/0506611}
}

@article{DEugenio2023,
    title = {{A fast-rotator post-starburst galaxy quenched by supermassive black-hole feedback at z=3}},
    year = {2023},
    journal = {Nature Astronomy 2024 8:11},
    author = {D'Eugenio, Francesco and Perez-Gonzalez, Pablo and Maiolino, Roberto and Scholtz, Jan and Perna, Michele and Circosta, Chiara and Uebler, Hannah and Arribas, Santiago and Boeker, Torsten and Bunker, Andrew and Carniani, Stefano and Charlot, Stephane and Chevallard, Jacopo and Cresci, Giovanni and Curtis-Lake, Emma and Jones, Gareth and Kumari, Nimisha and Lamperti, Isabella and Looser, Tobias and Parlanti, Eleonora and Rix, Hans-Walter and Robertson, Brant and Del Pino, Bruno Rodriguez and Tacchella, Sandro and Venturi, Giacomo and Willott, Chris},
    number = {11},
    month = {8},
    pages = {1443--1456},
    volume = {8},
    publisher = {Nature Publishing Group},
    url = {http://arxiv.org/abs/2308.06317},
    doi = {10.1038/S41550-024-02345-1;SUBJMETA=33,34,4120,639,863;KWRD=EARLY+UNIVERSE,GALAXIES+AND+CLUSTERS},
    issn = {23973366},
    arxivId = {2308.06317}
}

@article{Fromang2006,
    title = {{A high order Godunov scheme with constrained transport and adaptive mesh refinement for astrophysical magnetohydrodynamics}},
    year = {2006},
    journal = {Astronomy {\&} Astrophysics},
    author = {Fromang, S. and Hennebelle, P. and Teyssier, R.},
    number = {2},
    month = {10},
    pages = {371--384},
    volume = {457},
    publisher = {EDP Sciences},
    url = {http://www.aanda.org/10.1051/0004-6361:20065371},
    isbn = {0004-6361{\textbackslash}r1432-0746},
    doi = {10.1051/0004-6361:20065371},
    issn = {0004-6361},
    pmid = {9010224},
    arxivId = {astro-ph/0607230}
}

@article{Strickland2004,
    title = {{A High Spatial Resolution X‐Ray and H{$\alpha$} Study of Hot Gas in the Halos of Star‐forming Disk Galaxies. I. Spatial and Spectral Properties of the Diffuse X‐Ray Emission}},
    year = {2004},
    journal = {The Astrophysical Journal Supplement Series},
    author = {Strickland, David K. and Heckman, Timothy M. and Colbert, Edward J. M. and Hoopes, Charles G. and Weaver, Kimberly A.},
    number = {2},
    month = {4},
    pages = {193--236},
    volume = {151},
    publisher = {American Astronomical Society},
    url = {https://iopscience.iop.org/article/10.1086/382214 https://iopscience.iop.org/article/10.1086/382214/meta},
    doi = {10.1086/382214/XML},
    issn = {0067-0049},
    arxivId = {astro-ph/0306592}
}

@article{Mezcua2016,
    title = {{A POPULATION OF INTERMEDIATE-MASS BLACK HOLES IN DWARF STARBURST GALAXIES UP TO REDSHIFT = 1.5}},
    year = {2016},
    journal = {The Astrophysical Journal},
    author = {Mezcua, M. and Civano, F. and Fabbiano, G. and Miyaji, T. and Marchesi, S.},
    number = {1},
    month = {1},
    pages = {20},
    volume = {817},
    publisher = {IOP Publishing},
    url = {https://iopscience.iop.org/article/10.3847/0004-637X/817/1/20 https://iopscience.iop.org/article/10.3847/0004-637X/817/1/20/meta},
    doi = {10.3847/0004-637X/817/1/20},
    issn = {0004-637X},
    arxivId = {1511.05844}
}

@article{Looser2024,
    title = {{A recently quenched galaxy 700 million years after the Big Bang}},
    year = {2024},
    journal = {Nature},
    author = {Looser, Tobias J. and D’Eugenio, Francesco and Maiolino, Roberto and Witstok, Joris and Sandles, Lester and Curtis-Lake, Emma and Chevallard, Jacopo and Tacchella, Sandro and Johnson, Benjamin D. and Baker, William M. and Suess, Katherine A. and Carniani, Stefano and Ferruit, Pierre and Arribas, Santiago and Bonaventura, Nina and Bunker, Andrew J. and Cameron, Alex J. and Charlot, Stephane and Curti, Mirko and de Graaff, Anna and Maseda, Michael V. and Rawle, Tim and Rix, Hans Walter and Del Pino, Bruno Rodríguez and Smit, Renske and {\"{U}}bler, Hannah and Willott, Chris and Alberts, Stacey and Egami, Eiichi and Eisenstein, Daniel J. and Endsley, Ryan and Hausen, Ryan and Rieke, Marcia and Robertson, Brant and Shivaei, Irene and Williams, Christina C. and Boyett, Kristan and Chen, Zuyi and Ji, Zhiyuan and Jones, Gareth C. and Kumari, Nimisha and Nelson, Erica and Perna, Michele and Saxena, Aayush and Scholtz, Jan},
    number = {8010},
    month = {3},
    pages = {53--57},
    volume = {629},
    publisher = {Nature Publishing Group},
    url = {https://www.nature.com/articles/s41586-024-07227-0},
    doi = {10.1038/s41586-024-07227-0},
    issn = {14764687},
    pmid = {38447669},
    arxivId = {2302.14155}
}

@article{Rosdahl2015a,
    title = {{A scheme for radiation pressure and photon diffusion with the M1 closure in RAMSES-RT}},
    year = {2015},
    journal = {Monthly Notices of the Royal Astronomical Society},
    author = {Rosdahl, J. and Teyssier, R.},
    number = {4},
    month = {6},
    pages = {4380--4403},
    volume = {449},
    publisher = {Oxford University Press},
    url = {http://academic.oup.com/mnras/article/449/4/4380/1194692/A-scheme-for-radiation-pressure-and-photon},
    doi = {10.1093/mnras/stv567},
    issn = {13652966}
}

@article{Geha2012,
    title = {{A stellar mass threshold for quenching of field galaxies}},
    year = {2012},
    journal = {Astrophysical Journal},
    author = {Geha, M. and Blanton, M. R. and Yan, R. and Tinker, J. L.},
    number = {1},
    month = {9},
    pages = {85},
    volume = {757},
    publisher = {IOP Publishing},
    url = {https://iopscience.iop.org/article/10.1088/0004-637X/757/1/85 https://iopscience.iop.org/article/10.1088/0004-637X/757/1/85/meta},
    doi = {10.1088/0004-637X/757/1/85},
    issn = {15384357},
    arxivId = {1206.3573}
}

@article{Martin-Alvarez2018,
    title = {{A three-phase amplification of the cosmic magnetic field in galaxies}},
    year = {2018},
    journal = {Monthly Notices of the Royal Astronomical Society},
    author = {Martin-Alvarez, Sergio and Devriendt, Julien and Slyz, Adrianne and Teyssier, Romain},
    number = {3},
    month = {9},
    pages = {3343--3365},
    volume = {479},
    publisher = {Oxford University Press},
    url = {https://academic.oup.com/mnras/article/479/3/3343/5040247},
    doi = {10.1093/mnras/sty1623},
    issn = {13652966},
    arxivId = {1806.06866}
}

@article{Stern2016,
    title = {{A UNIVERSAL DENSITY STRUCTURE FOR CIRCUMGALACTIC GAS}},
    year = {2016},
    journal = {The Astrophysical Journal},
    author = {Stern, Jonathan and Hennawi, Joseph F. and Prochaska, J. Xavier and Werk, Jessica K.},
    number = {2},
    month = {10},
    pages = {87},
    volume = {830},
    publisher = {IOP Publishing},
    url = {https://iopscience.iop.org/article/10.3847/0004-637X/830/2/87 https://iopscience.iop.org/article/10.3847/0004-637X/830/2/87/meta},
    doi = {10.3847/0004-637X/830/2/87},
    issn = {0004-637X},
    arxivId = {1604.02168}
}

@article{Chen2010,
    title = {{ABSORPTION-LINE PROBES OF THE PREVALENCE AND PROPERTIES OF OUTFLOWS IN PRESENT-DAY STAR-FORMING GALAXIES}},
    year = {2010},
    journal = {The Astronomical Journal},
    author = {Chen, Yan Mei and Tremonti, Christy A. and Heckman, Timothy M. and Guinevere, Kauffmann and Weiner, Benjamin J. and Jarle, Brinchmann and Jing, Wang},
    number = {2},
    month = {6},
    pages = {445},
    volume = {140},
    publisher = {IOP Publishing},
    url = {https://iopscience.iop.org/article/10.1088/0004-6256/140/2/445 https://iopscience.iop.org/article/10.1088/0004-6256/140/2/445/meta},
    doi = {10.1088/0004-6256/140/2/445},
    issn = {1538-3881},
    arxivId = {1003.5425}
}

@article{Dubois2016,
    title = {{An implicit scheme for solving the anisotropic diffusion of heat and cosmic rays in the RAMSES code}},
    year = {2016},
    journal = {Astronomy {\&} Astrophysics},
    author = {Dubois, Yohan and Commer{\c{c}}on, Benoît},
    month = {1},
    pages = {A138},
    volume = {585},
    publisher = {EDP Sciences},
    url = {http://www.aanda.org/10.1051/0004-6361/201527126 http://arxiv.org/abs/1509.07037%0Ahttp://dx.doi.org/10.1051/0004-6361/201527126},
    doi = {10.1051/0004-6361/201527126},
    issn = {0004-6361},
    arxivId = {1509.07037}
}

@article{Ferland1992,
    title = {{Anisotropic line emission and the geometry of the broad-line region in active galactic nuclei}},
    year = {1992},
    journal = {The Astrophysical Journal},
    author = {Ferland, G. J. and Peterson, B. M. and Horne, K. and Welsh, W. F. and Nahar, S. N.},
    month = {3},
    pages = {95},
    volume = {387},
    publisher = {American Astronomical Society},
    url = {https://ui.adsabs.harvard.edu/abs/1992ApJ...387...95F/abstract},
    doi = {10.1086/171063},
    issn = {0004-637X}
}

@article{Concas2022,
    title = {{Being KLEVER at cosmic noon: Ionized gas outflows are inconspicuous in low-mass star-forming galaxies but prominent in massive AGN hosts}},
    year = {2022},
    journal = {Monthly Notices of the Royal Astronomical Society},
    author = {Concas, Alice and Maiolino, Roberto and Curti, Mirko and Hayden-Pawson, Connor and Cirasuolo, Michele and Jones, Gareth C. and Mercurio, Amata and Belfiore, Francesco and Cresci, Giovanni and Cullen, Fergus and Mannucci, Filippo and Marconi, Alessandro and Cappellari, Michele and Cicone, Claudia and Peng, Yingjie and Troncoso, Paulina},
    number = {2},
    month = {5},
    pages = {2535--2562},
    volume = {513},
    publisher = {Oxford Academic},
    url = {https://dx.doi.org/10.1093/mnras/stac1026},
    doi = {10.1093/mnras/stac1026},
    issn = {13652966},
    arxivId = {2203.11958}
}

@article{Ocvirk2008,
    title = {{Bimodal gas accretion in the Horizon-MareNostrum galaxy formation simulation}},
    year = {2008},
    journal = {Monthly Notices of the Royal Astronomical Society},
    author = {Ocvirk, P. and Pichon, C. and Teyssier, R.},
    number = {4},
    month = {10},
    pages = {1326--1338},
    volume = {390},
    publisher = {Narnia},
    url = {https://academic.oup.com/mnras/article-lookup/doi/10.1111/j.1365-2966.2008.13763.x},
    doi = {10.1111/j.1365-2966.2008.13763.x},
    issn = {00358711},
    arxivId = {arXiv:0803.4506v1}
}

@article{Yang2017b,
    title = {{Blueberry Galaxies: The Lowest Mass Young Starbursts}},
    year = {2017},
    journal = {The Astrophysical Journal},
    author = {Yang, Huan and Malhotra, Sangeeta and Rhoads, James E. and Wang, Junxian},
    number = {1},
    month = {9},
    pages = {38},
    volume = {847},
    publisher = {IOP Publishing},
    url = {https://iopscience.iop.org/article/10.3847/1538-4357/aa8809 https://iopscience.iop.org/article/10.3847/1538-4357/aa8809/meta},
    doi = {10.3847/1538-4357/AA8809},
    issn = {0004-637X},
    arxivId = {1706.02819}
}

@article{Tweed2009,
    title = {{Building Merger Trees from Cosmological N-body Simulations}},
    year = {2009},
    journal = {Astronomy {\&} Astrophysics},
    author = {Tweed, D. and Devriendt, J. and Blaizot, J. and Colombi, S. and Slyz, A.},
    number = {2},
    month = {11},
    pages = {647--660},
    volume = {506},
    publisher = {EDP Sciences},
    url = {http://www.aanda.org/10.1051/0004-6361/200911787 http://arxiv.org/abs/0902.0679%0Ahttp://dx.doi.org/10.1051/0004-6361/200911787},
    doi = {10.1051/0004-6361/200911787},
    issn = {0004-6361},
    arxivId = {0902.0679}
}

@article{Hopkins2020,
    title = {{But what about...: cosmic rays, magnetic fields, conduction, and viscosity in galaxy formation}},
    year = {2020},
    journal = {Monthly Notices of the Royal Astronomical Society},
    author = {Hopkins, Philip F. and Chan, T. K. and Garrison-Kimmel, Shea and Ji, Suoqing and Su, Kung Yi and Hummels, Cameron B. and Kere{\v{s}}, Dušan and Quataert, Eliot and Faucher-Gigu{\`{e}}re, Claude André},
    number = {3},
    month = {3},
    pages = {3465--3498},
    volume = {492},
    publisher = {Oxford Academic},
    url = {https://dx.doi.org/10.1093/mnras/stz3321},
    doi = {10.1093/MNRAS/STZ3321},
    issn = {0035-8711},
    arxivId = {1905.04321}
}

@article{Stein2020,
    title = {{CHANG-ES - XXI. Transport processes and the X-shaped magnetic field of NGC 4217: off-center superbubble structure}},
    year = {2020},
    journal = {Astronomy {\&} Astrophysics},
    author = {Stein, Y. and Dettmar, R. J. and Beck, R. and Irwin, J. and Wiegert, T. and Miskolczi, A. and Wang, Q. D. and English, J. and Henriksen, R. and Radica, M. and Li, J. T.},
    month = {7},
    pages = {A111},
    volume = {639},
    publisher = {EDP Sciences},
    url = {https://www.aanda.org/articles/aa/full_html/2020/07/aa37675-20/aa37675-20.html https://www.aanda.org/articles/aa/abs/2020/07/aa37675-20/aa37675-20.html},
    doi = {10.1051/0004-6361/202037675},
    issn = {0004-6361},
    arxivId = {2007.03002}
}

@article{Krause2020,
    title = {{CHANG-ES - XXII. Coherent magnetic fields in the halos of spiral galaxies}},
    year = {2020},
    journal = {Astronomy {\&} Astrophysics},
    author = {Krause, Marita and Irwin, Judith and Schmidt, Philip and Stein, Yelena and Miskolczi, Arpad and Carolina Mora-Partiarroyo, Silvia and Wiegert, Theresa and Beck, Rainer and Stil, Jeroen M. and Heald, George and Li, Jiang Tao and Damas-Segovia, Ancor and Vargas, Carlos J. and Rand, Richard J. and West, Jennifer and Walterbos, Rene A.M. and Dettmar, Ralf Jürgen and English, Jayanne and Woodfinden, Alex},
    month = {7},
    pages = {A112},
    volume = {639},
    publisher = {EDP Sciences},
    url = {https://www.aanda.org/articles/aa/full_html/2020/07/aa37780-20/aa37780-20.html https://www.aanda.org/articles/aa/abs/2020/07/aa37780-20/aa37780-20.html},
    doi = {10.1051/0004-6361/202037780},
    issn = {0004-6361},
    arxivId = {2004.14383}
}

@article{Stein2025,
    title = {{CHANG-ES - XXXIV. Magnetic field structure in edge-on galaxies: Characterising large-scale magnetic fields in galactic halos}},
    year = {2025},
    journal = {Astronomy {\&} Astrophysics},
    author = {Stein, M. and Kleimann, J. and Adebahr, B. and Dettmar, R. J. and Fichtner, H. and English, J. and Heesen, V. and Kamphuis, P. and Irwin, J. and Mele, C. and Bomans, D. J. and Li, J. and Skeggs, N. B. and Wang, Q. D. and Yang, Y.},
    month = {4},
    pages = {A112},
    volume = {696},
    publisher = {EDP Sciences},
    url = {https://www.aanda.org/articles/aa/full_html/2025/04/aa52322-24/aa52322-24.html https://www.aanda.org/articles/aa/abs/2025/04/aa52322-24/aa52322-24.html},
    doi = {10.1051/0004-6361/202452322},
    issn = {0004-6361}
}

@article{Heald2022,
    title = {{CHANG-ES XXIII: influence of a galactic wind in NGC 5775}},
    year = {2021},
    journal = {Monthly Notices of the Royal Astronomical Society},
    author = {Heald, G H and Heesen, V and Sridhar, S S and Beck, R and Bomans, D J and Br{\"{u}}ggen, M and Chy{\.{z}}y, K T and Damas-Segovia, A and Dettmar, R-j and English, J and Henriksen, R and Ideguchi, S and Irwin, J and Krause, M and Li, J-t and Murphy, E J and Nikiel-Wroczy{\'{n}}ski, B and Piotrowska, J and Rand, R J and Shimwell, T and Stein, Y and Vargas, C J and Wang, Q D and van Weeren, R J and Wiegert, T},
    number = {1},
    pages = {658--684},
    volume = {509},
    url = {https://doi.org/10.1093/mnras/stab2804},
    doi = {10.1093/mnras/stab2804},
    issn = {0035-8711},
    arxivId = {2109.12267}
}

@article{Fluetsch2019,
    title = {{Cold molecular outflows in the local Universe and their feedback effect on galaxies}},
    year = {2019},
    journal = {Monthly Notices of the Royal Astronomical Society},
    author = {Fluetsch, A. and Maiolino, R. and Carniani, S. and Marconi, A. and Cicone, C. and Bourne, M. A. and Costa, T. and Fabian, A. C. and Ishibashi, W. and Venturi, G.},
    number = {4},
    month = {3},
    pages = {4586--4614},
    volume = {483},
    publisher = {Oxford Academic},
    url = {https://dx.doi.org/10.1093/mnras/sty3449},
    doi = {10.1093/MNRAS/STY3449},
    issn = {0035-8711}
}

@article{Fichtner2024,
    title = {{Connecting stellar and galactic scales: Energetic feedback from stellar wind bubbles to supernova remnants}},
    year = {2024},
    journal = {Astronomy {\&} Astrophysics},
    author = {Fichtner, Yvonne A. and Mackey, Jonathan and Grassitelli, Luca and Romano-D{\'{i}}az, Emilio and Porciani, Cristiano},
    month = {10},
    pages = {A72},
    volume = {690},
    publisher = {EDP Sciences},
    url = {https://www.aanda.org/articles/aa/full_html/2024/10/aa49638-24/aa49638-24.html https://www.aanda.org/articles/aa/abs/2024/10/aa49638-24/aa49638-24.html},
    doi = {10.1051/0004-6361/202449638},
    issn = {0004-6361},
    arxivId = {2402.11008}
}

@article{Butsky2023,
    title = {{Constraining cosmic ray transport with observations of the circumgalactic medium}},
    year = {2023},
    journal = {Monthly Notices of the Royal Astronomical Society},
    author = {Butsky, Iryna S. and Nakum, Shreya and Ponnada, Sam B. and Hummels, Cameron B. and Ji, Suoqing and Hopkins, Philip F.},
    number = {2},
    month = {3},
    pages = {2477--2483},
    volume = {521},
    publisher = {Oxford Academic},
    url = {https://dx.doi.org/10.1093/mnras/stad671},
    doi = {10.1093/MNRAS/STAD671},
    issn = {0035-8711},
    arxivId = {2210.14232}
}

@article{Sanati2020,
    title = {{Constraining the primordial magnetic field with dwarf galaxy simulations}},
    year = {2020},
    journal = {Astronomy and Astrophysics},
    author = {Sanati, Mahsa and Revaz, Yves and Schober, Jennifer and Kunze, Kerstin E and Jablonka, Pascale},
    pages = {54},
    volume = {643},
    url = {https://doi.org/10.1051/0004-6361/202038382},
    doi = {10.1051/0004-6361/202038382},
    issn = {14320746},
    arxivId = {2005.05401}
}

@article{Pavicevic2025,
    title = {{Constraints on Primordial Magnetic Fields from the Lyman-<math xmlns="http://www.w3.org/1998/Math/MathML" display="inline"><mi>{$\alpha$}</mi></math> Forest}},
    year = {2025},
    journal = {Physical Review Letters},
    author = {Pavi{\v{c}}evi{\'{c}}, Mak and Ir{\v{s}}i{\v{c}}, Vid and Viel, Matteo and Bolton, James S. and Haehnelt, Martin G. and Martin-Alvarez, Sergio and Puchwein, Ewald and Ralegankar, Pranjal},
    number = {7},
    month = {8},
    pages = {071001},
    volume = {135},
    publisher = {American Physical Society},
    url = {https://journals.aps.org/prl/abstract/10.1103/77rd-vkpz},
    doi = {10.1103/77rd-vkpz},
    issn = {0031-9007},
    pmid = {40929213},
    arxivId = {2501.06299}
}

@article{Girichidis2018,
    title = {{Cooler and smoother - the impact of cosmic rays on the phase structure of galactic outflows}},
    year = {2018},
    journal = {Monthly Notices of the Royal Astronomical Society},
    author = {Girichidis, Philipp and Naab, Thorsten and Hanasz, Michał and Walch, Stefanie},
    number = {3},
    pages = {3042--3067},
    volume = {479},
    url = {http://flash.uchicago.edu/site/},
    doi = {10.1093/mnras/sty1653},
    issn = {13652966},
    arxivId = {1805.09333}
}

@article{White1978,
    title = {{Core condensation in heavy halos: a two-stage theory for galaxy formation and clustering}},
    year = {1978},
    journal = {Monthly Notices of the Royal Astronomical Society},
    author = {White, S. D. M. and Rees, M. J.},
    number = {3},
    month = {5},
    pages = {341--358},
    volume = {183},
    publisher = {Oxford University Press (OUP)},
    url = {https://ui.adsabs.harvard.edu/abs/1978MNRAS.183..341W/abstract},
    doi = {10.1093/mnras/183.3.341},
    issn = {0035-8711}
}

@article{Salem2014a,
    title = {{Cosmic ray driven outflows in global galaxy disc models}},
    year = {2014},
    journal = {Monthly Notices of the Royal Astronomical Society},
    author = {Salem, Munier and Bryan, Greg L.},
    number = {4},
    month = {2},
    pages = {3312--3330},
    volume = {437},
    publisher = {Oxford University Press},
    url = {http://academic.oup.com/mnras/article/437/4/3312/1001083/Cosmic-ray-driven-outflows-in-global-galaxy-disc},
    doi = {10.1093/mnras/stt2121},
    issn = {00358711}
}

@article{Hopkins2021a,
    title = {{Cosmic ray driven outflows to Mpc scales from L*galaxies}},
    year = {2021},
    journal = {Monthly Notices of the Royal Astronomical Society},
    author = {Hopkins, Philip F and Chan, T K and Ji, Suoqing and Hummels, Cameron B and Kere{\v{s}}, Dušan and Quataert, Eliot and Faucher-Gigu{\`{e}}re, Claude André},
    number = {3},
    pages = {3640--3662},
    volume = {501},
    url = {http://fire.northwestern.edu},
    doi = {10.1093/mnras/staa3690},
    issn = {13652966},
    arxivId = {2002.02462}
}

@article{Dashyan2020,
    title = {{Cosmic ray feedback from supernovae in dwarf galaxies}},
    year = {2020},
    journal = {Astronomy and Astrophysics},
    author = {Dashyan, Gohar and Dubois, Yohan},
    month = {6},
    pages = {A123},
    volume = {638},
    publisher = {EDP Sciences},
    url = {https://www.aanda.org/10.1051/0004-6361/201936339},
    doi = {10.1051/0004-6361/201936339},
    issn = {14320746},
    arxivId = {2003.09900}
}

@article{Jubelgas2008,
    title = {{Cosmic ray feedback in hydrodynamical simulations of galaxy formation}},
    year = {2008},
    journal = {Astronomy and Astrophysics},
    author = {Jubelgas, M. and Springel, V. and En{\ss}lin, T. and Pfrommer, C.},
    number = {1},
    month = {4},
    pages = {33--63},
    volume = {481},
    publisher = {EDP Sciences},
    url = {https://www.aanda.org/articles/aa/abs/2008/13/aa5295-06/aa5295-06.html},
    doi = {10.1051/0004-6361:20065295},
    issn = {00046361},
    arxivId = {astro-ph/0603485}
}

@article{Armillotta2024,
    title = {{Cosmic-Ray Acceleration of Galactic Outflows in Multiphase Gas}},
    year = {2024},
    journal = {The Astrophysical Journal},
    author = {Armillotta, Lucia and Ostriker, Eve C. and Kim, Chang-Goo and Jiang, Yan-Fei},
    number = {1},
    month = {3},
    pages = {99},
    volume = {964},
    publisher = {IOP Publishing},
    url = {https://iopscience.iop.org/article/10.3847/1538-4357/ad1e5c https://iopscience.iop.org/article/10.3847/1538-4357/ad1e5c/meta},
    doi = {10.3847/1538-4357/AD1E5C},
    issn = {0004-637X},
    arxivId = {2401.04169}
}

@article{Nunez-Castineyra2022,
    title = {{Cosmic-ray diffusion and the multi-phase interstellar medium in a dwarf galaxy. I. Large-scale properties and {\$}{\textbackslash}gamma{\$}-ray luminosities}},
    year = {2022},
    journal = {arXiv},
    author = {Nu{\~{n}}ez-Casti{\~{n}}eyra, A. and Grenier, I. A. and Bournaud, F. and Dubois, Y. and Youssef, F. R. Kamal and Hennebelle, P.},
    month = {5},
    url = {http://arxiv.org/abs/2205.08163},
    arxivId = {2205.08163}
}

@article{Commercon2019,
    title = {{Cosmic-ray propagation in the bi-stable interstellar medium}},
    year = {2019},
    journal = {Astronomy {\&} Astrophysics},
    author = {Commer{\c{c}}on, Benoît and Marcowith, Alexandre and Dubois, Yohan},
    month = {2},
    pages = {A143},
    volume = {622},
    publisher = {EDP Sciences},
    url = {https://www.aanda.org/10.1051/0004-6361/201833809},
    doi = {10.1051/0004-6361/201833809},
    issn = {0004-6361}
}

@article{Armillotta2021,
    title = {{Cosmic-Ray Transport in Simulations of Star-forming Galactic Disks}},
    year = {2021},
    journal = {The Astrophysical Journal},
    author = {Armillotta, Lucia and Ostriker, Eve C. and Jiang, Yan-Fei},
    number = {1},
    month = {11},
    pages = {11},
    volume = {922},
    publisher = {IOP Publishing},
    url = {https://iopscience.iop.org/article/10.3847/1538-4357/ac1db2 https://iopscience.iop.org/article/10.3847/1538-4357/ac1db2/meta},
    doi = {10.3847/1538-4357/AC1DB2},
    issn = {0004-637X},
    arxivId = {2108.09356}
}

@article{Teyssier2002,
    title = {{Cosmological Hydrodynamics with Adaptive Mesh Refinement: a new high resolution code called RAMSES}},
    year = {2002},
    journal = {Astronomy {\&} Astrophysics},
    author = {Teyssier, Romain},
    number = {1},
    month = {4},
    pages = {337--364},
    volume = {385},
    publisher = {EDP Sciences},
    url = {http://www.aanda.org/10.1051/0004-6361:20011817 http://arxiv.org/abs/astro-ph/0111367%0Ahttp://dx.doi.org/10.1051/0004-6361:20011817},
    isbn = {0902009192},
    doi = {10.1051/0004-6361:20011817},
    issn = {14052059},
    pmid = {9010224},
    arxivId = {astro-ph/0111367}
}

@article{Dubois2008,
    title = {{Cosmological MHD simulation of a cooling flow cluster}},
    year = {2008},
    journal = {Astronomy {\&} Astrophysics},
    author = {Dubois, Y. and Teyssier, R.},
    number = {2},
    month = {5},
    pages = {L13-L16},
    volume = {482},
    publisher = {EDP Sciences},
    url = {http://www.aanda.org/10.1051/0004-6361:200809513},
    doi = {10.1051/0004-6361:200809513},
    issn = {0004-6361}
}

@article{Smith2019,
    title = {{Cosmological simulations of dwarfs: the need for ISM physics beyond SN feedback alone}},
    year = {2019},
    journal = {Monthly Notices of the Royal Astronomical Society},
    author = {Smith, Matthew C and Sijacki, Debora and Shen, Sijing},
    number = {3},
    month = {5},
    pages = {3317--3333},
    volume = {485},
    publisher = {Narnia},
    url = {https://academic.oup.com/mnras/article/485/3/3317/5368355 http://arxiv.org/abs/1807.04288%0Ahttp://dx.doi.org/10.1093/mnras/stz599},
    doi = {10.1093/mnras/stz599},
    issn = {0035-8711},
    arxivId = {1807.04288}
}

@article{Vogelsberger2020,
    title = {{Cosmological simulations of galaxy formation}},
    year = {2020},
    journal = {Nature Reviews Physics},
    author = {Vogelsberger, Mark and Marinacci, Federico and Torrey, Paul and Puchwein, Ewald},
    number = {1},
    month = {1},
    pages = {42--66},
    volume = {2},
    publisher = {Nature Publishing Group},
    url = {https://www.nature.com/articles/s42254-019-0127-2},
    doi = {10.1038/s42254-019-0127-2},
    issn = {25225820},
    arxivId = {1909.07976}
}

@article{Teyssier2013,
    title = {{Cusp-core transformations in dwarf galaxies: Observational predictions}},
    year = {2013},
    journal = {Monthly Notices of the Royal Astronomical Society},
    author = {Teyssier, Romain and Pontzen, Andrew and Dubois, Yohan and Read, Justin I.},
    number = {4},
    month = {3},
    pages = {3068--3078},
    volume = {429},
    publisher = {Oxford University Press},
    url = {http://academic.oup.com/mnras/article/429/4/3068/1012066/Cuspcore-transformations-in-dwarf-galaxies},
    doi = {10.1093/mnras/sts563},
    issn = {00358711}
}

@article{Dubois2014,
    title = {{Dancing in the dark: Galactic properties trace spin swings along the cosmic web}},
    year = {2014},
    journal = {Monthly Notices of the Royal Astronomical Society},
    author = {Dubois, Y. and Pichon, C. and Welker, C. and Le Borgne, D. and Devriendt, J. and Laigle, C. and Codis, S. and Pogosyan, D. and Arnouts, S. and Benabed, K. and Bertin, E. and Blaizot, J. and Bouchet, F. and Cardoso, J. F. and Colombi, S. and De Lapparent, V. and Desjacques, V. and Gavazzi, R. and Kassin, S. and Kimm, T. and McCracken, H. and Milliard, B. and Peirani, S. and Prunet, S. and Rouberol, S. and Silk, J. and Slyz, A. and Sousbie, T. and Teyssier, R. and Tresse, L. and Treyer, M. and Vibert, D. and Volonteri, M.},
    number = {2},
    month = {10},
    pages = {1453--1468},
    volume = {444},
    publisher = {Oxford University Press},
    url = {https://academic.oup.com/mnras/article/444/2/1453/990507},
    doi = {10.1093/mnras/stu1227},
    issn = {13652966},
    arxivId = {1402.1165}
}

@article{Witten2024,
    title = {{Deciphering Lyman-{$\alpha$} emission deep into the epoch of reionization}},
    year = {2024},
    journal = {Nature Astronomy},
    author = {Witten, Callum and Laporte, Nicolas and Martin-Alvarez, Sergio and Sijacki, Debora and Yuan, Yuxuan and Haehnelt, Martin G. and Baker, William M. and Dunlop, James S. and Ellis, Richard S. and Grogin, Norman A. and Illingworth, Garth and Katz, Harley and Koekemoer, Anton M. and Magee, Daniel and Maiolino, Roberto and McClymont, William and P{\'{e}}rez-Gonz{\'{a}}lez, Pablo G. and Pusk{\'{a}}s, Dávid and Roberts-Borsani, Guido and Santini, Paola and Simmonds, Charlotte},
    number = {3},
    month = {1},
    pages = {384--396},
    volume = {8},
    publisher = {Nature Publishing Group},
    url = {https://www.nature.com/articles/s41550-023-02179-3},
    doi = {10.1038/s41550-023-02179-3},
    issn = {23973366},
    arxivId = {2303.16225}
}

@article{Agertz2009b,
    title = {{Disc formation and the origin of clumpy galaxies at high redshift}},
    year = {2009},
    journal = {Monthly Notices of the Royal Astronomical Society: Letters},
    author = {Agertz, Oscar and Teyssier, Romain and Moore, Ben},
    number = {1},
    month = {7},
    pages = {L64-L68},
    volume = {397},
    publisher = {Oxford Academic},
    url = {https://dx.doi.org/10.1111/j.1745-3933.2009.00685.x},
    doi = {10.1111/J.1745-3933.2009.00685.X},
    issn = {1745-3925},
    arxivId = {0901.2536}
}

@article{Katz1992,
    title = {{Dissipational galaxy formation. II - Effects of star formation}},
    year = {1992},
    journal = {The Astrophysical Journal},
    author = {Katz, Neal},
    month = {6},
    pages = {502},
    volume = {391},
    url = {http://adsabs.harvard.edu/doi/10.1086/171366},
    doi = {10.1086/171366},
    issn = {0004-637X}
}

@article{Sanati2024,
    title = {{Dwarf galaxies as a probe of a primordially magnetized Universe}},
    year = {2024},
    journal = {Astronomy {\&} Astrophysics},
    author = {Sanati, Mahsa and Martin-Alvarez, Sergio and Schober, Jennifer and Revaz, Yves and Slyz, Adrianne and Devriendt, Julien},
    month = {10},
    pages = {A59},
    volume = {690},
    publisher = {EDP Sciences},
    url = {https://www.aanda.org/articles/aa/full_html/2024/10/aa49822-24/aa49822-24.html https://www.aanda.org/articles/aa/abs/2024/10/aa49822-24/aa49822-24.html},
    doi = {10.1051/0004-6361/202449822},
    issn = {0004-6361},
    arxivId = {2403.05672}
}

@article{Rey2020,
    title = {{EDGE: From quiescent to gas-rich to star-forming low-mass dwarf galaxies}},
    year = {2020},
    journal = {Monthly Notices of the Royal Astronomical Society},
    author = {Rey, Martin P. and Pontzen, Andrew and Agertz, Oscar and Orkney, Matthew D.A. and Read, Justin I. and Rosdahl, Joakim},
    number = {2},
    month = {4},
    pages = {1508--1520},
    volume = {497},
    url = {http://arxiv.org/abs/2004.09530},
    doi = {10.1093/mnras/staa1640},
    issn = {13652966},
    arxivId = {2004.09530}
}

@article{Agertz2020,
    title = {{EDGE: The mass-metallicity relation as a critical test of galaxy formation physics}},
    year = {2020},
    journal = {Monthly Notices of the Royal Astronomical Society},
    author = {Agertz, Oscar and Pontzen, Andrew and Read, Justin I and Rey, Martin P and Orkney, Matthew and Rosdahl, Joakim and Teyssier, Romain and Verbeke, Robbert and Kretschmer, Michael and Nickerson, Sarah},
    number = {2},
    pages = {1656--1672},
    volume = {491},
    url = {https://bitbucket.org/rteyssie/ramses.},
    doi = {10.1093/mnras/stz3053},
    issn = {13652966},
    arxivId = {1904.02723}
}

@article{Diesing2018,
    title = {{Effect of Cosmic Rays on the Evolution and Momentum Deposition of Supernova Remnants}},
    year = {2018},
    journal = {Physical Review Letters},
    author = {Diesing, Rebecca and Caprioli, Damiano},
    number = {9},
    month = {8},
    volume = {121},
    publisher = {American Physical Society},
    doi = {10.1103/PhysRevLett.121.091101},
    issn = {10797114},
    pmid = {30230878},
    arxivId = {1804.09731}
}

@article{Smith2021,
    title = {{Efficient early stellar feedback can suppress galactic outflows by reducing supernova clustering}},
    year = {2021},
    journal = {Monthly Notices of the Royal Astronomical Society},
    author = {Smith, Matthew C. and Bryan, Greg L. and Somerville, Rachel S. and Hu, Chia Yu and Teyssier, Romain and Burkhart, Blakesley and Hernquist, Lars},
    number = {3},
    month = {8},
    pages = {3882--3915},
    volume = {506},
    publisher = {Oxford Academic},
    url = {https://dx.doi.org/10.1093/mnras/stab1896},
    doi = {10.1093/mnras/stab1896},
    issn = {13652966},
    arxivId = {2009.11309}
}

@article{Thornton1998,
    title = {{Energy Input and Mass Redistribution by Supernovae in the Interstellar Medium}},
    year = {1998},
    journal = {The Astrophysical Journal},
    author = {Thornton, K. and Gaudlitz, M. and Janka, H.‐Th. and Steinmetz, M.},
    number = {1},
    month = {6},
    pages = {95--119},
    volume = {500},
    publisher = {American Astronomical Society},
    url = {https://iopscience.iop.org/article/10.1086/305704 https://iopscience.iop.org/article/10.1086/305704/meta},
    doi = {10.1086/305704/FULLTEXT/},
    issn = {0004-637X}
}

@article{Kimm2014,
    title = {{Escape fraction of ionizing photons during reionization: Effects due to supernova feedback and runaway ob stars}},
    year = {2014},
    journal = {Astrophysical Journal},
    author = {Kimm, Taysun and Cen, Renyue},
    number = {2},
    month = {5},
    pages = {121},
    volume = {788},
    publisher = {IOP Publishing},
    url = {http://stacks.iop.org/0004-637X/788/i=2/a=121?key=crossref.0f0973026711d57f08aaf5f5ed24ac18},
    doi = {10.1088/0004-637X/788/2/121},
    issn = {15384357},
    arxivId = {1405.0552}
}

@article{Taziaux2025,
    title = {{Exploring magnetised galactic outflows in starburst dwarf galaxies NGC 3125 and IC 4662}},
    year = {2025},
    journal = {Astronomy {\&} Astrophysics},
    author = {Taziaux, Sam and M{\"{u}}ller, Ancla and Adebahr, Björn and Basu, Aritra and Pfrommer, Christoph and Stein, Michael and Chy{\.{z}}y, Krysztof T. and Bomans, Dominik J. and En{\ss}lin, Torsten and Heesen, Volker and Kamphuis, Peter and Soida, Marian and Wezgowiec, Marek and Dettmar, Ralf Jürgen and Das, Samata and Tjus, Julia},
    month = {4},
    pages = {A226},
    volume = {696},
    publisher = {EDP Sciences},
    url = {https://www.aanda.org/articles/aa/full_html/2025/04/aa53311-24/aa53311-24.html https://www.aanda.org/articles/aa/abs/2025/04/aa53311-24/aa53311-24.html},
    doi = {10.1051/0004-6361/202453311},
    issn = {0004-6361}
}

@article{Yuan2025,
    title = {{Extended red wings and the visibility of reionization-epoch Lyman-{$\alpha$} emitters}},
    year = {2025},
    journal = {Monthly Notices of the Royal Astronomical Society},
    author = {Yuan, Yuxuan and Martin-Alvarez, Sergio and Haehnelt, Martin G and Garel, Thibault and Keating, Laura and Witstok, Joris and Sijacki, Debora},
    number = {2},
    month = {8},
    pages = {762--789},
    volume = {542},
    publisher = {Oxford Academic},
    url = {https://dx.doi.org/10.1093/mnras/staf1252},
    doi = {10.1093/MNRAS/STAF1252},
    issn = {0035-8711},
    arxivId = {2412.07970}
}

@article{Borlaff2023,
    title = {{Extragalactic Magnetism with SOFIA (SALSA Legacy Program). V. First Results on the Magnetic Field Orientation of Galaxies}},
    year = {2023},
    journal = {The Astrophysical Journal},
    author = {Borlaff, Alejandro S. and Lopez-Rodriguez, Enrique and Beck, Rainer and Clark, Susan E. and Ntormousi, Evangelia and Tassis, Konstantinos and Martin-Alvarez, Sergio and Tahani, Mehrnoosh and Dale, Daniel A. and Moral-Castro, Ignacio del and Roman-Duval, Julia and Marcum, Pamela M. and Beckman, John E. and Subramanian, Kandaswamy and Eftekharzadeh, Sarah and Proudfit, Leslie},
    number = {1},
    month = {7},
    pages = {4},
    volume = {952},
    publisher = {IOP Publishing},
    url = {https://iopscience.iop.org/article/10.3847/1538-4357/acd934 https://iopscience.iop.org/article/10.3847/1538-4357/acd934/meta},
    doi = {10.3847/1538-4357/ACD934},
    issn = {0004-637X},
    arxivId = {2303.13586}
}

@article{Martin-Alvarez2024a,
    title = {{Extragalactic Magnetism with SOFIA (SALSA Legacy Program). VII. A tomographic view of far infrared and radio polarimetric observations through MHD simulations of galaxies}},
    year = {2024},
    journal = {The Astrophysical Journal},
    author = {Martin-Alvarez, Sergio and Lopez-Rodriguez, Enrique and Dacunha, Tara and Clark, Susan E. and Borlaff, Alejandro S. and Beck, Rainer and Montero, Francisco Rodríguez and Jung, Seoyoung Lyla and Devriendt, Julien and Slyz, Adrianne and Roman-Duval, Julia and Ntormousi, Evangelia and Tahani, Mehrnoosh and Subramanian, Kandaswamy and Dale, Daniel A. and Marcum, Pamela M. and Tassis, Konstantinos and del Moral-Castro, Ignacio and Tram, Le Ngoc and Jarvis, Matt J.},
    number = {1},
    month = {4},
    pages = {43},
    volume = {966},
    publisher = {IOP Publishing},
    url = {https://iopscience.iop.org/article/10.3847/1538-4357/ad2e9e https://iopscience.iop.org/article/10.3847/1538-4357/ad2e9e/meta http://arxiv.org/abs/2311.06356},
    doi = {10.3847/1538-4357/ad2e9e},
    issn = {15384357},
    arxivId = {2311.06356}
}

@article{Lopez-Rodriguez2023,
    title = {{Extragalactic Magnetism with SOFIA (SALSA Legacy Program): The Magnetic Fields in the Multiphase Interstellar Medium of the Antennae Galaxies*}},
    year = {2023},
    journal = {The Astrophysical Journal Letters},
    author = {Lopez-Rodriguez, Enrique and Borlaff, Alejandro S. and Beck, Rainer and Reach, William T. and Mao, Sui Ann and Ntormousi, Evangelia and Tassis, Konstantinos and Martin-Alvarez, Sergio and Clark, Susan E. and Dale, Daniel A. and Moral-Castro, Ignacio del},
    number = {1},
    month = {12},
    pages = {L13},
    volume = {942},
    publisher = {IOP Publishing},
    url = {https://iopscience.iop.org/article/10.3847/2041-8213/acaaa2 https://iopscience.iop.org/article/10.3847/2041-8213/acaaa2/meta},
    doi = {10.3847/2041-8213/acaaa2},
    issn = {2041-8205},
    arxivId = {2211.00012}
}

@article{Guo2008,
    title = {{Feedback heating by cosmic rays in clusters of galaxies}},
    year = {2008},
    journal = {Monthly Notices of the Royal Astronomical Society},
    author = {Guo, Fulai and Oh, S. Peng},
    number = {1},
    month = {2},
    pages = {251--266},
    volume = {384},
    publisher = {Oxford Academic},
    url = {https://academic.oup.com/mnras/article-lookup/doi/10.1111/j.1365-2966.2007.12692.x},
    doi = {10.1111/j.1365-2966.2007.12692.x},
    issn = {00358711}
}

@article{Kimm2017,
    title = {{Feedback-regulated star formation and escape of LyC photons from mini-haloes during reionization}},
    year = {2017},
    journal = {Monthly Notices of the Royal Astronomical Society},
    author = {Kimm, Taysun and Katz, Harley and Haehnelt, Martin and Rosdahl, Joakim and Devriendt, Julien and Slyz, Adrianne},
    number = {4},
    month = {5},
    pages = {4826--4846},
    volume = {466},
    publisher = {Oxford Academic},
    url = {https://dx.doi.org/10.1093/mnras/stx052},
    doi = {10.1093/MNRAS/STX052},
    issn = {0035-8711},
    arxivId = {1608.04762}
}

@article{Hopkins2022a,
    title = {{First predicted cosmic ray spectra, primary-to-secondary ratios, and ionization rates from MHD galaxy formation simulations}},
    year = {2022},
    journal = {Monthly Notices of the Royal Astronomical Society},
    author = {Hopkins, Philip F. and Butsky, Iryna S. and Panopoulou, Georgia V. and Ji, Suoqing and Quataert, Eliot and Faucher-Gigu{\`{e}}re, Claude André and Kere{\v{s}}, Dušan},
    number = {3},
    month = {9},
    pages = {3470--3514},
    volume = {516},
    publisher = {Oxford Academic},
    url = {https://dx.doi.org/10.1093/mnras/stac1791},
    doi = {10.1093/MNRAS/STAC1791},
    issn = {0035-8711},
    arxivId = {2109.09762}
}

@article{Onorbe2015,
    title = {{Forged in FIRE: Cusps, cores and baryons in low-mass dwarf galaxies}},
    year = {2015},
    journal = {Monthly Notices of the Royal Astronomical Society},
    author = {O{\~{n}}orbe, Jose and Boylan-Kolchin, Michael and Bullock, James S. and Hopkins, Philip F. and Kere{\v{s}}, Dušan and Faucher-Gigu{\`{e}}re, Claude André and Quataert, Eliot and Murray, Norman},
    number = {2},
    pages = {2092--2106},
    volume = {454},
    publisher = {Oxford University Press},
    doi = {10.1093/mnras/stv2072},
    issn = {13652966},
    arxivId = {1502.02036}
}

@article{Cummings2016,
    title = {{GALACTIC COSMIC RAYS IN THE LOCAL INTERSTELLAR MEDIUM: VOYAGER 1 OBSERVATIONS AND MODEL RESULTS}},
    year = {2016},
    journal = {The Astrophysical Journal},
    author = {Cummings, A C and Stone, E C and Heikkila, B C and Lal, N and Webber, W R and J{\'{o}}hannesson, G and Moskalenko, I V and Orlando, E and Porter, T A},
    number = {1},
    pages = {18},
    volume = {831},
    doi = {10.3847/0004-637x/831/1/18}
}

@article{Veilleux2005,
    title = {{Galactic winds}},
    year = {2005},
    journal = {Annual Review of Astronomy and Astrophysics},
    author = {Veilleux, Sylvain and Cecil, Gerald and Bland-Hawthorn, Joss},
    number = {Volume 43, 2005},
    month = {8},
    pages = {769--826},
    volume = {43},
    publisher = {Annual Reviews},
    url = {https://www.annualreviews.org/content/journals/10.1146/annurev.astro.43.072103.150610},
    doi = {10.1146/ANNUREV.ASTRO.43.072103.150610/CITE/REFWORKS},
    issn = {00664146},
    arxivId = {astro-ph/0504435}
}

@article{Pakmor2016,
    title = {{GALACTIC WINDS DRIVEN BY ISOTROPIC AND ANISOTROPIC COSMIC-RAY DIFFUSION IN DISK GALAXIES}},
    year = {2016},
    journal = {The Astrophysical Journal},
    author = {Pakmor, R. and Pfrommer, C. and Simpson, C. M. and Springel, V.},
    number = {2},
    month = {6},
    pages = {L30},
    volume = {824},
    publisher = {IOP Publishing},
    url = {http://stacks.iop.org/2041-8205/824/i=2/a=L30?key=crossref.83d7a5946d0b7ce2d7f44bf1c4f293c9},
    doi = {10.3847/2041-8205/824/2/l30},
    issn = {2041-8213}
}

@misc{McQuinn2019,
    title = {{Galactic Winds in Low-Mass Galaxies}},
    year = {2019},
    booktitle = {arXiv},
    author = {McQuinn, Kristen B.W. and van Zee, Liese and Skillman, Evan D},
    url = {https://doi.org/10.3847/1538-4357/ab4c37},
    doi = {10.1017/s1743921319000085},
    issn = {23318422},
    arxivId = {1910.04167}
}

@article{Rosdahl2015b,
    title = {{Galaxies that shine: Radiation-hydrodynamical simulations of disc galaxies}},
    year = {2015},
    journal = {Monthly Notices of the Royal Astronomical Society},
    author = {Rosdahl, Joakim and Schaye, Joop and Teyssier, Romain and Agertz, Oscar},
    number = {1},
    month = {5},
    pages = {34--58},
    volume = {451},
    publisher = {Oxford University Press},
    url = {https://academic.oup.com/mnras/article/451/1/34/1362527},
    doi = {10.1093/mnras/stv937},
    issn = {13652966},
    arxivId = {1501.04632}
}

@article{DeRossi2017,
    title = {{Galaxy metallicity scaling relations in the EAGLE simulations}},
    year = {2017},
    journal = {Monthly Notices of the Royal Astronomical Society},
    author = {De Rossi, María Emilia and Bower, Richard G. and Font, Andreea S. and Schaye, Joop and Theuns, Tom},
    number = {3},
    month = {12},
    pages = {3354--3377},
    volume = {472},
    publisher = {Oxford Academic},
    url = {https://dx.doi.org/10.1093/mnras/stx2158},
    doi = {10.1093/MNRAS/STX2158},
    issn = {0035-8711},
    arxivId = {1704.00006}
}

@article{Ackermann2012,
    title = {{GeV observations of star-forming galaxies with the fermi large area telescope}},
    year = {2012},
    journal = {Astrophysical Journal},
    author = {Ackermann, M. and Ajello, M. and Allafort, A. and Baldini, L. and Ballet, J. and Bastieri, D. and Bechtol, K. and Bellazzini, R. and Berenji, B. and Bloom, E. D. and Bonamente, E. and Borgland, A. W. and Bouvier, A. and Bregeon, J. and Brigida, M. and Bruel, P. and Buehler, R. and Buson, S. and Caliandro, G. A. and Cameron, R. A. and Caraveo, P. A. and Casandjian, J. M. and Cecchi, C. and Charles, E. and Chekhtman, A. and Cheung, C. C. and Chiang, J. and Cillis, A. N. and Ciprini, S. and Claus, R. and Cohen-Tanugi, J. and Conrad, J. and Cutini, S. and De Palma, F. and Dermer, C. D. and Digel, S. W. and Do Couto E Silva, E. and Drell, P. S. and Drlica-Wagner, A. and Favuzzi, C. and Fegan, S. J. and Fortin, P. and Fukazawa, Y. and Funk, S. and Fusco, P. and Gargano, F. and Gasparrini, D. and Germani, S. and Giglietto, N. and Giordano, F. and Glanzman, T. and Godfrey, G. and Grenier, I. A. and Guiriec, S. and Gustafsson, M. and Hadasch, D. and Hayashida, M. and Hays, E. and Hughes, R. E. and J{\'{o}}hannesson, G. and Johnson, A. S. and Kamae, T. and Katagiri, H. and Kataoka, J. and Kn{\"{o}}dlseder, J. and Kuss, M. and Lande, J. and Longo, F. and Loparco, F. and Lott, B. and Lovellette, M. N. and Lubrano, P. and Madejski, G. M. and Martin, P. and Mazziotta, M. N. and McEnery, J. E. and Michelson, P. F. and Mizuno, T. and Monte, C. and Monzani, M. E. and Morselli, A. and Moskalenko, I. V. and Murgia, S. and Nishino, S. and Norris, J. P. and Nuss, E. and Ohno, M. and Ohsugi, T. and Okumura, A. and Omodei, N. and Orlando, E. and Ozaki, M. and Parent, D. and Persic, M. and Pesce-Rollins, M. and Petrosian, V. and Pierbattista, M. and Piron, F. and Pivato, G. and Porter, T. A. and Rain{\`{o}}, S. and Rando, R. and Razzano, M. and Reimer, A. and Reimer, O. and Ritz, S. and Roth, M. and Sbarra, C. and Sgr{\`{o}}, C. and Siskind, E. J. and Spandre, G. and Spinelli, P. and Stawarz, Łukasz and Strong, A. W. and Takahashi, H. and Tanaka, T. and Thayer, J. B. and Tibaldo, L. and Tinivella, M. and Torres, D. F. and Tosti, G. and Troja, E. and Uchiyama, Y. and Vandenbroucke, J. and Vianello, G. and Vitale, V. and Waite, A. P. and Wood, M. and Yang, Z.},
    number = {2},
    month = {8},
    pages = {164},
    volume = {755},
    publisher = {Institute of Physics Publishing},
    url = {https://iopscience.iop.org/article/10.1088/0004-637X/755/2/164 https://iopscience.iop.org/article/10.1088/0004-637X/755/2/164/meta},
    doi = {10.1088/0004-637X/755/2/164},
    issn = {15384357}
}

@article{Kortgen2019,
    title = {{Global dynamics of the interstellar medium in magnetized disc galaxies}},
    year = {2019},
    journal = {Monthly Notices of the Royal Astronomical Society},
    author = {K{\"{o}}rtgen, Bastian and Banerjee, Robi and Pudritz, Ralph E. and Schmidt, Wolfram},
    number = {4},
    month = {11},
    pages = {5004--5021},
    volume = {489},
    publisher = {Oxford University Press},
    url = {https://academic.oup.com/mnras/article/489/4/5004/5567193},
    doi = {10.1093/mnras/stz2491},
    issn = {13652966},
    arxivId = {1909.01623}
}

@article{Gorski2005,
    title = {{HEALPix: A Framework for High‐Resolution Discretization and Fast Analysis of Data Distributed on the Sphere}},
    year = {2005},
    journal = {The Astrophysical Journal},
    author = {Gorski, K. M. and Hivon, E. and Banday, A. J. and Wandelt, B. D. and Hansen, F. K. and Reinecke, M. and Bartelmann, M.},
    number = {2},
    month = {4},
    pages = {759--771},
    volume = {622},
    publisher = {IOP Publishing},
    url = {http://www.eso.org/},
    doi = {10.1086/427976},
    issn = {0004-637X},
    arxivId = {astro-ph/0409513}
}

@article{Kocjan2024,
    title = {{Hot gas accretion fuels star formation faster than cold accretion in high-redshift galaxies}},
    year = {2024},
    journal = {Monthly Notices of the Royal Astronomical Society},
    author = {Kocjan, Zuzanna and Cadiou, Corentin and Agertz, Oscar and Pontzen, Andrew},
    number = {1},
    month = {9},
    pages = {918--929},
    volume = {534},
    publisher = {Oxford Academic},
    url = {https://dx.doi.org/10.1093/mnras/stae2128},
    doi = {10.1093/MNRAS/STAE2128},
    issn = {0035-8711},
    arxivId = {2311.04961}
}

@article{Bogdan2013,
    title = {{HOT X-RAY CORONAE AROUND MASSIVE SPIRAL GALAXIES: A UNIQUE PROBE OF STRUCTURE FORMATION MODELS}},
    year = {2013},
    journal = {The Astrophysical Journal},
    author = {Bogdan, Akos and Forman, William R. and Vogelsberger, Mark and Bourdin, Hervé and Sijacki, Debora and Mazzotta, Pasquale and Kraft, Ralph P. and Jones, Christine and Gilfanov, Marat and Churazov, Eugene and David, Laurence P.},
    number = {2},
    month = {7},
    pages = {97},
    volume = {772},
    publisher = {IOP Publishing},
    url = {https://iopscience.iop.org/article/10.1088/0004-637X/772/2/97 https://iopscience.iop.org/article/10.1088/0004-637X/772/2/97/meta},
    isbn = {0673170101},
    doi = {10.1088/0004-637X/772/2/97},
    issn = {0004-637X}
}

@article{Keres2005,
    title = {{How do galaxies get their gas?}},
    year = {2005},
    journal = {Monthly Notices of the Royal Astronomical Society},
    author = {Kere{\v{s}}, Dušan and Katz, Neal and Weinberg, David H. and Dav{\'{e}}, Romeel},
    number = {1},
    month = {10},
    pages = {2--28},
    volume = {363},
    publisher = {Oxford University Press},
    url = {https://academic.oup.com/mnras/article-lookup/doi/10.1111/j.1365-2966.2005.09451.x},
    doi = {10.1111/j.1365-2966.2005.09451.x},
    issn = {00358711}
}

@article{Martin-Alvarez2020,
    title = {{How primordial magnetic fields shrink galaxies}},
    year = {2020},
    journal = {Monthly Notices of the Royal Astronomical Society},
    author = {Martin-Alvarez, Sergio and Slyz, Adrianne and Devriendt, Julien and G{\'{o}}mez-Guijarro, Carlos},
    pages = {4475--4495},
    volume = {495},
    url = {https://academic.oup.com/mnras/article-abstract/495/4/4475/5843277},
    doi = {10.1093/mnras/staa1438},
    issn = {0035-8711},
    arxivId = {2005.10269}
}

@article{Hayward2017,
    title = {{How stellar feedback simultaneously regulates star formation and drives outflows}},
    year = {2017},
    journal = {Monthly Notices of the Royal Astronomical Society},
    author = {Hayward, Christopher C. and Hopkins, Philip F.},
    number = {2},
    month = {2},
    pages = {1682--1698},
    volume = {465},
    publisher = {Narnia},
    url = {https://academic.oup.com/mnras/article-lookup/doi/10.1093/mnras/stw2888 http://arxiv.org/abs/1510.05650%0Ahttp://dx.doi.org/10.1093/mnras/stw2888},
    doi = {10.1093/mnras/stw2888},
    issn = {0035-8711},
    arxivId = {1510.05650}
}

@article{Hopkins2018a,
    title = {{How to model supernovae in simulations of star and galaxy formation}},
    year = {2018},
    journal = {Monthly Notices of the Royal Astronomical Society},
    author = {Hopkins, Philip F and Wetzel, Andrew and Kere{\v{s}}, Dušan and Faucher-Gigu{\`{e}}re, Claude André and Quataert, Eliot and Boylan-Kolchin, Michael and Murray, Norman and Hayward, Christopher C and El-Badry, Kareem},
    number = {2},
    month = {6},
    pages = {1578--1603},
    volume = {477},
    publisher = {Narnia},
    url = {https://academic.oup.com/mnras/article/477/2/1578/4935189},
    doi = {10.1093/mnras/sty674},
    issn = {13652966}
}

@article{Katz2020,
    title = {{How to quench a dwarf galaxy: The impact of inhomogeneous reionization on dwarf galaxies and cosmic filaments}},
    year = {2020},
    journal = {Monthly Notices of the Royal Astronomical Society},
    author = {Katz, Harley and Ramsoy, Marius and Rosdahl, Joakim and Kimm, Taysun and Blaizot, Jérémy and Haehnelt, Martin G and Michel-Dansac, Léo and Garel, Thibault and Laigle, Clotilde and Devriendt, Julien and Slyz, Adrianne},
    number = {2},
    month = {5},
    pages = {2200--2220},
    volume = {494},
    publisher = {Oxford University Press (OUP)},
    url = {https://academic.oup.com/mnras/article/494/2/2200/5780248},
    doi = {10.1093/mnras/staa639},
    issn = {0035-8711},
    arxivId = {1905.11414}
}

@article{Dome2025,
    title = {{Increased burstiness at high redshift in multiphysics models combining supernova feedback, radiative transfer, and cosmic rays}},
    year = {2025},
    journal = {Monthly Notices of the Royal Astronomical Society},
    author = {Dome, Tibor and Martin-Alvarez, Sergio and Tacchella, Sandro and Yuan, Yuxuan and Sijacki, Debora},
    number = {2},
    month = {1},
    pages = {629--639},
    volume = {537},
    publisher = {Oxford Academic},
    url = {https://dx.doi.org/10.1093/mnras/staf006},
    doi = {10.1093/MNRAS/STAF006},
    issn = {0035-8711}
}

@article{Geen2017,
    title = {{Interpreting the star formation efficiency of nearby molecular clouds with ionizing radiation}},
    year = {2017},
    journal = {Monthly Notices of the Royal Astronomical Society},
    author = {Geen, Sam and Soler, Juan D. and Hennebelle, Patrick},
    number = {4},
    month = {11},
    pages = {4844--4855},
    volume = {471},
    publisher = {Oxford Academic},
    url = {https://dx.doi.org/10.1093/mnras/stx1765},
    doi = {10.1093/MNRAS/STX1765},
    issn = {13652966}
}

@article{KMA2021,
    title = {{Introducing SPHINX-MHD: the impact of primordial magnetic fields on the first galaxies, reionization, and the global 21-cm signal}},
    year = {2021},
    journal = {Monthly Notices of the Royal Astronomical Society},
    author = {Katz, Harley and Martin-Alvarez, Sergio and Rosdahl, Joakim and Kimm, Taysun and Blaizot, Jérémy and Haehnelt, Martin G and Michel-Dansac, Léo and Garel, Thibault and O{\~{n}}orbe, Jose and Devriendt, Julien and Slyz, Adrianne and Attia, Omar and Teyssier, Romain},
    number = {1},
    month = {8},
    pages = {1254--1282},
    volume = {507},
    publisher = {Oxford Academic},
    url = {https://academic.oup.com/mnras/article/507/1/1254/6329052},
    doi = {10.1093/mnras/stab2148},
    issn = {0035-8711},
    arxivId = {2101.11624}
}

@article{Dubois2021,
    title = {{Introducing the NEWHORIZON simulation: Galaxy properties with resolved internal dynamics across cosmic time}},
    year = {2021},
    journal = {Astronomy {\&} Astrophysics},
    author = {Dubois, Yohan and Beckmann, Ricarda and Bournaud, Frédéric and Choi, Hoseung and Devriendt, Julien and Jackson, Ryan and Kaviraj, Sugata and Kimm, Taysun and Kraljic, Katarina and Laigle, Clotilde and Martin, Garreth and Park, Min Jung and Peirani, Sébastien and Pichon, Christophe and Volonteri, Marta and Yi, Sukyoung K.},
    month = {7},
    pages = {A109},
    volume = {651},
    publisher = {EDP Sciences},
    url = {https://www.aanda.org/articles/aa/full_html/2021/07/aa39429-20/aa39429-20.html https://www.aanda.org/articles/aa/abs/2021/07/aa39429-20/aa39429-20.html},
    doi = {10.1051/0004-6361/202039429},
    issn = {0004-6361},
    arxivId = {2009.10578}
}

@article{Llerena2023,
    title = {{Ionized gas kinematics and chemical abundances of low-mass star-forming galaxies at z  ∼  3}},
    year = {2023},
    journal = {Astronomy {\&} Astrophysics},
    author = {Llerena, M. and Amor{\'{i}}n, R. and Pentericci, L. and Calabr{\`{o}}, A. and Shapley, A. E. and Boutsia, K. and P{\'{e}}rez-Montero, E. and V{\'{i}}lchez, J. M. and Nakajima, K.},
    month = {8},
    pages = {A53},
    volume = {676},
    publisher = {EDP Sciences},
    url = {https://www.aanda.org/articles/aa/full_html/2023/08/aa46232-23/aa46232-23.html https://www.aanda.org/articles/aa/abs/2023/08/aa46232-23/aa46232-23.html},
    doi = {10.1051/0004-6361/202346232},
    issn = {0004-6361},
    arxivId = {2303.01536}
}

@article{Dale2012,
    title = {{Ionizing feedback from massive stars in massive clusters - II. Disruption of bound clusters by photoionization}},
    year = {2012},
    journal = {Monthly Notices of the Royal Astronomical Society},
    author = {Dale, J. E. and Ercolano, B. and Bonnell, I. A.},
    number = {1},
    month = {7},
    pages = {377--392},
    volume = {424},
    publisher = {Oxford University Press},
    url = {https://dx.doi.org/10.1111/j.1365-2966.2012.21205.x},
    doi = {10.1111/J.1365-2966.2012.21205.X/2/MNRAS0424-0377-F17.JPEG},
    issn = {13652966}
}

@article{Curti2024,
    title = {{JADES: Insights into the low-mass end of the mass–metallicity–SFR relation at 3 < z < 10 from deep JWST/NIRSpec spectroscopy}},
    year = {2024},
    journal = {Astronomy {\&} Astrophysics},
    author = {Curti, Mirko and Maiolino, Roberto and Curtis-Lake, Emma and Chevallard, Jacopo and Carniani, Stefano and D'Eugenio, Francesco and Looser, Tobias J. and Scholtz, Jan and Charlot, Stephane and Cameron, Alex and {\"{U}}bler, Hannah and Witstok, Joris and Boyett, Kristian and Laseter, Isaac and Sandles, Lester and Arribas, Santiago and Bunker, Andrew and Giardino, Giovanna and Maseda, Michael V. and Rawle, Tim and Del Pino, Bruno Rodríguez and Smit, Renske and Willott, Chris J. and Eisenstein, Daniel J. and Hausen, Ryan and Johnson, Benjamin and Rieke, Marcia and Robertson, Brant and Tacchella, Sandro and Williams, Christina C. and Willmer, Christopher and Baker, William M. and Bhatawdekar, Rachana and Egami, Eiichi and Helton, Jakob M. and Ji, Zhiyuan and Kumari, Nimisha and Perna, Michele and Shivaei, Irene and Sun, Fengwu},
    month = {4},
    pages = {A75},
    volume = {684},
    publisher = {EDP Sciences},
    url = {https://www.aanda.org/articles/aa/full_html/2024/04/aa46698-23/aa46698-23.html https://www.aanda.org/articles/aa/abs/2024/04/aa46698-23/aa46698-23.html},
    doi = {10.1051/0004-6361/202346698},
    issn = {0004-6361}
}

@article{Carniani2024,
    title = {{JADES: The incidence rate and properties of galactic outflows in low-mass galaxies across 3 < z < 9}},
    year = {2024},
    journal = {Astronomy {\&} Astrophysics},
    author = {Carniani, Stefano and Venturi, Giacomo and Parlanti, Eleonora and de Graaff, Anna and Maiolino, Roberto and Arribas, Santiago and Bonaventura, Nina and Boyett, Kristan and Bunker, Andrew J. and Cameron, Alex J. and Charlot, Stephane and Chevallard, Jacopo and Curti, Mirko and Curtis-Lake, Emma and Eisenstein, Daniel J. and Giardino, Giovanna and Hausen, Ryan and Kumari, Nimisha and Maseda, Michael V. and Nelson, Erica and Perna, Michele and Rix, Hans Walter and Robertson, Brant and Rodr{\'{i}}guez Del Pino, Bruno and Sandles, Lester and Scholtz, Jan and Simmonds, Charlotte and Smit, Renske and Tacchella, Sandro and {\"{U}}bler, Hannah and Williams, Christina C. and Willott, Chris and Witstok, Joris},
    month = {5},
    pages = {A99},
    volume = {685},
    publisher = {EDP Sciences},
    url = {https://www.aanda.org/articles/aa/full_html/2024/05/aa47230-23/aa47230-23.html https://www.aanda.org/articles/aa/abs/2024/05/aa47230-23/aa47230-23.html},
    doi = {10.1051/0004-6361/202347230},
    issn = {0004-6361},
    arxivId = {2306.11801}
}

@article{Teyssier2006,
    title = {{Kinematic dynamos using constrained transport with high order Godunov schemes and adaptive mesh refinement}},
    year = {2006},
    journal = {Journal of Computational Physics},
    author = {Teyssier, Romain and Fromang, Sébastien and Dormy, Emmanuel},
    number = {1},
    month = {10},
    pages = {44--67},
    volume = {218},
    publisher = {Academic Press},
    url = {https://www.sciencedirect.com/science/article/pii/S0021999106000593?via%3Dihub},
    doi = {10.1016/j.jcp.2006.01.042},
    issn = {00219991},
    arxivId = {astro-ph/0601715}
}

@article{Schruba2012,
    title = {{LOW CO LUMINOSITIES IN DWARF GALAXIES}},
    year = {2012},
    journal = {The Astronomical Journal},
    author = {Schruba, Andreas and Leroy, Adam K. and Walter, Fabian and Bigiel, Frank and Brinks, Elias and De Blok, W. J.G. and Kramer, Carsten and Rosolowsky, Erik and Sandstrom, Karin and Schuster, Karl and Usero, Antonio and Weiss, Axel and Wiesemeyer, Helmut},
    number = {6},
    month = {5},
    pages = {138},
    volume = {143},
    publisher = {IOP Publishing},
    url = {https://iopscience.iop.org/article/10.1088/0004-6256/143/6/138 https://iopscience.iop.org/article/10.1088/0004-6256/143/6/138/meta},
    doi = {10.1088/0004-6256/143/6/138},
    issn = {1538-3881},
    arxivId = {1203.4231}
}

@article{Rosdahl2022,
    title = {{LyC escape from sphinx galaxies in the Epoch of Reionization}},
    year = {2022},
    journal = {Monthly Notices of the Royal Astronomical Society},
    author = {Rosdahl, Joakim and Blaizot, Jérémy and Katz, Harley and Kimm, Taysun and Garel, Thibault and Haehnelt, Martin and Keating, Laura C. and Martin-Alvarez, Sergion and Michel-Dansac, Léo and Ocvirk, Pierre},
    number = {2},
    month = {7},
    pages = {2386--2414},
    volume = {515},
    url = {http://arxiv.org/abs/2207.03232},
    doi = {10.1093/mnras/stac1942},
    issn = {0035-8711},
    arxivId = {2207.03232}
}

@article{Gutcke2022,
    title = {{LYRA. III. The Smallest Reionization Survivors}},
    year = {2022},
    journal = {The Astrophysical Journal},
    author = {Gutcke, Thales A. and Pfrommer, Christoph and Bryan, Greg L. and Pakmor, Rüdiger and Springel, Volker and Naab, Thorsten},
    number = {2},
    month = {9},
    pages = {120},
    volume = {941},
    url = {http://arxiv.org/abs/2209.03366},
    doi = {10.3847/1538-4357/aca1b4},
    issn = {0004-637X},
    arxivId = {2209.03366}
}

@article{Yuan2024,
    title = {{Ly{$\alpha$} emission as a sensitive probe of feedback-regulated LyC escape from dwarf galaxies}},
    year = {2024},
    journal = {Monthly Notices of the Royal Astronomical Society},
    author = {Yuan, Yuxuan and Martin-Alvarez, Sergio and Haehnelt, Martin G. and Garel, Thibault and Sijacki, Debora},
    number = {4},
    month = {7},
    pages = {3643--3668},
    volume = {532},
    publisher = {Oxford Academic},
    url = {https://dx.doi.org/10.1093/mnras/stae1606},
    doi = {10.1093/MNRAS/STAE1606},
    issn = {0035-8711},
    arxivId = {2401.02572}
}

@article{Yang2017a,
    title = {{Ly{$\alpha$} Profile, Dust, and Prediction of Ly{$\alpha$} Escape Fraction in Green Pea Galaxies}},
    year = {2017},
    journal = {The Astrophysical Journal},
    author = {Yang, Huan and Malhotra, Sangeeta and Gronke, Max and Rhoads, James E. and Leitherer, Claus and Wofford, Aida and Jiang, Tianxing and Dijkstra, Mark and Tilvi, V. and Wang, Junxian},
    number = {2},
    month = {8},
    pages = {171},
    volume = {844},
    publisher = {IOP Publishing},
    url = {https://iopscience.iop.org/article/10.3847/1538-4357/aa7d4d https://iopscience.iop.org/article/10.3847/1538-4357/aa7d4d/meta},
    doi = {10.3847/1538-4357/AA7D4D},
    issn = {0004-637X}
}

@techreport{Shukurov2018,
    title = {{Magnetic field effects on the ISM structure and galactic outflows}},
    year = {2018},
    booktitle = {XXIXth IAU General Assembly},
    author = {Shukurov, A. and Evirgen, C. C. and Fletcher, A. and Bushby, P. J. and Gent, F. A.},
    month = {10},
    volume = {1},
    url = {http://arxiv.org/abs/1810.01202 https://arxiv.org/pdf/1810.01202.pdf},
    institution = {International Astronomical Union},
    arxivId = {arXiv:1810.01202v1}
}

@article{Gronnow2018,
    title = {{Magnetic Fields in the Galactic Halo Restrict Fountain-driven Recycling and Accretion}},
    year = {2018},
    journal = {The Astrophysical Journal},
    author = {Gr{\o}nnow, Asger and Tepper-Garc{\'{i}}a, Thor and Bland-Hawthorn, Joss},
    number = {1},
    month = {9},
    pages = {64},
    volume = {865},
    publisher = {IOP Publishing},
    url = {http://stacks.iop.org/0004-637X/865/i=1/a=64?key=crossref.f4df477e60a718854be452ec165b3405},
    doi = {10.3847/1538-4357/aada0e},
    issn = {1538-4357},
    arxivId = {1805.03903}
}

@article{Zamora-Aviles2018,
    title = {{Magnetic suppression of turbulence and the star formation activity of molecular clouds}},
    year = {2018},
    journal = {Monthly Notices of the Royal Astronomical Society},
    author = {Zamora-Avil{\'{e}}s, Manuel and V{\'{a}}zquez-Semadeni, Enrique and K{\"{o}}rtgen, Bastian and Banerjee, Robi and Hartmann, Lee},
    number = {4},
    month = {3},
    pages = {4824--4836},
    volume = {474},
    publisher = {Oxford University Press},
    url = {http://academic.oup.com/mnras/article/474/4/4824/4675228},
    doi = {10.1093/mnras/stx3080},
    issn = {13652966}
}

@article{Choi2024,
    title = {{Metallicity Mapping of the Ionized Diffuse Gas at the Milky Way Disk–Halo Interface}},
    year = {2024},
    journal = {The Astrophysical Journal},
    author = {Choi, Bo-Eun and Werk, Jessica K. and Tchernyshyov, Kirill and Prochaska, J. Xavier and Zheng, Yong and Putman, Mary E. and Fielding, Drummond B. and Strader, Jay},
    number = {2},
    month = {11},
    pages = {222},
    volume = {976},
    publisher = {IOP Publishing},
    url = {https://iopscience.iop.org/article/10.3847/1538-4357/ad84f8 https://iopscience.iop.org/article/10.3847/1538-4357/ad84f8/meta},
    isbn = {212734.4+115714},
    doi = {10.3847/1538-4357/AD84F8},
    issn = {0004-637X},
    arxivId = {2410.06286}
}

@article{Katz2022b,
    title = {{Mg ii in the JWST era: a probe of Lyman continuum escape?}},
    year = {2022},
    journal = {Monthly Notices of the Royal Astronomical Society},
    author = {Katz, Harley and Garel, Thibault and Rosdahl, Joakim and Mauerhofer, Valentin and Kimm, Taysun and Blaizot, Jérémy and Michel-Dansac, Léo and Devriendt, Julien and Slyz, Adrianne and Haehnelt, Martin},
    number = {3},
    month = {8},
    pages = {4265--4286},
    volume = {515},
    publisher = {Oxford Academic},
    url = {https://dx.doi.org/10.1093/mnras/stac1437},
    doi = {10.1093/MNRAS/STAC1437},
    issn = {0035-8711},
    arxivId = {2205.11534}
}

@article{Broderick2018,
    title = {{Missing Gamma-ray Halos and the Need for New Physics in the Gamma-ray Sky}},
    year = {2018},
    journal = {The Astrophysical Journal},
    author = {Broderick, Avery E. and Tiede, Paul and Chang, Philip and Lamberts, Astrid and Pfrommer, Christoph and Puchwein, Ewald and Shalaby, Mohamad and Werhahn, Maria},
    number = {2},
    month = {11},
    pages = {87},
    volume = {868},
    publisher = {IOP Publishing},
    url = {http://stacks.iop.org/0004-637X/868/i=2/a=87?key=crossref.df4f9bc30658360344bfa0ad53519a58 http://arxiv.org/abs/1808.02959},
    doi = {10.3847/1538-4357/aae5f2},
    issn = {1538-4357},
    arxivId = {1808.02959}
}

@article{Curro2022,
    title = {{Momentum deposition of supernovae with cosmic rays}},
    year = {2022},
    journal = {Monthly Notices of the Royal Astronomical Society},
    author = {Rodr{\'{i}}guez Montero, Francisco and Martin-Alvarez, Sergio and Sijacki, Debora and Slyz, Adrianne and Devriendt, Julien and Dubois, Yohan},
    number = {1},
    month = {2},
    pages = {1247--1264},
    volume = {511},
    publisher = {Oxford Academic},
    url = {https://academic.oup.com/mnras/article/511/1/1247/6479141},
    doi = {10.1093/mnras/stab3716},
    issn = {0035-8711},
    arxivId = {2110.09862}
}

@article{Rodriguez-Montero2022,
    title = {{Momentum deposition of supernovae with cosmic rays}},
    year = {2022},
    journal = {Monthly Notices of the Royal Astronomical Society},
    author = {Rodr{\'{i}}guez Montero, Francisco and Martin-Alvarez, Sergio and Sijacki, Debora and Slyz, Adrianne and Devriendt, Julien and Dubois, Yohan},
    number = {1},
    month = {2},
    pages = {1247--1264},
    volume = {511},
    publisher = {Oxford Academic},
    url = {https://dx.doi.org/10.1093/mnras/stab3716},
    doi = {10.1093/mnras/stab3716},
    issn = {13652966},
    arxivId = {2110.09862}
}

@article{Gentry2020,
    title = {{Momentum injection by clustered supernovae: testing subgrid feedback prescriptions}},
    year = {2020},
    journal = {Monthly Notices of the Royal Astronomical Society},
    author = {Gentry, Eric S. and Madau, Piero and Krumholz, Mark R.},
    number = {1},
    month = {2},
    pages = {1243--1256},
    volume = {492},
    publisher = {Oxford Academic},
    url = {https://dx.doi.org/10.1093/mnras/stz3440},
    doi = {10.1093/MNRAS/STZ3440},
    issn = {0035-8711},
    arxivId = {1912.01141}
}

@article{Kim2015,
    title = {{MOMENTUM INJECTION BY SUPERNOVAE IN THE INTERSTELLAR MEDIUM}},
    year = {2015},
    journal = {The Astrophysical Journal},
    author = {Kim, Chang Goo and Ostriker, Eve C.},
    number = {2},
    month = {3},
    pages = {99},
    volume = {802},
    publisher = {IOP Publishing},
    url = {https://iopscience.iop.org/article/10.1088/0004-637X/802/2/99 https://iopscience.iop.org/article/10.1088/0004-637X/802/2/99/meta},
    doi = {10.1088/0004-637X/802/2/99},
    issn = {0004-637X},
    arxivId = {1410.1537}
}

@article{Dave2017,
    title = {{Mufasa: galaxy star formation, gas and metal properties across cosmic time}},
    year = {2017},
    journal = {Monthly Notices of the Royal Astronomical Society},
    author = {Dav{\'{e}}, Romeel and Rafieferantsoa, Mika H. and Thompson, Robert J. and Hopkins, Philip F.},
    number = {1},
    month = {5},
    pages = {115--132},
    volume = {467},
    publisher = {Oxford Academic},
    url = {https://dx.doi.org/10.1093/mnras/stx108},
    doi = {10.1093/MNRAS/STX108},
    issn = {0035-8711},
    arxivId = {1610.01626}
}

@article{Cherrey2025,
    title = {{MusE GAs FLOw and Wind (MEGAFLOW) - XIII. Cool gas traced by Mg II around isolated galaxies}},
    year = {2025},
    journal = {Astronomy {\&} Astrophysics},
    author = {Cherrey, Maxime and Bouch{\'{e}}, Nicolas F. and Zabl, Johannes and Schroetter, Ilane and Wendt, Martin and Langan, Ivanna and Schaye, Joop and Wisotzki, Lutz and Guo, Yucheng and Pessa, Ismael},
    month = {2},
    pages = {A117},
    volume = {694},
    publisher = {EDP Sciences},
    url = {https://www.aanda.org/articles/aa/full_html/2025/02/aa51165-24/aa51165-24.html https://www.aanda.org/articles/aa/abs/2025/02/aa51165-24/aa51165-24.html},
    doi = {10.1051/0004-6361/202451165},
    issn = {0004-6361}
}

@article{Baron2024,
    title = {{Not So Windy After All: MUSE Disentangles AGN-driven Winds from Merger-induced Flows in Galaxies along the Starburst Sequence}},
    year = {2024},
    journal = {The Astrophysical Journal},
    author = {Baron, Dalya and Netzer, Hagai and Lutz, Dieter and Davies, Ric I. and Prochaska, J. Xavier},
    number = {1},
    month = {6},
    pages = {23},
    volume = {968},
    publisher = {IOP Publishing},
    url = {https://iopscience.iop.org/article/10.3847/1538-4357/ad39e9 https://iopscience.iop.org/article/10.3847/1538-4357/ad39e9/meta},
    doi = {10.3847/1538-4357/AD39E9},
    issn = {0004-637X}
}

@article{Kim2018,
    title = {{Numerical Simulations of Multiphase Winds and Fountains from Star-forming Galactic Disks. I. Solar Neighborhood TIGRESS Model}},
    year = {2018},
    journal = {The Astrophysical Journal},
    author = {Kim, Chang-Goo and Ostriker, Eve C.},
    number = {2},
    month = {2},
    pages = {173},
    volume = {853},
    publisher = {IOP Publishing},
    url = {https://iopscience.iop.org/article/10.3847/1538-4357/aaa5ff https://iopscience.iop.org/article/10.3847/1538-4357/aaa5ff/meta},
    doi = {10.3847/1538-4357/AAA5FF},
    issn = {0004-637X},
    arxivId = {1801.03952}
}

@article{Parizot2006,
    title = {{Observational constraints on energetic particle diffusion in young supernovae remnants: Amplified magnetic field and maximum energy}},
    year = {2006},
    journal = {Astronomy and Astrophysics},
    author = {Parizot, E and Marcowith, A and Ballet, J and Gallant, Y A},
    number = {2},
    pages = {387--395},
    volume = {453},
    url = {http://dx.doi.org/10.1051/0004-6361:20064985},
    doi = {10.1051/0004-6361:20064985},
    issn = {14320746}
}

@article{Shimizu2019,
    title = {{Osaka feedback model: isolated disc galaxy simulations}},
    year = {2019},
    journal = {Monthly Notices of the Royal Astronomical Society},
    author = {Shimizu, Ikkoh and Todoroki, Keita and Yajima, Hidenobu and Nagamine, Kentaro},
    number = {2},
    month = {4},
    pages = {2632--2655},
    volume = {484},
    publisher = {Oxford Academic},
    url = {https://dx.doi.org/10.1093/mnras/stz098},
    doi = {10.1093/MNRAS/STZ098},
    issn = {0035-8711},
    arxivId = {1901.03815}
}

@article{Menon2023,
    title = {{Outflows driven by direct and reprocessed radiation pressure in massive star clusters}},
    year = {2023},
    journal = {Monthly Notices of the Royal Astronomical Society},
    author = {Menon, Shyam H. and Federrath, Christoph and Krumholz, Mark R.},
    number = {4},
    month = {3},
    pages = {5160--5176},
    volume = {521},
    publisher = {Oxford Academic},
    url = {https://dx.doi.org/10.1093/mnras/stad856},
    doi = {10.1093/MNRAS/STAD856},
    issn = {0035-8711},
    arxivId = {2210.02818}
}

@article{Barfety2025,
    title = {{PHIBSS: Searching for Molecular Gas Outflows in Star-forming Galaxies at z = 0.5–2.6}},
    year = {2025},
    journal = {The Astrophysical Journal},
    author = {Barfety, Capucine and Jolly, Jean-Baptiste and Schreiber, Natascha M. Förster and Tacconi, Linda J. and Genzel, Reinhard and Tozzi, Giulia and Burkert, Andreas and Chen, Jianhang and Combes, Françoise and Davies, Ric and Eisenhauer, Frank and Salcedo, Juan M. Espejo and Herrera-Camus, Rodrigo and Lee, Lilian L. and Lee, Minju M. and Liu, Daizhong and Lutz, Dieter and Naab, Thorsten and Neri, Roberto and Shachar, Amit Nestor and Pastras, Stavros and Pulsoni, Claudia and Price, Sedona H. and Renzini, Alvio and Schuster, Karl and Shimizu, Taro T. and Sternberg, Amiel and Sturm, Eckhard and {\"{U}}bler, Hannah and Wuyts, Stijn},
    number = {1},
    month = {7},
    pages = {55},
    volume = {988},
    publisher = {IOP Publishing},
    url = {https://iopscience.iop.org/article/10.3847/1538-4357/addc6f https://iopscience.iop.org/article/10.3847/1538-4357/addc6f/meta},
    doi = {10.3847/1538-4357/ADDC6F},
    issn = {0004-637X}
}

@article{Geen2015b,
    title = {{Photoionization feedback in a self-gravitating, magnetized, turbulent cloud}},
    year = {2015},
    journal = {Monthly Notices of the Royal Astronomical Society},
    author = {Geen, Sam and Hennebelle, Patrick and Tremblin, Pascal and Rosdahl, Joakim},
    number = {4},
    month = {12},
    pages = {4484--4502},
    volume = {454},
    publisher = {Oxford University Press},
    url = {https://academic.oup.com/mnras/article/454/4/4484/1001540},
    doi = {10.1093/mnras/stv2272},
    issn = {13652966},
    arxivId = {1507.02981}
}

@article{Sartorio2021,
    title = {{Photoionization feedback in turbulent molecular clouds}},
    year = {2020},
    journal = {Monthly Notices of the Royal Astronomical Society},
    author = {Sartorio, Nina S. and Vandenbroucke, Bert and Falceta-Goncalves, Diego and Wood, Kenneth},
    number = {2},
    month = {12},
    pages = {1833--1843},
    volume = {500},
    publisher = {Oxford Academic},
    url = {https://dx.doi.org/10.1093/mnras/staa3380},
    doi = {10.1093/MNRAS/STAA3380},
    issn = {0035-8711},
    arxivId = {2011.00020}
}

@article{Bennett2025,
    title = {{Prevention is better than cure? Feedback from high specific energy winds in cosmological simulations with Arkenstone}},
    year = {2025},
    journal = {Monthly Notices of the Royal Astronomical Society},
    author = {Bennett, Jake S and Smith, Matthew C and Fielding, Drummond B and Bryan, Greg L and Kim, Chang-Goo and Springel, Volker and Hernquist, Lars and Somerville, Rachel S and Sommovigo, Laura},
    number = {2},
    month = {9},
    pages = {1456--1478},
    volume = {543},
    publisher = {Oxford Academic},
    url = {https://dx.doi.org/10.1093/mnras/staf1440},
    doi = {10.1093/MNRAS/STAF1440},
    issn = {0035-8711}
}

@article{Maglione2025,
    title = {{PRIMA Vista: far-infrared polarimetry to unveil small-scale magnetohydrodynamics in extragalactic observations}},
    year = {2025},
    journal = {Journal of Astronomical Telescopes, Instruments, and Systems},
    author = {Maglione, Diego and Martin-Alvarez, Sergio and Lopez-Rodriguez, Enrique and Clark, Susan E. and Karpovich, Kaitlyn},
    number = {03},
    month = {9},
    volume = {11},
    url = {https://www.spiedigitallibrary.org/journals/Journal-of-Astronomical-Telescopes-Instruments-and-Systems/volume-11/issue-03/031620/PRIMA-Vista--far-infrared-polarimetry-to-unveil-small-scale/10.1117/1.JATIS.11.3.031620.full},
    doi = {10.1117/1.JATIS.11.3.031620},
    issn = {2329-4124}
}

@article{Li2020,
    title = {{Probing the CGM of low-redshift dwarf galaxies using FIRE simulations}},
    year = {2020},
    journal = {Monthly Notices of the Royal Astronomical Society},
    author = {Li, Fei and Rahman, Mubdi and Murray, Norman and Hafen, Zachary and Faucher-Gigu{\`{e}}re, Claude André and Stern, Jonathan and Hummels, Cameron B. and Hopkins, Philip F. and El-Badry, Kareem and Kere{\v{s}}, Dušan},
    number = {1},
    month = {12},
    pages = {1038--1053},
    volume = {500},
    publisher = {Oxford Academic},
    url = {https://dx.doi.org/10.1093/mnras/staa3322},
    doi = {10.1093/MNRAS/STAA3322},
    issn = {0035-8711},
    arxivId = {2010.13606}
}

@article{Helder2013,
    title = {{Proper motions of H{$\alpha$} filaments in the supernova remnant RCW 86}},
    year = {2013},
    journal = {Monthly Notices of the Royal Astronomical Society},
    author = {Helder, E. A. and Vink, J. and Bamba, A. and Bleeker, J. A.M. and Burrows, D. N. and Ghavamian, P. and Yamazaki, R.},
    number = {2},
    month = {10},
    pages = {910--916},
    volume = {435},
    publisher = {Oxford Academic},
    url = {https://academic.oup.com/mnras/article/435/2/910/1047123},
    doi = {10.1093/mnras/stt993},
    issn = {00358711}
}

@article{Fluetsch2021,
    title = {{Properties of the multiphase outflows in local (ultra)luminous infrared galaxies}},
    year = {2021},
    journal = {Monthly Notices of the Royal Astronomical Society},
    author = {Fluetsch, A. and Maiolino, R. and Carniani, S. and Arribas, S. and Belfiore, F. and Bellocchi, E. and Cazzoli, S. and Cicone, C. and Cresci, G. and Fabian, A. C. and Gallagher, R. and Ishibashi, W. and Mannucci, F. and Marconi, A. and Perna, M. and Sturm, E. and Venturi, G.},
    number = {4},
    month = {7},
    pages = {5753--5783},
    volume = {505},
    publisher = {Oxford Academic},
    url = {https://dx.doi.org/10.1093/mnras/stab1666},
    doi = {10.1093/MNRAS/STAB1666},
    issn = {0035-8711},
    arxivId = {2006.13232}
}

@article{Farcy2022,
    title = {{Radiation-magnetohydrodynamics simulations of cosmic ray feedback in disc galaxies}},
    year = {2022},
    journal = {Monthly Notices of the Royal Astronomical Society},
    author = {Farcy, Marion and Rosdahl, Joakim and Dubois, Yohan and Blaizot, Jérémy and Martin-Alvarez, Sergio},
    number = {4},
    month = {2},
    pages = {5000--5019},
    volume = {513},
    url = {http://arxiv.org/abs/2202.01245},
    isbn = {2202.01245v1},
    doi = {10.1093/mnras/stac1196},
    issn = {0035-8711},
    arxivId = {2202.01245}
}

@article{Rosdahl2013,
    title = {{Ramses-rt: Radiation hydrodynamics in the cosmological context}},
    year = {2013},
    journal = {Monthly Notices of the Royal Astronomical Society},
    author = {Rosdahl, J. and Blaizot, J. and Aubert, D. and Stranex, T. and Teyssier, R.},
    number = {3},
    month = {12},
    pages = {2188--2231},
    volume = {436},
    publisher = {Oxford Academic},
    url = {https://academic.oup.com/mnras/article/436/3/2188/1247446},
    doi = {10.1093/mnras/stt1722},
    issn = {00358711}
}

@article{Robinson2024,
    title = {{Regulating star formation in a magnetized disc galaxy}},
    year = {2024},
    journal = {Monthly Notices of the Royal Astronomical Society},
    author = {Robinson, Hector and Wadsley, James},
    number = {2},
    month = {9},
    pages = {1420--1432},
    volume = {534},
    publisher = {Oxford Academic},
    url = {https://dx.doi.org/10.1093/mnras/stae2132},
    doi = {10.1093/MNRAS/STAE2132},
    issn = {0035-8711},
    arxivId = {2310.15244}
}

@article{Witten2025,
    title = {{Rising from the ashes: evidence of old stellar populations and rejuvenation events in the very early Universe}},
    year = {2025},
    journal = {Monthly Notices of the Royal Astronomical Society},
    author = {Witten, Callum and McClymont, William and Laporte, Nicolas and Roberts-Borsani, Guido and Sijacki, Debora and Tacchella, Sandro and Simmonds, Charlotte and Katz, Harley and Ellis, Richard S. and Witstok, Joris and Maiolino, Roberto and Ji, Xihan and Hayes, Billy R. and Looser, Tobias J. and D'Eugenio, Francesco},
    number = {1},
    month = {1},
    pages = {112--126},
    volume = {537},
    publisher = {Oxford Academic},
    url = {https://dx.doi.org/10.1093/mnras/staf001},
    doi = {10.1093/MNRAS/STAF001},
    issn = {0035-8711},
    arxivId = {2407.07937}
}

@article{Salem2016,
    title = {{Role of cosmic rays in the circumgalactic medium}},
    year = {2016},
    journal = {Monthly Notices of the Royal Astronomical Society},
    author = {Salem, Munier and Bryan, Greg L. and Corlies, Lauren},
    number = {1},
    month = {2},
    pages = {582--601},
    volume = {456},
    publisher = {Oxford University Press},
    url = {https://academic.oup.com/mnras/article/456/1/582/1066363},
    doi = {10.1093/mnras/stv2641},
    issn = {13652966},
    arxivId = {1511.05144}
}

@article{Kado-Fong2024,
    title = {{SAGAbg. I. A Near-unity Mass-loading Factor in Low-mass Galaxies via Their Low-redshift Evolution in Stellar Mass, Oxygen Abundance, and Star Formation Rate}},
    year = {2024},
    journal = {The Astrophysical Journal},
    author = {Kado-Fong, Erin and Geha, Marla and Mao, Yao-Yuan and Reyes, Mithi A. C. de los and Wechsler, Risa H. and Asali, Yasmeen and Kallivayalil, Nitya and Nadler, Ethan O. and Tollerud, Erik J. and Weiner, Benjamin},
    number = {1},
    month = {5},
    pages = {129},
    volume = {966},
    publisher = {IOP Publishing},
    url = {https://iopscience.iop.org/article/10.3847/1538-4357/ad3042 https://iopscience.iop.org/article/10.3847/1538-4357/ad3042/meta},
    doi = {10.3847/1538-4357/AD3042},
    issn = {0004-637X},
    arxivId = {2401.16469}
}

@article{Chisholm2015,
    title = {{Scaling relations between warm galactic outflows and their host galaxies}},
    year = {2015},
    journal = {Astrophysical Journal},
    author = {Chisholm, John and Tremonti, Christy A. and Leitherer, Claus and Chen, Yanmei and Wofford, Aida and Lundgren, Britt},
    number = {2},
    month = {9},
    pages = {149},
    volume = {811},
    publisher = {IOP Publishing},
    url = {https://iopscience.iop.org/article/10.1088/0004-637X/811/2/149 https://iopscience.iop.org/article/10.1088/0004-637X/811/2/149/meta},
    doi = {10.1088/0004-637X/811/2/149},
    issn = {15384357},
    arxivId = {1412.2139}
}

@article{Marasco2023,
    title = {{Shaken, but not expelled: Gentle baryonic feedback from nearby starburst dwarf galaxies}},
    year = {2023},
    journal = {Astronomy {\&} Astrophysics},
    author = {Marasco, A. and Belfiore, F. and Cresci, G. and Lelli, F. and Venturi, G. and Hunt, L. K. and Concas, A. and Marconi, A. and Mannucci, F. and Mingozzi, M. and McLeod, A. F. and Kumari, N. and Carniani, S. and Vanzi, L. and Ginolfi, M.},
    month = {2},
    pages = {A92},
    volume = {670},
    publisher = {EDP Sciences},
    url = {https://www.aanda.org/articles/aa/full_html/2023/02/aa44895-22/aa44895-22.html https://www.aanda.org/articles/aa/abs/2023/02/aa44895-22/aa44895-22.html},
    doi = {10.1051/0004-6361/202244895},
    issn = {0004-6361}
}

@article{Dubois2019,
    title = {{Shock-accelerated cosmic rays and streaming instability in the adaptive mesh refinement code Ramses}},
    year = {2019},
    journal = {Astronomy and Astrophysics},
    author = {Dubois, Yohan and Commer{\c{c}}on, Benoît and Marcowith, Alexandre and Brahimi, Loann},
    month = {11},
    pages = {A121},
    volume = {631},
    url = {https://www.aanda.org/10.1051/0004-6361/201936275},
    doi = {10.1051/0004-6361/201936275},
    issn = {14320746},
    arxivId = {1907.04300}
}

@article{Dave2019,
    title = {{simba: Cosmological simulations with black hole growth and feedback}},
    year = {2019},
    journal = {Monthly Notices of the Royal Astronomical Society},
    author = {Dav{\'{e}}, Romeel and Angl{\'{e}}s-Alc{\'{a}}zar, Daniel and Narayanan, Desika and Li, Qi and Rafieferantsoa, Mika H. and Appleby, Sarah},
    number = {2},
    month = {6},
    pages = {2827--2849},
    volume = {486},
    publisher = {Oxford Academic},
    url = {https://dx.doi.org/10.1093/mnras/stz937},
    doi = {10.1093/MNRAS/STZ937},
    issn = {0035-8711},
    arxivId = {1901.10203}
}

@article{Pfrommer2017b,
    title = {{Simulating cosmic ray physics on a moving mesh}},
    year = {2017},
    journal = {Monthly Notices of the Royal Astronomical Society},
    author = {Pfrommer, C. and Pakmor, R. and Schaal, K. and Simpson, C. M. and Springel, V.},
    number = {4},
    month = {3},
    pages = {4500--4529},
    volume = {465},
    publisher = {Oxford University Press},
    url = {https://academic.oup.com/mnras/article-lookup/doi/10.1093/mnras/stw2941},
    doi = {10.1093/mnras/stw2941},
    issn = {0035-8711}
}

@article{Pfrommer2007,
    title = {{Simulating cosmic rays in clusters of galaxies - I. Effects on the Sunyaev-Zel'dovich effect and the X-ray emission}},
    year = {2007},
    journal = {Monthly Notices of the Royal Astronomical Society},
    author = {Pfrommer, C. and En{\ss}lin, T. A. and Springel, V. and Jubelgas, M. and Dolag, K.},
    number = {2},
    month = {6},
    pages = {385--408},
    volume = {378},
    publisher = {Blackwell Publishing Ltd},
    url = {https://dx.doi.org/10.1111/j.1365-2966.2007.11732.x},
    doi = {10.1111/J.1365-2966.2007.11732.X/2/M{\_}MNRAS0378-0385-M13.GIF},
    issn = {13652966}
}

@article{Pillepich2018a,
    title = {{Simulating galaxy formation with the IllustrisTNG model}},
    year = {2018},
    journal = {Monthly Notices of the Royal Astronomical Society},
    author = {Pillepich, Annalisa and Springel, Volker and Nelson, Dylan and Genel, Shy and Naiman, Jill and Pakmor, Rüdiger and Hernquist, Lars and Torrey, Paul and Vogelsberger, Mark and Weinberger, Rainer and Marinacci, Federico},
    number = {3},
    month = {1},
    pages = {4077--4106},
    volume = {473},
    publisher = {Oxford University Press},
    url = {http://academic.oup.com/mnras/article/473/3/4077/4494369},
    doi = {10.1093/mnras/stx2656},
    issn = {13652966},
    arxivId = {1703.02970}
}

@article{Pfrommer2017a,
    title = {{Simulating Gamma-Ray Emission in Star-forming Galaxies}},
    year = {2017},
    journal = {The Astrophysical Journal},
    author = {Pfrommer, Christoph and Pakmor, Rüdiger and Simpson, Christine M. and Springel, Volker},
    number = {2},
    month = {9},
    pages = {L13},
    volume = {847},
    publisher = {American Astronomical Society},
    url = {https://doi.org/10.3847/2041-8213/aa8bb1},
    doi = {10.3847/2041-8213/aa8bb1},
    issn = {2041-8213},
    arxivId = {1709.05343}
}

@article{Emerick2020,
    title = {{Simulating Metal Mixing of Both Common and Rare Enrichment Sources in a Low-mass Dwarf Galaxy}},
    year = {2020},
    journal = {The Astrophysical Journal},
    author = {Emerick, Andrew and Bryan, Greg L. and Low, Mordecai-Mark Mac},
    number = {2},
    month = {2},
    pages = {155},
    volume = {890},
    publisher = {IOP Publishing},
    url = {https://iopscience.iop.org/article/10.3847/1538-4357/ab6efc https://iopscience.iop.org/article/10.3847/1538-4357/ab6efc/meta},
    doi = {10.3847/1538-4357/AB6EFC},
    issn = {0004-637X},
    arxivId = {1909.04695}
}

@article{Booth2013,
    title = {{Simulations of disk galaxies with cosmic ray driven galactic winds}},
    year = {2013},
    journal = {Astrophysical Journal Letters},
    author = {Booth, C. M. and Agertz, Oscar and Kravtsov, Andrey V. and Gnedin, Nickolay Y.},
    number = {1},
    month = {10},
    pages = {L16},
    volume = {777},
    publisher = {IOP Publishing},
    url = {https://iopscience.iop.org/article/10.1088/2041-8205/777/1/L16 https://iopscience.iop.org/article/10.1088/2041-8205/777/1/L16/meta},
    doi = {10.1088/2041-8205/777/1/L16},
    issn = {20418205},
    arxivId = {1308.4974}
}

@article{Bullock2017,
    title = {{Small-Scale Challenges to the CDM Paradigm}},
    year = {2017},
    journal = {Annual Review of Astronomy and Astrophysics},
    author = {Bullock, James S and Boylan-Kolchin, Michael},
    pages = {343--387},
    volume = {55},
    url = {https://doi.org/10.1146/annurev-astro-091916-},
    doi = {10.1146/annurev-astro-091916}
}

@article{Rosdahl2017,
    title = {{Snap, crackle, pop: sub-grid supernova feedback in AMR simulations of disc galaxies}},
    year = {2017},
    journal = {Monthly Notices of the Royal Astronomical Society},
    author = {Rosdahl, Joakim and Schaye, Joop and Dubois, Yohan and Kimm, Taysun and Teyssier, Romain},
    number = {1},
    month = {4},
    pages = {11--33},
    volume = {466},
    publisher = {Oxford University Press},
    url = {https://academic.oup.com/mnras/article-lookup/doi/10.1093/mnras/stw3034},
    doi = {10.1093/mnras/stw3034},
    issn = {0035-8711}
}

@article{Girichidis2022,
    title = {{Spectrally resolved cosmic rays – II. Momentum-dependent cosmic ray diffusion drives powerful galactic winds}},
    year = {2022},
    journal = {Monthly Notices of the Royal Astronomical Society},
    author = {Girichidis, Philipp and Pfrommer, Christoph and Pakmor, Rüdiger and Springel, Volker},
    number = {3},
    month = {1},
    pages = {3917--3938},
    volume = {510},
    publisher = {Oxford Academic},
    url = {https://academic.oup.com/mnras/article/510/3/3917/6446005},
    doi = {10.1093/mnras/stab3462},
    issn = {0035-8711},
    arxivId = {2109.13250}
}

@article{Hopkins2022b,
    title = {{Standard self-confinement and extrinsic turbulence models for cosmic ray transport are fundamentally incompatible with observations}},
    year = {2022},
    journal = {Monthly Notices of the Royal Astronomical Society},
    author = {Hopkins, Philip F. and Squire, Jonathan and Butsky, Iryna S. and Ji, Suoqing},
    number = {4},
    month = {11},
    pages = {5413--5448},
    volume = {517},
    publisher = {Oxford Academic},
    url = {https://dx.doi.org/10.1093/mnras/stac2909},
    doi = {10.1093/MNRAS/STAC2909},
    issn = {0035-8711},
    arxivId = {2112.02153}
}

@article{Romano2023,
    title = {{Star-formation-driven outflows in local dwarf galaxies as revealed from [CII] observations by Herschel}},
    year = {2023},
    journal = {Astronomy {\&} Astrophysics},
    author = {Romano, M. and Nanni, A. and Donevski, D. and Ginolfi, M. and Jones, G. C. and Shivaei, I. and {Junais} and Salak, D. and Sawant, P.},
    month = {9},
    pages = {A44},
    volume = {677},
    publisher = {EDP Sciences},
    url = {https://www.aanda.org/articles/aa/full_html/2023/09/aa46143-23/aa46143-23.html https://www.aanda.org/articles/aa/abs/2023/09/aa46143-23/aa46143-23.html},
    doi = {10.1051/0004-6361/202346143},
    issn = {0004-6361},
    arxivId = {2306.10433}
}

@article{Strickland2000,
    title = {{Starburst-driven galactic winds - I. Energetics and intrinsic X-ray emission}},
    year = {2000},
    journal = {Monthly Notices of the Royal Astronomical Society},
    author = {Strickland, David K. and Stevens, Ian R.},
    number = {3},
    month = {5},
    pages = {511--545},
    volume = {314},
    publisher = {Blackwell Publishing Ltd},
    url = {https://dx.doi.org/10.1046/j.1365-8711.2000.03391.x},
    doi = {10.1046/J.1365-8711.2000.03391.X/2/314-3-511-FIG021.JPEG},
    issn = {00358711},
    arxivId = {astro-ph/0001395}
}

@article{Grudic2021,
    title = {{STARFORGE: Towards a comprehensive numerical model of star cluster formation and feedback}},
    year = {2021},
    journal = {Monthly Notices of the Royal Astronomical Society},
    author = {Grudi{\'{c}}, Michael Y. and Guszejnov, Dávid and Hopkins, Philip F. and Offner, Stella S.R. and Faucher-Gigu{\`{e}}re, Claude André},
    number = {2},
    month = {7},
    pages = {2199--2231},
    volume = {506},
    publisher = {Oxford Academic},
    url = {https://dx.doi.org/10.1093/mnras/stab1347},
    doi = {10.1093/mnras/stab1347},
    issn = {13652966},
    arxivId = {2010.11254}
}

@article{Hopkins2012a,
    title = {{Stellar feedback in galaxies and the origin of galaxy-scale winds}},
    year = {2012},
    journal = {Monthly Notices of the Royal Astronomical Society},
    author = {Hopkins, Philip F. and Quataert, Eliot and Murray, Norman},
    number = {4},
    month = {4},
    pages = {3522--3537},
    volume = {421},
    publisher = {Narnia},
    url = {https://academic.oup.com/mnras/article-lookup/doi/10.1111/j.1365-2966.2012.20593.x},
    doi = {10.1111/j.1365-2966.2012.20593.x},
    issn = {00358711}
}

@article{Fu2024b,
    title = {{Stellar Metallicities and Gradients in the Faint M31 Satellites Andromeda XVI and Andromeda XXVIII}},
    year = {2024},
    journal = {The Astrophysical Journal},
    author = {Fu, Sal Wanying and Weisz, Daniel R. and Starkenburg, Else and Martin, Nicolas and Collins, Michelle L. M. and Savino, Alessandro and Boylan-Kolchin, Michael and C{\^{o}}t{\'{e}}, Patrick and Dolphin, Andrew E. and Longeard, Nicolas and Mateo, Mario L. and Mercado, Francisco J. and Sandford, Nathan R. and Skillman, Evan D.},
    number = {1},
    month = {10},
    pages = {2},
    volume = {975},
    publisher = {IOP Publishing},
    url = {https://iopscience.iop.org/article/10.3847/1538-4357/ad76a2 https://iopscience.iop.org/article/10.3847/1538-4357/ad76a2/meta},
    doi = {10.3847/1538-4357/AD76A2},
    issn = {0004-637X}
}

@article{Fu2024a,
    title = {{Stellar Metallicities and Gradients in the Isolated, Quenched Low-mass Galaxy Tucana}},
    year = {2024},
    journal = {The Astrophysical Journal},
    author = {Fu, Sal Wanying and Weisz, Daniel R. and Starkenburg, Else and Martin, Nicolas and Mercado, Francisco J. and Savino, Alessandro and Boylan-Kolchin, Michael and C{\^{o}}t{\'{e}}, Patrick and Dolphin, Andrew E. and Longeard, Nicolas and Mateo, Mario L. and Samuel, Jenna and Sandford, Nathan R.},
    number = {1},
    month = {4},
    pages = {36},
    volume = {965},
    publisher = {IOP Publishing},
    url = {https://iopscience.iop.org/article/10.3847/1538-4357/ad25ed https://iopscience.iop.org/article/10.3847/1538-4357/ad25ed/meta},
    doi = {10.3847/1538-4357/AD25ED},
    issn = {0004-637X}
}

@article{Stanway2016,
    title = {{Stellar population effects on the inferred photon density at reionization}},
    year = {2016},
    journal = {Monthly Notices of the Royal Astronomical Society},
    author = {Stanway, Elizabeth R. and Eldridge, J. J. and Becker, George D.},
    number = {1},
    month = {2},
    pages = {485--499},
    volume = {456},
    publisher = {Oxford University Press},
    url = {https://academic.oup.com/mnras/article/456/1/485/1068024},
    doi = {10.1093/mnras/stv2661},
    issn = {13652966},
    arxivId = {1511.03268}
}

@article{Emerick2018,
    title = {{Stellar Radiation Is Critical for Regulating Star Formation and Driving Outflows in Low-mass Dwarf Galaxies}},
    year = {2018},
    journal = {The Astrophysical Journal},
    author = {Emerick, Andrew and Bryan, Greg L. and Mac Low, Mordecai-Mark},
    number = {2},
    month = {10},
    pages = {L22},
    volume = {865},
    publisher = {IOP Publishing},
    url = {https://iopscience.iop.org/article/10.3847/2041-8213/aae315 https://iopscience.iop.org/article/10.3847/2041-8213/aae315/meta},
    doi = {10.3847/2041-8213/aae315},
    issn = {2041-8205},
    arxivId = {1808.00468}
}

@article{Morlino2012,
    title = {{Strong evidence for hadron acceleration in Tycho's supernova remnant}},
    year = {2012},
    journal = {Astronomy and Astrophysics},
    author = {Morlino, G. and Caprioli, D.},
    month = {2},
    pages = {A81},
    volume = {538},
    publisher = {EDP Sciences},
    url = {http://www.aanda.org/10.1051/0004-6361/201117855},
    doi = {10.1051/0004-6361/201117855},
    issn = {00046361}
}

@article{Martizzi2015,
    title = {{Supernova feedback in an inhomogeneous interstellar medium}},
    year = {2015},
    journal = {MNRAS},
    author = {Martizzi, Davide and Faucher, Claude-André and Ere, Gigù and Quataert, Eliot},
    pages = {504--522},
    volume = {450},
    url = {http://fire.northwestern.edu.},
    isbn = {450/1/504/998665},
    doi = {10.1093/mnras/stv562}
}

@article{Smith2018,
    title = {{Supernova feedback in numerical simulations of galaxy formation: Separating physics from numerics}},
    year = {2018},
    journal = {Monthly Notices of the Royal Astronomical Society},
    author = {Smith, Matthew C and Sijacki, Debora and Shen, Sijing},
    number = {1},
    month = {7},
    pages = {302--331},
    volume = {478},
    publisher = {Oxford University Press},
    url = {https://academic.oup.com/mnras/article/478/1/302/4980956},
    doi = {10.1093/mnras/sty994},
    issn = {13652966},
    pmid = {21630094},
    arxivId = {1709.03515}
}

@article{Ponnada2024,
    title = {{Synchrotron signatures of cosmic ray transport physics in galaxies}},
    year = {2024},
    journal = {Monthly Notices of the Royal Astronomical Society: Letters},
    author = {Ponnada, Sam B. and Butsky, Iryna S. and Skalidis, Raphael and Hopkins, Philip F. and Panopoulou, Georgia V. and Hummels, Cameron and Kere{\v{s}}, Dušan and Quataert, Eliot and Faucher-Gigu{\'{e}}re, Claude André and Su, Kung Yi},
    number = {1},
    month = {3},
    pages = {L1-L6},
    volume = {530},
    publisher = {Oxford Academic},
    url = {https://dx.doi.org/10.1093/mnrasl/slae017},
    doi = {10.1093/MNRASL/SLAE017},
    issn = {1745-3925},
    arxivId = {2309.16752}
}

@article{Sawala2016,
    title = {{The APOSTLE simulations: Solutions to the Local Group's cosmic puzzles}},
    year = {2016},
    journal = {Monthly Notices of the Royal Astronomical Society},
    author = {Sawala, Till and Frenk, Carlos S. and Fattahi, Azadeh and Navarro, Julio F. and Bower, Richard G. and Crain, Robert A. and Dalla Vecchia, Claudio and Furlong, Michelle and Helly, John C. and Jenkins, Adrian and Oman, Kyle A. and Schaller, Matthieu and Schaye, Joop and Theuns, Tom and Trayford, James and White, Simon D.M.},
    number = {2},
    month = {4},
    pages = {1931--1943},
    volume = {457},
    publisher = {Oxford University Press},
    doi = {10.1093/mnras/stw145},
    issn = {13652966}
}

@article{Wise2012,
    title = {{The birth of a galaxy: Primordial metal enrichment and stellar populations}},
    year = {2012},
    journal = {Astrophysical Journal},
    author = {Wise, John H. and Turk, Matthew J. and Norman, Michael L. and Abel, Tom},
    number = {1},
    month = {1},
    pages = {50},
    volume = {745},
    publisher = {IOP Publishing},
    url = {http://stacks.iop.org/0004-637X/745/i=1/a=50?key=crossref.12736ba7b909322f6f02152f3faf89bd},
    doi = {10.1088/0004-637X/745/1/50},
    issn = {15384357},
    arxivId = {1011.2632}
}

@article{Zahid2013,
    title = {{THE CHEMICAL EVOLUTION OF STAR-FORMING GALAXIES OVER THE LAST 11 BILLION YEARS}},
    year = {2013},
    journal = {The Astrophysical Journal Letters},
    author = {Zahid, H. Jabran and Geller, Margaret J. and Kewley, Lisa J. and Hwang, Ho Seong and Fabricant, Daniel G. and Kurtz, Michael J.},
    number = {2},
    month = {6},
    pages = {L19},
    volume = {771},
    publisher = {IOP Publishing},
    url = {https://iopscience.iop.org/article/10.1088/2041-8205/771/2/L19 https://iopscience.iop.org/article/10.1088/2041-8205/771/2/L19/meta},
    doi = {10.1088/2041-8205/771/2/L19},
    issn = {2041-8205},
    arxivId = {1303.5987}
}

@article{Tumlinson2017,
    title = {{The Circumgalactic Medium}},
    year = {2017},
    journal = {Annual Review of Astronomy and Astrophysics},
    author = {Tumlinson, Jason and Peeples, Molly S. and Werk, Jessica K.},
    number = {Volume 55, 2017},
    month = {8},
    pages = {389--432},
    volume = {55},
    publisher = {Annual Reviews Inc.},
    url = {https://www.annualreviews.org/content/journals/10.1146/annurev-astro-091916-055240},
    doi = {10.1146/annurev-astro-091916-055240},
    issn = {00664146},
    arxivId = {1709.09180}
}

@article{Werk2016,
    title = {{THE COS-HALOS SURVEY: ORIGINS OF THE HIGHLY IONIZED CIRCUMGALACTIC MEDIUM OF STAR-FORMING GALAXIES}},
    year = {2016},
    journal = {The Astrophysical Journal},
    author = {Werk, Jessica K. and Prochaska, J. Xavier and Cantalupo, Sebastiano and Fox, Andrew J. and Oppenheimer, Benjamin and Tumlinson, Jason and Tripp, Todd M. and Lehner, Nicolas and McQuinn, Matthew},
    number = {1},
    month = {12},
    pages = {54},
    volume = {833},
    publisher = {IOP Publishing},
    url = {https://iopscience.iop.org/article/10.3847/1538-4357/833/1/54 https://iopscience.iop.org/article/10.3847/1538-4357/833/1/54/meta},
    doi = {10.3847/1538-4357/833/1/54},
    issn = {0004-637X},
    arxivId = {1609.00012}
}

@article{Grudic2022,
    title = {{The dynamics and outcome of star formation with jets, radiation, winds, and supernovae in concert}},
    year = {2022},
    journal = {Monthly Notices of the Royal Astronomical Society},
    author = {Grudi{\'{c}}, Michael Y. and Guszejnov, Dávid and Offner, Stella S.R. and Rosen, Anna L. and Raju, Aman N. and Faucher-Gigu{\`{e}}re, Claude André and Hopkins, Philip F.},
    number = {1},
    month = {3},
    pages = {216--232},
    volume = {512},
    publisher = {Oxford Academic},
    url = {https://dx.doi.org/10.1093/mnras/stac526},
    doi = {10.1093/MNRAS/STAC526},
    issn = {0035-8711},
    arxivId = {2201.00882}
}

@article{Schaye2015,
    title = {{The EAGLE project: Simulating the evolution and assembly of galaxies and their environments}},
    year = {2015},
    journal = {Monthly Notices of the Royal Astronomical Society},
    author = {Schaye, Joop and Crain, Robert A. and Bower, Richard G. and Furlong, Michelle and Schaller, Matthieu and Theuns, Tom and Dalla Vecchia, Claudio and Frenk, Carlos S. and Mccarthy, I. G. and Helly, John C. and Jenkins, Adrian and Rosas-Guevara, Y. M. and White, Simon D.M. and Baes, Maarten and Booth, C. M. and Camps, Peter and Navarro, Julio F. and Qu, Yan and Rahmati, Alireza and Sawala, Till and Thomas, Peter A. and Trayford, James},
    number = {1},
    month = {1},
    pages = {521--554},
    volume = {446},
    publisher = {Oxford Academic},
    url = {https://dx.doi.org/10.1093/mnras/stu2058},
    doi = {10.1093/mnras/stu2058},
    issn = {13652966},
    arxivId = {1407.7040}
}

@article{vandeVoort2021,
    title = {{The effect of magnetic fields on properties of the circumgalactic medium}},
    year = {2021},
    journal = {Monthly Notices of the Royal Astronomical Society},
    author = {Van De Voort, Freeke and Bieri, Rebekka and Pakmor, Rüdiger and G{\'{o}}mez, Facundo A and Grand, Robert J.J. and Marinacci, Federico},
    number = {4},
    pages = {4888--4902},
    volume = {501},
    url = {https://wwwmpa.mpa-garching.mpg.de/auriga/},
    doi = {10.1093/mnras/staa3938},
    issn = {13652966},
    arxivId = {2008.07537}
}

@article{Eldridge2008,
    title = {{The effect of massive binaries on stellar populations and supernova progenitors}},
    year = {2008},
    journal = {Monthly Notices of the Royal Astronomical Society},
    author = {Eldridge, John J. and Izzard, Robert G. and Tout, Christopher A.},
    number = {3},
    month = {3},
    pages = {1109--1118},
    volume = {384},
    publisher = {Oxford Academic},
    url = {https://academic.oup.com/mnras/article-lookup/doi/10.1111/j.1365-2966.2007.12738.x},
    doi = {10.1111/j.1365-2966.2007.12738.x},
    issn = {00358711}
}

@article{Buck2020,
    title = {{The effects of cosmic rays on the formation of Milky Way-mass galaxies in a cosmological context}},
    year = {2020},
    journal = {Monthly Notices of the Royal Astronomical Society},
    author = {Buck, Tobias and Pfrommer, Christoph and Pakmor, Rüdiger and Grand, Robert J.J. and Springel, Volker},
    number = {2},
    month = {10},
    pages = {1712--1737},
    volume = {497},
    url = {http://arxiv.org/abs/1911.00019},
    doi = {10.1093/mnras/staa1960},
    issn = {13652966},
    arxivId = {1911.00019}
}

@article{Walch2015,
    title = {{The energy and momentum input of supernova explosions in structured and ionized molecular clouds}},
    year = {2015},
    journal = {Monthly Notices of the Royal Astronomical Society},
    author = {Walch, Stefanie and Naab, Thorsten},
    number = {3},
    month = {8},
    pages = {2757--2771},
    volume = {451},
    publisher = {Oxford Academic},
    url = {https://academic.oup.com/mnras/article/451/3/2757/1193887},
    doi = {10.1093/mnras/stv1155},
    issn = {13652966},
    arxivId = {1410.0011}
}

@article{Torrey2019,
    title = {{The evolution of the mass-metallicity relation and its scatter in IllustrisTNG}},
    year = {2019},
    journal = {Monthly Notices of the Royal Astronomical Society},
    author = {Torrey, Paul and Vogelsberger, Mark and Marinacci, Federico and Pakmor, Rüdiger and Springel, Volker and Nelson, Dylan and Naiman, Jill and Pillepich, Annalisa and Genel, Shy and Weinberger, Rainer and Hernquist, Lars},
    pages = {5587--5607},
    volume = {484},
    url = {https://academic.oup.com/mnras/article-abstract/484/4/5587/5299581},
    doi = {10.1093/mnras/stz243},
    issn = {0035-8711}
}

@article{Chemerynska2024,
    title = {{The Extreme Low-mass End of the Mass–Metallicity Relation at z ∼ 7}},
    year = {2024},
    journal = {The Astrophysical Journal Letters},
    author = {Chemerynska, Iryna and Atek, Hakim and Dayal, Pratika and Furtak, Lukas J. and Feldmann, Robert and Greene, Jenny E. and Maseda, Michael V. and Nanayakkara, Themiya and Oesch, Pascal A. and Fujimoto, Seiji and Labb{\'{e}}, Ivo and Bezanson, Rachel and Brammer, Gabriel and Cutler, Sam E. and Leja, Joel and Pan, Richard and Price, Sedona H. and Wang, Bingjie and Weaver, John R. and Whitaker, Katherine E.},
    number = {1},
    month = {11},
    pages = {L15},
    volume = {976},
    publisher = {IOP Publishing},
    url = {https://iopscience.iop.org/article/10.3847/2041-8213/ad8dc9 https://iopscience.iop.org/article/10.3847/2041-8213/ad8dc9/meta},
    doi = {10.3847/2041-8213/AD8DC9},
    issn = {2041-8205},
    arxivId = {2407.17110}
}

@article{Dacunha2025,
    title = {{The Fallibility of Equipartition Magnetic Field Strengths from Synchrotron Emission Using Synthetically Observed Galaxies}},
    year = {2025},
    journal = {The Astrophysical Journal},
    author = {Dacunha, Tara and Martin-Alvarez, Sergio and Clark, Susan E. and Lopez-Rodriguez, Enrique},
    number = {2},
    month = {2},
    pages = {197},
    volume = {980},
    publisher = {IOP Publishing},
    url = {https://iopscience.iop.org/article/10.3847/1538-4357/adab72 https://iopscience.iop.org/article/10.3847/1538-4357/adab72/meta},
    doi = {10.3847/1538-4357/ADAB72},
    issn = {0004-637X},
    arxivId = {2409.08437}
}

@article{Schneider2012,
    title = {{The first low-mass stars: Critical metallicity or dust-to-gas ratio?}},
    year = {2012},
    journal = {Monthly Notices of the Royal Astronomical Society},
    author = {Schneider, Raffaella and Omukai, Kazuyuki and Bianchi, Simone and Valiante, Rosa},
    number = {2},
    month = {1},
    pages = {1566--1575},
    volume = {419},
    publisher = {Oxford Academic},
    url = {https://academic.oup.com/mnras/article/419/2/1566/989396},
    doi = {10.1111/j.1365-2966.2011.19818.x},
    issn = {00358711}
}

@article{Santistevan2020,
    title = {{The formation times and building blocks of Milky Way-mass galaxies in the FIRE simulations}},
    year = {2020},
    journal = {Monthly Notices of the Royal Astronomical Society},
    author = {Santistevan, Isaiah B. and Wetzel, Andrew and El-Badry, Kareem and Bland-Hawthorn, Joss and Boylan-Kolchin, Michael and Bailin, Jeremy and Faucher-Gigu{\`{e}}re, Claude Andre and Benincasa, Samantha},
    number = {1},
    month = {9},
    pages = {747--764},
    volume = {497},
    publisher = {Oxford Academic},
    url = {https://dx.doi.org/10.1093/mnras/staa1923},
    doi = {10.1093/MNRAS/STAA1923},
    issn = {0035-8711},
    arxivId = {2001.03178}
}

@article{Springel2003b,
    title = {{The history of star formation in a {$\Lambda$} cold dark matter universe}},
    year = {2003},
    journal = {Monthly Notices of the Royal Astronomical Society},
    author = {Springel, Volker and Hernquist, Lars},
    number = {2},
    month = {2},
    pages = {312--334},
    volume = {339},
    publisher = {Oxford Academic},
    url = {https://academic.oup.com/mnras/article/339/2/312/1003930},
    doi = {10.1046/j.1365-8711.2003.06207.x},
    issn = {00358711}
}

@article{Rasera2006,
    title = {{The history of the baryon budget}},
    year = {2006},
    journal = {Astronomy {\&} Astrophysics},
    author = {Rasera, Y. and Teyssier, R.},
    number = {1},
    month = {1},
    pages = {1--27},
    volume = {445},
    publisher = {EDP Sciences},
    url = {http://www.aanda.org/10.1051/0004-6361:20053116},
    doi = {10.1051/0004-6361:20053116},
    issn = {0004-6361}
}

@article{Bahe2017,
    title = {{The Hydrangea simulations: galaxy formation in and around massive clusters}},
    year = {2017},
    journal = {Monthly Notices of the Royal Astronomical Society},
    author = {Bah{\'{e}}, Yannick M. and Barnes, David J. and Vecchia, Claudio Dalla and Kay, Scott T. and White, Simon D.M. and McCarthy, Ian G. and Schaye, Joop and Bower, Richard G. and Crain, Robert A. and Theuns, Tom and Jenkins, Adrian and McGee, Sean L. and Schaller, Matthieu and Thomas, Peter A. and Trayford, James W.},
    number = {4},
    month = {10},
    pages = {4186--4208},
    volume = {470},
    publisher = {Oxford Academic},
    url = {https://dx.doi.org/10.1093/mnras/stx1403},
    doi = {10.1093/MNRAS/STX1403},
    issn = {0035-8711},
    arxivId = {1703.10610}
}

@article{Farcy2025,
    title = {{The impact of cosmic ray feedback during the epoch of reionisation}},
    year = {2025},
    journal = {Astronomy {\&} Astrophysics},
    author = {Farcy, Marion and Rosdahl, Joakim and Dubois, Yohan and Blaizot, Jérémy and Martin-Alvarez, Sergio and Haehnelt, Martin and Kimm, Taysun and Teyssier, Romain},
    month = {6},
    pages = {A89},
    volume = {698},
    publisher = {EDP Sciences},
    url = {https://www.aanda.org/articles/aa/full_html/2025/06/aa53924-25/aa53924-25.html https://www.aanda.org/articles/aa/abs/2025/06/aa53924-25/aa53924-25.html},
    doi = {10.1051/0004-6361/202553924},
    issn = {0004-6361},
    arxivId = {2501.17239}
}

@article{Curro2024,
    title = {{The impact of cosmic rays on the interstellar medium and galactic outflows of Milky Way analogues}},
    year = {2024},
    journal = {Monthly Notices of the Royal Astronomical Society},
    author = {Rodr{\'{i}}guez Montero, Francisco and Martin-Alvarez, Sergio and Slyz, Adrianne and Devriendt, Julien and Dubois, Yohan and Sijacki, Debora},
    number = {4},
    month = {5},
    pages = {3617--3640},
    volume = {530},
    publisher = {Oxford Academic},
    url = {https://dx.doi.org/10.1093/mnras/stae1083},
    doi = {10.1093/MNRAS/STAE1083},
    issn = {0035-8711},
    arxivId = {2307.13733}
}

@article{Ji2018,
    title = {{The impact of magnetic fields on thermal instability}},
    year = {2018},
    journal = {Monthly Notices of the Royal Astronomical Society},
    author = {Ji, Suoqing and Oh, O. Peng and McCourt, Michael},
    number = {1},
    month = {5},
    pages = {852--867},
    volume = {476},
    publisher = {Oxford University Press},
    url = {https://academic.oup.com/mnras/article/476/1/852/4839008},
    doi = {10.1093/mnras/sty293},
    issn = {13652966}
}

@article{Power2003,
    title = {{The inner structure of CDM haloes -- I. A numerical convergence study}},
    year = {2003},
    journal = {Monthly Notices of the Royal Astronomical Society},
    author = {Power, C. and Navarro, J. F. and Jenkins, A. and Frenk, C. S. and White, S. D. M. and Springel, V. and Stadel, J. and Quinn, T.},
    number = {1},
    month = {1},
    pages = {14--34},
    volume = {338},
    publisher = {Oxford University Press},
    url = {https://academic.oup.com/mnras/article-lookup/doi/10.1046/j.1365-8711.2003.05925.x},
    doi = {10.1046/j.1365-8711.2003.05925.x},
    issn = {0035-8711}
}

@article{Curti2020,
    title = {{The KLEVER Survey: Spatially resolved metallicity maps and gradients in a sample of 1.2 < z < 2.5 lensed galaxies}},
    year = {2020},
    journal = {Monthly Notices of the Royal Astronomical Society},
    author = {Curti, Mirko and Maiolino, Roberto and Cirasuolo, Michele and Mannucci, Filippo and Williams, Rebecca J. and Auger, Matt and Mercurio, Amata and Hayden-Pawson, Connor and Cresci, Giovanni and Marconi, Alessandro and Belfiore, Francesco and Cappellari, Michele and Cicone, Claudia and Cullen, Fergus and Meneghetti, Massimo and Ota, Kazuaki and Peng, Yingjie and Pettini, Max and Swinbank, Mark and Troncoso, Paulina},
    number = {1},
    month = {2},
    pages = {821--842},
    volume = {492},
    publisher = {Oxford University Press},
    url = {https://academic.oup.com/mnras/article/492/1/821/5681399},
    doi = {10.1093/mnras/stz3379},
    issn = {13652966},
    arxivId = {1910.13451}
}

@article{ForsterSchreiber2019,
    title = {{The KMOS3D Survey: Demographics and Properties of Galactic Outflows at z = 0.6–2.7*}},
    year = {2019},
    journal = {The Astrophysical Journal},
    author = {Forster Schreiber, N. M. Förster and {\"{U}}bler, H. and Davies, R. L. and Genzel, R. and Wisnioski, E. and Belli, S. and Shimizu, T. and Lutz, D. and Fossati, M. and Herrera-Camus, R. and Mendel, J. T. and Tacconi, L. J. and Wilman, D. and Beifiori, A. and Brammer, G. B. and Burkert, A. and Carollo, C. M. and Davies, R. I. and Eisenhauer, F. and Fabricius, M. and Lilly, S. J. and Momcheva, I. and Naab, T. and Nelson, E. J. and Price, S. H. and Renzini, A. and Saglia, R. and Sternberg, A. and Dokkum, P. van and Wuyts, S.},
    number = {1},
    month = {4},
    pages = {21},
    volume = {875},
    publisher = {IOP Publishing},
    url = {https://iopscience.iop.org/article/10.3847/1538-4357/ab0ca2 https://iopscience.iop.org/article/10.3847/1538-4357/ab0ca2/meta},
    doi = {10.3847/1538-4357/AB0CA2},
    issn = {0004-637X},
    arxivId = {1807.04738}
}

@article{Tumlinson2011,
    title = {{The large, oxygen-rich halos of star-forming galaxies are a major reservoir of galactic metals}},
    year = {2011},
    journal = {Science},
    author = {Tumlinson, J. and Thom, C. and Werk, J. K. and Prochaska, J. X. and Tripp, T. M. and Weinberg, D. H. and Peeples, M. S. and O'Meara, J. M. and Oppenheimer, B. D. and Meiring, J. D. and Katz, N. S. and Dav{\'{e}}, R. and Ford, A. B. and Sembach, K. R.},
    number = {6058},
    month = {11},
    pages = {948--952},
    volume = {334},
    publisher = {American Association for the Advancement of Science},
    url = {https://www.science.org/doi/10.1126/science.1209840},
    doi = {10.1126/science.1209840},
    issn = {10959203},
    arxivId = {1111.3980}
}

@article{Chisholm2017,
    title = {{The mass and momentum outflow rates of photoionized galactic outflows}},
    year = {2017},
    journal = {Monthly Notices of the Royal Astronomical Society},
    author = {Chisholm, John and Tremonti, Christy A. and Leitherer, Claus and Chen, Yanmei},
    number = {4},
    month = {8},
    pages = {4831--4849},
    volume = {469},
    publisher = {Oxford University Press},
    url = {https://academic.oup.com/mnras/article-lookup/doi/10.1093/mnras/stx1164},
    doi = {10.1093/mnras/stx1164},
    issn = {0035-8711}
}

@article{Erb2006,
    title = {{The Mass‐Metallicity Relation at z ≳2}},
    year = {2006},
    journal = {The Astrophysical Journal},
    author = {Erb, Dawn K. and Shapley, Alice E. and Pettini, Max and Steidel, Charles C. and Reddy, Naveen A. and Adelberger, Kurt L.},
    number = {2},
    month = {6},
    pages = {813--828},
    volume = {644},
    publisher = {IOP Publishing},
    doi = {10.1086/503623},
    issn = {0004-637X}
}

@article{Chisholm2016,
    title = {{THE MOLECULAR BARYON CYCLE OF M82}},
    year = {2016},
    journal = {The Astrophysical Journal},
    author = {Chisholm, John and Matsushita, Satoki},
    number = {2},
    month = {10},
    pages = {72},
    volume = {830},
    publisher = {IOP Publishing},
    url = {https://iopscience.iop.org/article/10.3847/0004-637X/830/2/72 https://iopscience.iop.org/article/10.3847/0004-637X/830/2/72/meta},
    doi = {10.3847/0004-637X/830/2/72},
    issn = {0004-637X},
    arxivId = {1608.00974}
}

@article{Leroy2015,
    title = {{THE MULTI-PHASE COLD FOUNTAIN IN M82 REVEALED BY A WIDE, SENSITIVE MAP OF THE MOLECULAR INTERSTELLAR MEDIUM}},
    year = {2015},
    journal = {The Astrophysical Journal},
    author = {Leroy, Adam K. and Walter, Fabian and Martini, Paul and Roussel, Hélene and Sandstrom, Karin and Ott, Jürgen and Weiss, Axel and Bolatto, Alberto D. and Schuster, Karl and Dessauges-Zavadsky, Miroslava},
    number = {2},
    month = {11},
    pages = {83},
    volume = {814},
    publisher = {IOP Publishing},
    url = {https://iopscience.iop.org/article/10.1088/0004-637X/814/2/83 https://iopscience.iop.org/article/10.1088/0004-637X/814/2/83/meta},
    doi = {10.1088/0004-637X/814/2/83},
    issn = {0004-637X}
}

@article{Brooks2007,
    title = {{The Origin and Evolution of the Mass-Metallicity Relationship for Galaxies: Results from Cosmological N -Body Simulations}},
    year = {2007},
    journal = {The Astrophysical Journal},
    author = {Brooks, A. M. and Governato, F. and Booth, C. M. and Willman, B. and Gardner, J. P. and Wadsley, J. and Stinson, G. and Quinn, T.},
    number = {1},
    month = {1},
    pages = {L17-L20},
    volume = {655},
    publisher = {American Astronomical Society},
    url = {https://iopscience.iop.org/article/10.1086/511765 https://iopscience.iop.org/article/10.1086/511765/meta},
    doi = {10.1086/511765},
    issn = {0004-637X},
    arxivId = {astro-ph/0609620}
}

@article{Tremonti2004,
    title = {{The Origin of the Mass‐Metallicity Relation: Insights from 53,000 Star‐forming Galaxies in the Sloan Digital Sky Survey}},
    year = {2004},
    journal = {The Astrophysical Journal},
    author = {Tremonti, Christy A. and Heckman, Timothy M. and Kauffmann, Guinevere and Brinchmann, Jarle and Charlot, Stephane and White, Simon D. M. and Seibert, Mark and Peng, Eric W. and Schlegel, David J. and Uomoto, Alan and Fukugita, Masataka and Brinkmann, Jon},
    number = {2},
    month = {10},
    pages = {898--913},
    volume = {613},
    publisher = {IOP Publishing},
    doi = {10.1086/423264},
    issn = {0004-637X}
}

@article{Schmidt1959,
    title = {{The Rate of Star Formation.}},
    year = {1959},
    journal = {The Astrophysical Journal},
    author = {Schmidt, Maarten},
    month = {3},
    pages = {243},
    volume = {129},
    url = {http://adsabs.harvard.edu/doi/10.1086/146614},
    doi = {10.1086/146614},
    issn = {0004-637X}
}

\appendix
\section{On the computation of thermal and non-thermal pressure components}
\label{ap:pressures}

We compute the standard thermal pressure assuming an ideal gas law: 
\begin{equation} 
P_\text{th} = (\gamma - 1)\,\rho_\text{gas} \epsilon, 
\label{eq:pthermal}
\end{equation}
where $\epsilon$ represents the internal energy per unit mass, and assuming a $\gamma = 5/3$ corresponding to a monoatomic ideal gas.

For the magnetic pressure, we assume an isotropic force distribution, estimated simply as
\begin{equation} P_\text{B} = \frac{B^2}{8\pi},
\label{eq:pmagnetic}
\end{equation}
and where $B$ is the total strength of the magnetic field. 

The CR pressure is also accessible in our models with CRs the simulation, computed as
\begin{equation}
P_\text{CR} = (\gamma_\text{CR} - 1)\, e_\text{CR}.
\label{eq:pcr} 
\end{equation}
Here, $e_\text{CR}$ is the CR energy density, with $\gamma_\text{CR} = 4/3$ corresponding to a relativistic fluid. To explore the importance of radiation, we follow \citet{Rosdahl2015b} and estimate the photo-heating pressure $p_\text{photoheat}$ due to the photoionisation of the hydrogen gas
\begin{equation}
\frac{p_\text{photoheat}}{\rho_\text{gas} k_\text{B}} \sim T^\text{ph} \sim 2 \cdot 10^4 \Kelvin \,\text{min}\left(f_\text{vol}, 1\right)
\label{eq:photoheat}    
\end{equation}
where $f_\text{vol}$ is the fraction of the cell volume that is covered by the Strömgrem sphere with Strömgrem radius $r_S$
\begin{equation}
f_\text{vol} = \frac{1}{(\Delta x)^3}\frac{4 \pi r_S^3}{3} = \frac{1}{(\Delta x)^3}\frac{4 \pi}{3} \frac{3 \hat{L}_\text{UV}}{4 \pi \alpha^B n_H^2},
\label{eq:fracvol}    
\end{equation}
with $n_H$ the hydrogen number density and $\alpha^B = 2.6 \cdot 10^{-13}\,\cm^3\,\sec^{-1}$ the case B recombination of hydrogen for $T \sim 2 \cdot 10^4\,\Kelvin$ \citep{Ferland1992}. To measure $f_\text{vol}$ for a given cell, we approximate the UV photon luminosity $\hat{L}_\text{UV}$ by using the ionising photons density $\hat{\epsilon}_\text{UV}$ in flux units. Thus eq. (\ref{eq:fracvol}) becomes
\begin{equation}
f_\text{vol} = \frac{1}{\Delta x}\frac{4 \pi}{3} \frac{3\,c\,\hat{\epsilon}_\text{UV}}{4 \pi \alpha^B n_H^2},
\label{eq:fracvol2}    
\end{equation}
where $c$ is the speed of light. We also follow \citet[see their Section 4.2.2;][]{Rosdahl2015b} to estimate the direct pressure from photoionisation ($p_\text{UV}$) using again the same approximation for $\hat{L}_\text{UV}$. However, we do not include it in our profiles as we find it to be negligible at these radial scales ($p_\text{UV} \ll 10^{-30}\,\erg\,\cm^{-3}$), in agreement with previous works \citep{Rosdahl2015b,Emerick2020}. Note that this pressure might still be important in the regulation of star formation at smaller scales \citep{Wise2012}.

\section{On the variability of properties across time}
\label{ap:profile_variability}
Given the variability of galaxy internal properties across their evolution, and specifically with their star formation activity, in this appendix we provide additional context for interpreting the conditions described in the main text. Specifically, we compare the radial profiles at a given time to time-aggregated profiles. While these aggregates do not represent any specific single physical state, they describe the range and variability of conditions over time. This is particularly insightful when separated onto quiescent and star-forming subsets, with the star-forming being of particular interest for our study.

\begin{figure}
    \centering
    \includegraphics[width=\columnwidth]{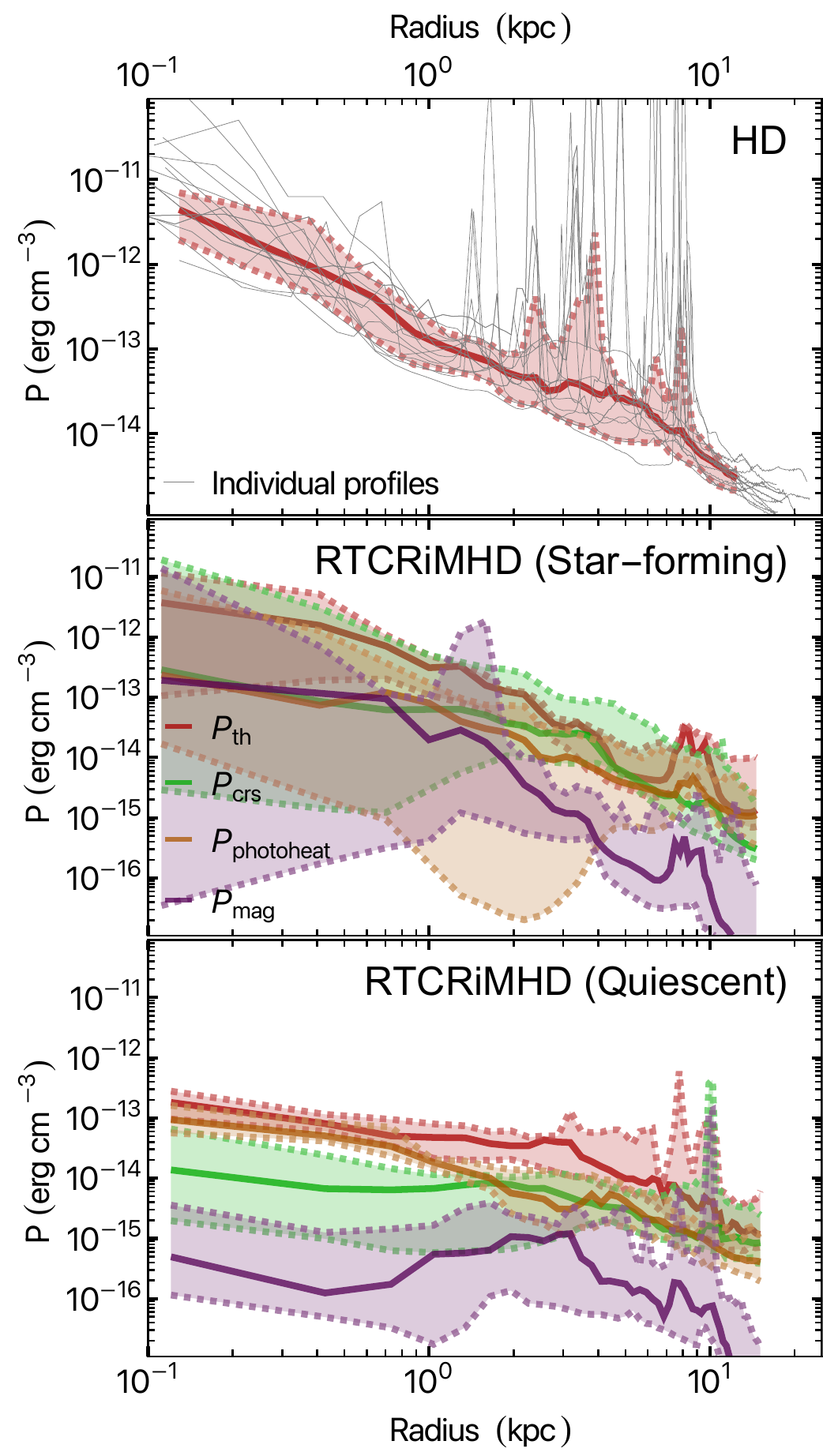}\\
    \caption{Median radial pressure profiles within the interval $z\,\in\,[5.0,\,3.5]$, shown as thick coloured curves. Shaded regions indicate the 16{--}84$^\text{th}$ percentiles. 
    The panels, from top to bottom, show: (top) the \HDSfFb~model, (centre) the full-physics model during high star formation periods, and (bottom) the full-physics run during quiescent periods. Pressure components are thermal ($P_{\rm th}$, red), cosmic-ray ($P_{\rm crs}$, green), photo-heating ($P_{\rm photoheat}$, orange), and magnetic ($P_{\rm mag}$, purple). We include as thin grey lines in the \HDSfFb~panel, $P_{\rm th}$ profiles for each snapshot. The Fig.~\ref{fig:OutflowProfs} snapshots are within the 1~$\sigma$ bands of the \HDSfFb~panel, and of the star-forming \RTCRiMHDSfFb~panel. The only deviations appear in peaks corresponding to satellite galaxies.}
    \label{fig:profile_variability}
\end{figure}

Figure~\ref{fig:profile_variability} shows the median radial pressure profiles averaging over time (solid curves) and their 1~$\sigma$ (16{--}84$^\text{th}$, shaded bands) envelopes. The top panel shows the standard hydrodynamics model (\HDSfFb) across the redshift interval $z\!\in\![5.0,\,3.4]$. The panel also shows the individual thermal pressure profiles as thin grey lines. Overall, the shaded band is relatively narrow, and captures well the variability of the profiles over time, with only individual peaks corresponding to satellites and companions deviating from this range. This is in agreement with a relatively continued star formation activity over time, approximately self-regulated by SN feedback.
Due to the large burstiness of the \RTCRiMHDSfFb~model, the model exhibits different conditions depending on the star formation rate of the galaxy. We show the profiles during either high sSFR (star-forming, sSFR $> 1\,\Gyr^{-1}$; central panel) or low sSFR (quiescent; bottom panel) periods. Overall, and focusing on the high sSFR period, we recover the main trends described in the main text. The non-thermal profiles exhibit wider envelopes than the thermal pressure, reflecting the higher variability of these components. This is specially notable for the magnetic energy, which only globally dominates at times in the innermost regions of the galaxy. Photoheating and cosmic rays remain subdominant in the galaxy, with their upper end of the 1~$\sigma$ bands comparable to the thermal pressure distribution. CRs and photoheating pressures are comparable to the thermal pressure at $r > 1\,\text{kpc}$, with photoheating dominating at large radii. The CR pressure bands reflect the discussed trend for this component to, at times, dominate the pressure of outflowing gas. Effectively, the CR pressure is comparable to (and overtakes at times) the thermal pressure in the $1\,\kpc\lesssim r \lesssim 4\,\text{kpc}$ range. The scaling of the median CR pressure in the $r \in [2, 15]\,\kpc$ regime is relatively insensitive to removing outflowing gas in the computation of this pressure, and remains approximately $\propto r^{-2} - r^{-3}$ 
Finally, during times of relative quiescence (sSFR $< 1\,\Gyr^{-1}$), all pressures in the galaxy fall well below the \HDSfFb~profiles. While this drop is partially driven by lower gas densities, we also find a relative drop in the non-thermal pressures contribution with respect to the thermal component. During these episodes, only the photoheating pressure remains important in the galaxy. At large radii, photoheating and CRs are comparable, and can contribute a non-negligible fraction of the total pressure. Magnetic energies remain subdominant globally, although we note that they are still locally important in some dense regions within the galaxy. 
Overall, while individual snapshots may feature profiles with small variations of the pressure distributions, the profiles studied in Fig.~\ref{fig:OutflowProfs}, and their main results are representative of the typical radial distribution of pressures, specifically during times of non-negligible star formation.

\section{Broad velocity component measurements as a proxy for galaxy outflows}
\label{ap:obs_like_measurements}

In this appendix, we provide additional information regarding our observation-like measurements of galaxy outflows. We adopt a simple approach to illustrate the potential impact of more sophisticated forward-modelling. Our method, following the observational procedure, is based on extracting information from the velocity distribution, which can be recovered observationally by studying emission line velocity profiles. Specifically, a common technique is to decompose emission line profiles into two Gaussian contributions, with the narrow component associated with the galaxy, and the broad component associated the galaxy's outflow \citep[e.g.,][]{Rupke2005, Marasco2023, Romano2023, Carniani2024}. The information about this broad component can be used to calculate the two main parameters required to estimate an outflow rate using Equation~(\ref{eq:obs_outflow_rate}). These are the outflow velocity, $\vbroad$, which is primarily connected with the FWHM of the Gaussian component, and the outflow mass, $\Moutflow$, which is affected by both its FWHM and its amplitude. 

To extract the broad component information from our simulations, we obtain for each snapshot the cumulative mass (or $L_{\text{H}\alpha}$) distribution function for the line-of-sight velocities across 12 directions. These are equidistantly distributed following a HEALPix sphere discretisation \citep{Gorski2005}. When computing the mass distribution function, gas cells contributing to this mass distribution function are filtered accordingly to whether the WIM or WNM component is being considered. We include in this calculation only cells within a distance less than $r < 0.1\,\rhalo$. We attempt both a fit with a single and a double Gaussian function. For the gas mass distribution functions, we determine whether the double Gaussian decomposition is valid following a simple Akaike Information Criterion (AIC), and discard the double Gaussian fit when it provides:
\begin{equation}
    \text{AIC}_\text{double} < \text{AIC}_\text{single} - 2.
\end{equation}
Due to the lower fraction of valid fits in our H$\alpha$ analysis, we relax this assumption to simply yielding at least a $10\%$ reduction of sum of squared residuals (SSR). We discard snapshots where the double Gaussian decomposition is invalid.

\begin{figure}
   \centering
   \includegraphics[width=\columnwidth]{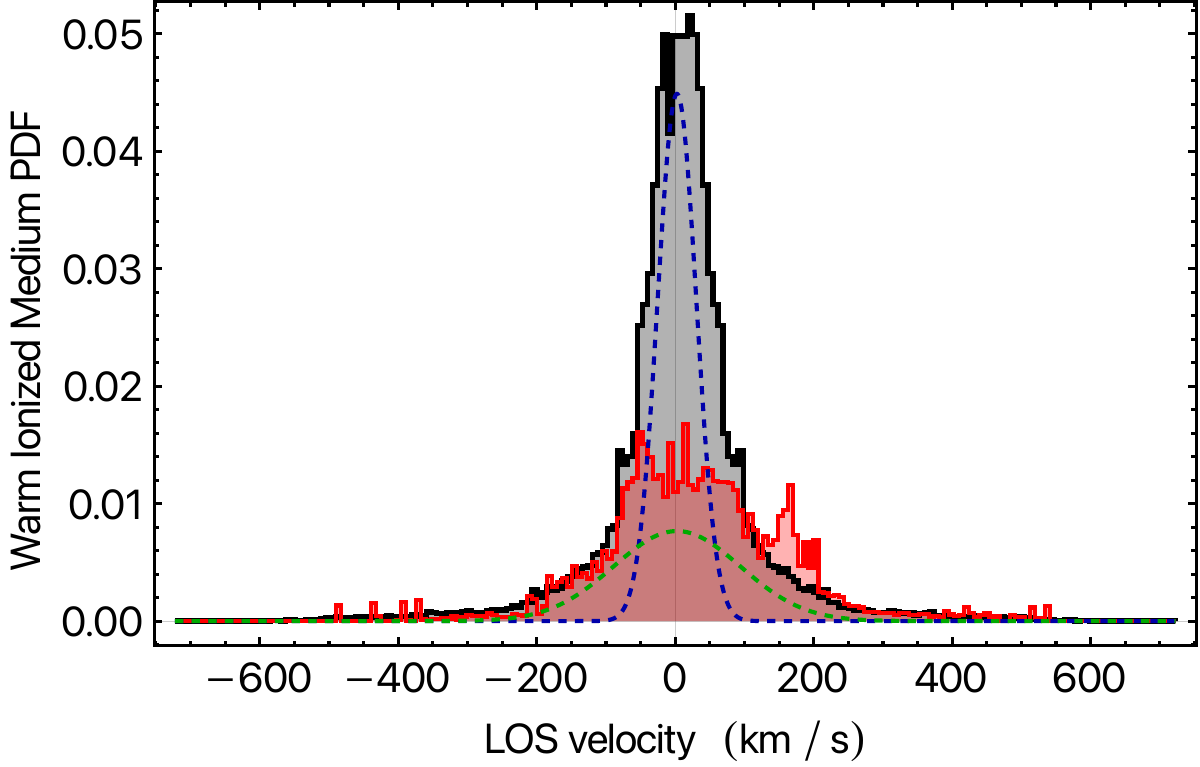}\\
   \caption{Sample LOS velocity distribution of WIM mass for a \RTCRiMHDSfFb~model output (also shown in Fig.~\ref{fig:front_image}) with a valid double Gaussian fit. Black and red histograms show all WIM gas and the outflowing component, respectively. The dashed blue and green lines show the narrow and broad Gaussian fits. The broad component effectively captures the outflowing gas, due to its roughly isotropic morphology.}
   \label{fig:OutflowPDF_2gauss}
\end{figure}

In Fig.~\ref{fig:OutflowPDF_2gauss} we show an example of the double Gaussian fit for the WIM velocity distribution of the \RTCRiMHDSfFb~model. It reflects how the broad component of the double Gaussian fit is able to capture the outflowing WIM mass (red histogram). This decomposition appears to be valid for our dwarf galaxies due to the relative isotropy of their outflows, although we note that inclination with respect to a face-on viewing line may be important for galaxies with coherent rotation \citep{Concas2022} and/or more complex outflow morphologies.

\begin{figure}
   \centering
   \includegraphics[width=\columnwidth]{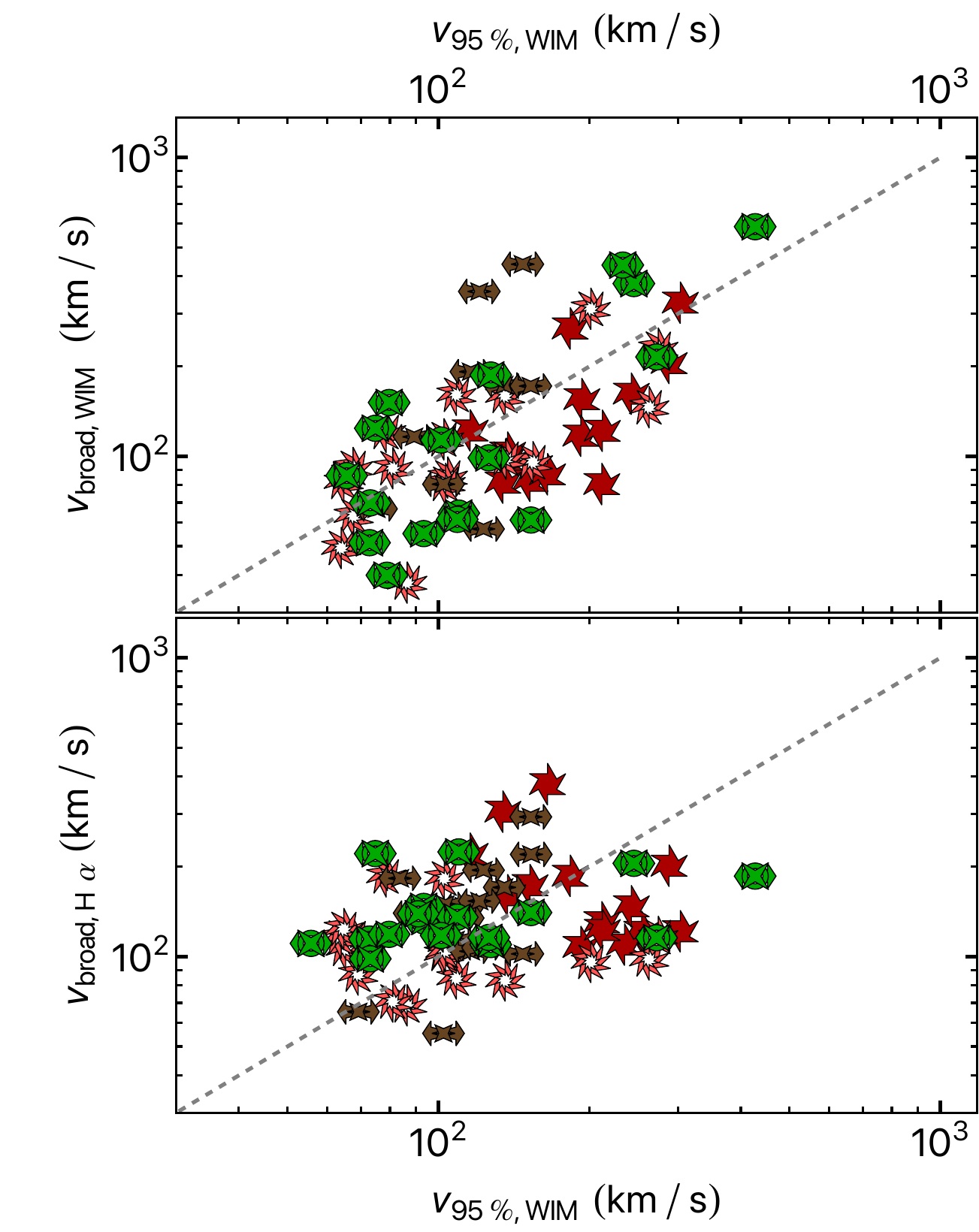}\\
   \rowlegend
   \caption{{\bf(Top panel)} Relation between the broad velocity estimate for high-velocity outflows, $v_\text{broad}$, and the mass-weighted velocity of gas with speeds above the $95^\text{th}$ percentile. {\bf(Bottom panel)} Same as the top panel, now employing the H$\alpha$ luminosity distribution of velocities to measure the broad velocity estimate. The relatively tight correlation between these two quantities suggests that $v_\text{broad}$ is a reasonable tracer of the $v_{95\%}$.}
   \label{fig:BroadVs95}
\end{figure}

The first quantity we extract from the broad component is the outflow velocity, $\vbroad$. This is typically interpreted as an observational proxy for the $95^\text{th}$ percentile of the velocity distribution. Still following the methodology from observational studies, we compute $\vbroad$ as described by Equation~(\ref{eq:Rupke2005}). To determine how this velocity relates to the high-velocity tail of the outflows in our simulations, we compare it with $v_{95\%}$. We measure $v_{95\%}$ directly, as the velocity marking the $95^\text{th}$ percentile of the mass-weighted velocity distribution. These two velocities are compared in the top panel of Fig.~\ref{fig:BroadVs95} for the WIM component, although we note that the WNM displays a similar relation. Overall, the $\vbroad$ traces the $v_{95\%}$ reasonably well, and does not display any particular bias for any of our models. As a result, we employ $\vbroad$ in our main analysis, which enables a more direct comparison across our observation-like results as well as with the observational ones. The bottom panel of the same figure compares $v_{95\%}$ with $\vHalpha$, which displays a similar correlation, although with a somewhat larger scatter, and with the highest velocities somewhat underestimated. Understanding why this is the case will require a larger sample of simulated galaxies. Velocity measurement variations are relatively small across our models, suggesting that for an individual galaxy, the uncertainty associated with the broad velocity does not significantly bias the measurement of the outflow rate, even for potentially large variations in physical modelling. Despite this, we note that, as discussed in the main text, $\vbroad$ (and $v_\text{95\%}$) is biased high when compared with the mass-weighted velocity of the outflowing gas frequently employed by simulations. Employing a luminosity-weighted value for the outflow velocity instead of using $v_\text{95\%}$ would likely decrease the estimated outflow rate by $\sim 1$ dex.

Estimating the outflowing mass is more challenging observationally, even when assuming that the broad component amplitude is well-recovered. Specifically, estimating the outflow mass requires of approximations such as those adopted in Equation~(\ref{eq:obs_outflow_mass}), where the electron number densities frequently cannot be directly inferred from (dwarf) galaxy observations. Instead, this is often set to $n_e \sim 300\,\text{cm}^{-3}$ as an extrapolation of the values retrieved from more massive galaxies and/or AGN-outflows. This measurement often relies on the [SII]$\lambda$6717\AA/[SII]$\lambda$6731\AA\ intensity ratio narrow sensitivity range ($100 - 2000\,\text{cm}^{-3}$) \citep{Baron2024}. Dwarf galaxy formation simulations such as those studied here typically feature lower average SN-driven outflow densities, on the order of a few $\text{H}\,\cm^{-3}$, and have clear density gradients. Assuming a comparable electron number density would lead to an increase of outflow mass rates estimates of $\sim1-3$~dex. 

From the observational side, a potential avenue to further bridge the gap between the methods typically employed by numerical simulations and observations would be assuming $n_e \sim 1 \cm^{-3}$, and measuring the outflow velocity as the amplitude-weighted norm of the velocity relative to the mean of the broad component. From the numerical side, our findings indicate that additional work addressing detailed forward-modelling of the outflow properties of simulated dwarf galaxies {--} especially leveraging a larger galaxy samples {--} will be fundamental to interpret the importance of these caveats, and establish a direct connection between the properties of simulated and observed outflows.

\section{The mass - metallicity relation across all \pandora~models}
\label{ap:metallicities_allmodels}

\begin{figure}
    \centering
    \includegraphics[width=\columnwidth]{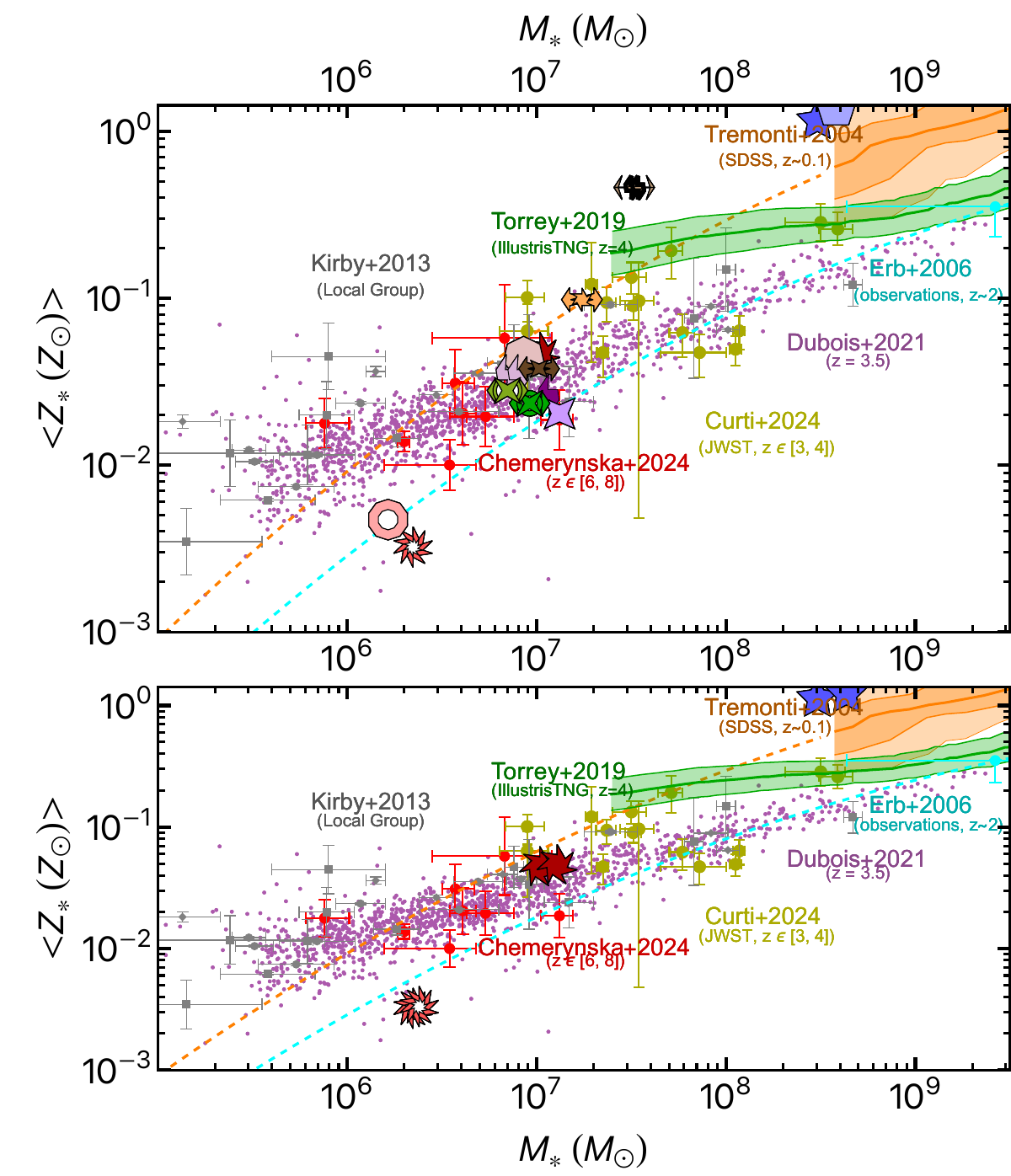}\\
    \allrowlegendtwoshort\\
    \caption{{\bf (Top panel)} Stellar mass - stellar metallicity relation, as shown in Fig.~\ref{fig:Metallicities2}, now for all the \pandora~dwarf galaxy simulated models.
    As in Fig.~\ref{fig:Metallicities2}, we include for comparison local dwarf galaxies stellar metallicity observations (gray points) by \citet{Kirby2013} and higher redshift observations employing \textit{JWST} by \citet{Curti2024} at $z \in [3, 4]$ (yellow points) and \citet{Chemerynska2024} at $z \in [6, 8]$ (red points). We include larger stellar masses SDSS observations at low \citep{Tremonti2004} (orange bands) and high \citep[$z \sim 2$,][]{Erb2006} redshift (cyan points and cyan dashed line). We also show data for the high-resolution NewHorizon \citep[$z = 3.5$, violet data points; ][]{Dubois2021} and IllustrisTNG \citep[$z = 4$, green band; ][]{Torrey2019} simulations for comparison.
    Most models provide a reasonable match to observations, except the `boosted' feedback model \HDSfFbBoost.
    {\bf (Bottom panel)} Same as the top panel, but now for the \HDSfNoFb, \HDSfFb~and \HDSfFbBoost~models at both $z \sim 3.5$ and $z \sim 0.5$, to showcase the lack of evolution after the main studied period.}
    \label{fig:all_metallicities}
\end{figure}

In this appendix, we provide additional results for the mass - metallicity relation in the \pandora~simulations in Fig.~\ref{fig:all_metallicities}. The top panel of the figure displays, for completeness, the mass - metallicity relation as measured for all the \pandora~models. The bottom panel shows the models that are evolved closer to $z = 0$ to underscore how the studied galaxy does not evolve further in the mass - metallicity space. Overall, the mass - metallicity relation for the additional models reinforces the conclusions discussed above. While the simulations employing a density threshold star formation model (\HDthSfNoFb, \HDthSfFb, and \HDthSfFbBoost) have important effects across various other galaxy properties \citep{Martin-Alvarez2023}, their final metallicities are comparable to their MTT star formation model counterparts (\HDSfNoFb, \HDSfFb, and \HDSfFbBoost, respectively). The inclusion of magnetic fields (\MHDSfFb, \iMHDSfFb) leads only to a minor variation in metallicity with respect to the standard hydrodynamical models (\HDthSfFb~and \HDSfFb). Their overall metallicity is slightly reduced as star formation becomes more concentrated \citep{Martin-Alvarez2020}, and the subsequent galactic outflows have a higher proportion of entrained metals, in an analogous manner to the \HDthSfFbBoost~and \HDSfFbBoost~models. In the model with stellar radiation in the absence of magnetic fields (\RTSfFb), a lower concentration of star formation only weakens the role of stellar feedback and leads to both a higher mass and stellar metallicity. \citet{Martin-Alvarez2023} described how various relations show that extreme primordial magnetisations (comparable or 1~dex below non-CMB constraints; \citealt{KMA2021, Pavicevic2025} {--} see also \citealt{Broderick2018}) are disfavoured by the \pandora~simulations in the presence of stellar radiation. We find this same apparent disagreement with both local and high redshift dwarf galaxies \citep{Kirby2013, Curti2024}. Finally, the {\it`full-physics'} model without CR streaming (\RTnsCRiMHDSfFb) features a comparable metallicity to the studied \RTCRiMHDSfFb~model, further supporting our main findings. 
Finally, focusing on the bottom panel of Fig.~\ref{fig:all_metallicities}, we confirm that when evolved closer to $z = 0$, the \pandora~galaxy undergoes very minor evolution in its stellar mass - metallicity relation. This lack of evolution holds regardless of the strength of stellar feedback, with the three displayed models spanning from `boosted' SN feedback to a model with no stellar feedback (only gas and metals mass return to the ISM), and matches the expectation for a very isolated dwarf galaxy with minor evolution and no additional mergers (see Appendix~A by \citealt{Martin-Alvarez2023})   

\bsp	
\label{lastpage}
\end{document}